\documentclass[a4paper,11pt]{article}
\usepackage{geometry}
\usepackage{a4wide}
\usepackage{graphicx}
\usepackage{epsf}
\usepackage{amsmath}
\usepackage{amssymb}

\usepackage{cite}
\usepackage{multirow,tabularx}
\usepackage{appendix}
\usepackage{tikz}
\usepackage{graphicx,amsmath,amsfonts,amssymb,amsthm,euscript,braket,xcolor}
\newcommand{\be}{\begin{equation}}
\newcommand{\ee}{\end{equation}}

\newcommand{\Rmnum}[1]{\expandafter\@slowromancap\romannumeral #1@}
\newcommand{\bea}{\begin{eqnarray}}
\newcommand{\eea}{\end{eqnarray}}

\numberwithin{equation}{section}

\newcommand*\circled[1]{\tikz[baseline=(char.base)]{
            \node[shape=circle,draw,inner sep=2pt] (char) {#1};}}

\begin{document}

\title{\bf Mixed state entanglement measures as probe for confinement}
\author{\textbf {Parul Jain$^{a}$}\thanks{paruljain@iitb.ac.in},
\textbf{Subhash Mahapatra$^{b}$}\thanks{mahapatrasub@nitrkl.ac.in},
 \\\\
\textit{{\small $^a$  Department of Physics, Indian Institute of Technology, Bombay - 400076, India}}\\
\textit{{\small $^b$ Department of Physics, National Institute of Technology Rourkela, Rourkela - 769008, India}}}
\date{}

\maketitle
\abstract{We study holographic aspects of mixed state entanglement measures in various large $N$ top-down as well as bottom-up confining models. For the top-down models, we consider wrapped $D3$ and $D4$ branes gravity solutions whereas, for the bottom-up confining model, the Einstein-Maxwell-dilaton gravity solution is considered. We study entanglement entropy, mutual information, entanglement wedge cross-section and entanglement negativity for the strip subsystems and find model independent features of these entanglement measures in all confining theories. The entanglement negativity and entropy exhibit a phase transition at the same critical strip length $L_{crit}$, at which the order of these measures changes from $\mathcal{O}(N^2)$ to $\mathcal{O}(N^0)$. The entanglement wedge cross-section similarly shows an order change at $L_{crit}$ and exhibits a discontinuous behaviour each time a phase transition between different entangling surfaces occur. We further test the inequality involving mutual information and entanglement wedge cross-section, and find that the latter always exceeds half of the former.}


\section{Introduction}
\label{sec1}
Quantum entanglement has recently emerged as an interesting and powerful tool to investigate diverse aspects in theoretical physics, extending from condensed matter to quantum gravity. One of the most commonly used entanglement measures is the entanglement entropy. In the last few decades, entanglement entropy  has been extensively studied, for example, in condensed matter physics to characterize different quantum phases \cite{Vidal:2002rm,Osborne:2002zz}, in black hole physics to better understand the Bekenstein-Hawking entropy \cite{Bombelli:1986rw,Srednicki:1993im}, in quantum communications \cite{Hoi,Karpov} etc. Perhaps, the most striking development in the discussion of entanglement entropy appeared from the advent of the gauge/gravity duality, in particular, from the Ryu-Takayanagi (RT) proposal for the entanglement entropy \cite{Ryu:2006bv,Ryu:2006ef}.

The RT proposal of the entanglement entropy is fundamental to providing an intriguing and deep connection between spacetime geometry, quantum field theories and quantum information notions. This proposal geometrizes the concept of entanglement entropy and relates the entanglement entropy of the boundary theory to a minimal area of certain bulk codimension two surface, whose boundary is homologous to the boundary of the subsystem. This proposal has been applied in a variety of systems to probe various physics, such as confinement/deconfinement transitions \cite{Klebanov:2007ws,Nishioka:2006gr}, large $N$ phase transitions \cite{Johnson:2013dka,Dey:2015ytd,Dey:2014voa}, quench dynamics \cite{Balasubramanian:2011ur,Liu:2013iza,Dey:2015poa}, quantum gravity \cite{VanRaamsdonk:2010pw,Balasubramanian:2013lsa}, holographic quantum error-correcting codes and tensor networks \cite{Pastawski:2015qua,Hayden:2016cfa}  etc, with recent proof of the proposal also appearing in \cite{Casini:2011kv,Lewkowycz:2013nqa}. It is fair to say that the entanglement entropy proposal is one of the most significant and useful suggestions that has emerged from the gauge/gravity duality. It not only provides a deep connection between geometry and quantum information but also provides an elegant way to compute and understand other information theoretic quantities, such as the mutual information, entanglement negativity, entanglement of purifications etc.

The entanglement entropy, however, unlike for the pure state, is not a good measure of entanglement for the mixed and multipartite states, as it mixes both classical and quantum correlations. For mixed states, new entanglement measures, such as (logarithmic) entanglement negativity, entanglement of purification, entanglement of formation etc,  have been proposed \cite{Vidal:2002zz,Terhal,Horodecki1996,Horodecki,Peres:1996dw,Eisert}. Unfortunately, computation of these measures in quantum field and many-body systems, which generally have a large Hilbert space, is notoriously difficult. From the holographic perspective, there have been a few proposals for these measures. For example, the entanglement of purification has been suggested to be dual to the area of the minimal cross-section on the entanglement wedge  \cite{Takayanagi:2017knl,Nguyen:2017yqw}. Whereas, there have been two different proposals for the entanglement negativity. In the first proposal, the logarithmic negativity is suggested to be given by the area of an extremal cosmic brane that terminates on the boundary of the entanglement wedge \cite{Kudler-Flam:2018qjo,Kusuki:2019zsp} and, in the second proposal, it is suggested to be given by certain combinations of the areas of codimension-two minimal bulk surfaces \cite{Chaturvedi:2016rft,Chaturvedi:2016rcn,Jain:2017aqk,Jain:2017xsu,Jain:2017uhe,Jain:2018bai,Malvimat:2018txq,Malvimat:2018izs,Basak:2020bot,Malvimat:2018cfe}.

It is important to emphasize that the entanglement wedge cross-section appears in the holographic proposals of many information theoretic quantities. Apart from the above mentioned entanglement of purification proposal as well as in the first proposal of the entanglement negativity, recently, it has also appeared in the holographic proposal of the reflected entropy (the entanglement entropy associated to a canonical purification) \cite{Dutta:2019gen}. Further, these different proposals for the entanglement wedge, in particular, as the holographic dual of the entanglement of purification and reflected entropy, have their own merits and demerits and are in tension with each other \cite{Akers:2019gcv}. Therefore, it appears that more caution is required while associating an information theoretic notion to the entanglement wedge cross-section and more work is needed to correctly establish the same.  In this work, we will not dwell and try to resolve the interpretational issues of the entanglement wedge cross-section. Instead, we will compute it in a variety of holographic confining backgrounds and try to investigate whether, like the entanglement entropy, can it also provide signatures and universal results for the (de)-confinement.

Due to severe technical difficulties, present both at analytical as well as at numerical level, it is generally very hard to obtain any reliable nonperturbative estimate of the entanglement measures in interacting quantum field theories.  For these reasons, the analysis of entanglement measures in quantum chromodynamics (QCD) like theories is rather limited. With the exception of few lattice related papers \cite{Buividovich:2008kq,Buividovich:2008gq,Itou:2015cyu,Rabenstein:2018bri}, most of the discussions are based on the holographic proposals, and that too is limited to the entanglement entropy. In \cite{Klebanov:2007ws,Nishioka:2006gr}, the holographic entanglement entropy was computed in the top-down confining models of the gauge/gravity duality and a phase transition from connected to disconnected entangling surface was found as the size of the entangling region varied. This phase transition, since it causes a change in the order of entanglement entropy (from $\mathcal{O}(N^2)$ to $\mathcal{O}(N^0)$ or vice versa), was suggested as a reminiscent of (de)confinement. Importantly, the phase transition and non-analyticity in the structure of entanglement entropy have also been observed in lattice $SU(N)$ gauge theories \cite{Buividovich:2008kq,Buividovich:2008gq,Itou:2015cyu,Rabenstein:2018bri}, see also \cite{Ramos:2020kyc}. The idea of \cite{Klebanov:2007ws,Nishioka:2006gr} was then applied to many different top-down as well as bottom-up confining systems and results similar to those were obtained \cite{Dudal:2016joz,Dudal:2018ztm,Mahapatra:2019uql,Gursoy:2018ydr,Fujita0806,Kola1403,Georgiou:2015pia,Ben-Ami:2014gsa,Kim,Lewkowycz,Ghodrati,Knaute:2017lll,Ali-Akbari:2017vtb,Anber:2018ohz,Arefeva:2020uec,Slepov:2019guc,
Liu:2019npm,Fujita:2020qvp,Fu:2020oep}.

Interactions in quantum field theories (QFT) via the entanglement in quantum states cause quantum information to be dispersed non-locally across space. To understand better this non-local quantum spreading in QCD like theories, it is not only important to investigate how this structure of shared information changes with the size of the subsystem but is also necessary to examine the entire structure of the entanglement spectrum, including its mixed and multipartite state measures. However, the discussion of mixed state entanglement measures in confining theories is relativity new. A partial discussion appeared in \cite{Jokela:2019ebz}, where the entanglement wedge cross-section in a potentially limited top-down confining model was discussed, whereas no such investigation has been done for the entanglement negativity. In this work, we would like to do a comprehensive analysis of mixed state entanglement measure, including both entanglement wedge cross-section and negativity, in a variety of top-down as well as bottom-up confining QCD models. The top-down QCD models although usually face several limitations in mimicking real QCD and contain undesirable features such as conformal symmetries, the non-running coupling constant, additional Hilbert space sector etc, however, have a well-defined gauge/gravity foundations. Whereas the phenomenological bottom-up QCD models, although lack solid gauge/gravity duality foundations and are generally formatted in an ad-hoc way to reproduce desirable features for the boundary QCD, however, can overcome most of the difficulties present in top-down models. By thoroughly investigating pure and mixed state entanglement measures in both top-down as well as bottom-up models, it might not only be possible to obtain universal features of the entanglement in confining theories but one might also get new predictions from holography, which can be tested via lattice calculations.

In this work, we consider two top-down and one bottom-up confining models. The top-down models are obtained by compactifying $D4$ and $D3$ branes on a circle \cite{Witten:1998zw}, whereas for the bottom-up model, we considered the Einstein-Maxwell-dilaton holographic QCD model constructed in \cite{Dudal:2017max,Dudal:2018ztm}. In all cases, the entanglement entropy with one strip goes through a phase transition from a connected to a disconnected surface at the critical strip length $L_{crit}$. With two equal size disjoint strips, depending on their length $L$ and separation $X$, four different types of  minimal area surfaces $\{S_1, S_2, S_3, S_4 \}$ appear, which lead to an interesting phase diagram. The mutual information is non-zero only in $S_2$ and $S_3$ phases and is always a monotonic function of $L$ and $X$. We then study the entanglement wedge cross-section and find that it is again non-zero only in $S_2$ and $S_3$ phases. However, unlike the mutual information, the entanglement wedge cross-section not only vanishes discontinuously for large values of $X$ and $L$ but also exhibits a non-analytic behaviour every time a phase transition between different entangling surfaces takes place. A further comparison reveals that the entanglement wedge cross-section always exceeds half of the mutual information \textit{i.e.} the holographically suggested inequality \cite{Takayanagi:2017knl} is satisfied in all confining theories.

We further investigate the entanglement negativity in confining theories with one and two disjoint intervals. For this purpose, we  use the second holographic entanglement negativity proposal \cite{Chaturvedi:2016rft,Chaturvedi:2016rcn,Jain:2017aqk,Jain:2017xsu,Jain:2017uhe,Jain:2018bai,Malvimat:2018txq,Malvimat:2018izs,Basak:2020bot,Malvimat:2018cfe}. The reason for choosing the second proposal is threefold (i) the first entanglement negativity proposal is closely related to the entanglement wedge cross-section, which we will anyhow compute (ii) it is also computationally easier to implement, as opposed to the first proposal which requires non-trivial cosmic brane backreaction calculation, and (iii) a direct non-trivial outcome of the second proposal is that it conveys a universal result for the entanglement negativity in all confining theories, which might be possible to test via lattice calculations in the near future. In particular, it suggests that the entanglement negativity is just $3/2$ times of the entanglement entropy. This implies that, just like the entanglement entropy, the  entanglement negativity also displays a discontinuous behaviour at $L_{crit}$ and undergoes a change in order from $\mathcal{O}(N^2)$ to $\mathcal{O}(N^0)$.  Moreover, the proposal also implies an interesting result with two disjoint strips. In particular, unlike the entanglement wedge cross-section, it can be non-zero in the $S_1$ phase as well.

The paper layout is as follows. In Section \ref{sec2}, we review various holographic notions of pure and mixed state entanglement measures. In Section \ref{D4system}, we present the calculations for entanglement entropy, mutual information, entanglement wedge cross-section and entanglement negativity for a top-down confining model obtain by compacting $D4$-branes on a circle. We repeat the computations of section \ref{D4system} in another top-down confining model, this time by compacting $D3$-branes on a circle, in Section \ref{D3system}. In Section \ref{bottomupsystem}, we further discuss these entanglement measures in a phenomenological bottom-up Einstein-Maxwell-dilaton confining model.  We end the paper with a discussion and conclusion in Section \ref{DiscussionandConclusion}.

\section{Entanglement measures}\label{sec2}
 Our main aim in this section is to briefly elucidate the mathematical procedure that is used to calculate various entanglement measures in holographic settings. The list includes entanglement entropy, mutual information, entanglement wedge cross-section and entanglement negativity.
\subsection{Entanglement entropy: one strip}
Entanglement entropy, which is given by the von Neumann entropy
\begin{equation}
S_A = -\mathrm{Tr}_A\rho_A\mathrm{ln}\,\rho_A\ ,
\label{vne}
\end{equation}
provides the measure of entanglement between pure states.  Generally, one uses the non-trivial replica trick to calculate it in quantum field theories \cite{Calabrese:2009qy}. However, implementing the replica technique in non-trivial field theories, such as those containing interactions, is much more tedious. The holographic idea, on the other hand, provides another pathway to calculate the entanglement entropy in field theories \cite{Ryu:2006bv,Ryu:2006ef}. For the case of $AdS_{d+1}/CFT_{d}$, the entanglement entropy $S_{A}$ of the subsystem $A$ in $CFT_{d}$ is given by the Ryu-Takayanagi prescription \cite{Ryu:2006bv,Ryu:2006ef},
\begin{equation}
S_A = \frac{\mathcal{A}(\Gamma_A^{\text{min}})}{4G_N^{(d+1)}}\ ,
\label{hee}
\end{equation}
where $\mathcal{A}(\Gamma_A^{\text{min}})$ is the area of $(d-1)$-dimensional static minimal surface ($\Gamma_A$) in the bulk whose boundary is homologous to the boundary $\partial A$ of the subsystem $A$. $G_N^{(d+1)}$ is the $(d+1)$-dimensional Newton constant. Equivalently, we can also recast the above holographic entanglement entropy formula in the following way,
\begin{equation}
S_A = \frac{1}{4G_N^{(d+1)}}\int_{\Gamma} d^{d-1}\sigma\sqrt{G^{d-1}_{\mathrm{ind}}}\ ,
\label{heestring}
\end{equation}
where $G^{d-1}_{\mathrm{ind}}$ is the induced metric on the surface $\Gamma$, which further needs to be appropriately minimised \cite{Klebanov:2007ws,Nishioka:2006gr}. Since in this work we are interested in computing the entanglement entropy and other information related quantities in top-down holographic confining theories, where the dual bulk spacetime metric is usually written down in string frame, it is also useful to write down the entanglement entropy expression in string frame metric \cite{Klebanov:2007ws,Nishioka:2006gr},
\begin{equation}
S_A = \frac{1}{4G_N^{(d+1)}}\int_{\Gamma} d^{d-1}\sigma e^{-2\phi}\sqrt{G^{d-1}_{\mathrm{ind}}}\ .
\label{heenc}
\end{equation}
where the dilaton field $\phi$ arises because of the frame change.

\subsection{Mutual information: two strip}
Since the entanglement structure of one interval subsystem is known to display interesting behaviour in the confined phases, it is also natural to ask and investigate the entanglement structure with many disjoint intervals. One such natural entanglement measure that appears with two disjoint intervals is the mutual information \cite{Hayden:2011ag,Headrick:2010zt,Calabrese:2009ez}. For two subsystems ($A_1$ and $A_2$), the mutual information is defined as the amount of information that $A_1$ and $A_2$ can share and is defined in terms of entanglement entropy as,
\begin{equation}\label{MIdef}
I(A_1,A_2)=S_{A_1} + S_{A_2} - S_{A_1 \cup A_2} \ ,
\end{equation}
where $S_{A_1}$, $S_{A_2}$ and $S_{A_1 \cup S_{A_2}}$ are as usual the entanglement entropies of subsystems $A_1$, $A_2$ and their union respectively. It is quite obvious from Eq.~(\ref{MIdef}) that the mutual information is zero for two uncorrelated systems. Importantly, the mutual information does not suffer from the ambiguities associated with the entanglement entropy and can provide more information than the entanglement entropy alone. In particular, mutual information is UV finite and does not contain the usual UV divergences. Moreover, the subadditivity property of the entanglement entropy also ensures that the mutual information is non-negative, i.e. $I(A_1, A_2)$ provides an upper bound on the correlation functions between operators in $A_1$ and $A_2$. For more discussion on the mutual information, see \cite{Fischler:2012uv,Kundu:2016dyk,Casini:2015woa,Alishahiha:2014jxa,MolinaVilaplana:2011xt,Balasubramanian:2018qqx,Cardy:2013nua,Larkoski:2014pca}. For the discussion on mutual information and two strip entanglement phase diagram in top-down and bottom-up confining models, see \cite{Mahapatra:2019uql,Ben-Ami:2014gsa}. Similarly, we can also define other entanglement measures, such as $n$-partite information, with more disjoint intervals
\begin{eqnarray}
I^{n}(A_{\{i\}})=  \sum_{i=1}^{n}  S_{A_{i} }-  \sum_{i<j}^{n} S_{A_{i} \cup A_{j}} + \sum_{i<j<k}^{n} S_{A_{i} \cup A_{j} \cup A_{k}} - \dots \nonumber \\  - (-1)^n S_{A_1 \cup A_2 \cup ... \cup A_n}
\label{npartitedefi}
\end{eqnarray}
In this work, we will concentrate only on mutual information as the results for the $n$-partite information can be analogously obtained.

\subsection{Entanglement wedge cross-section}
Entanglement entropy serves as a good measure for the entanglement in the case of pure states but it fails in
the case of mixed states. Therefore, it would be an interesting question to investigate how the mixed state entanglement measures behave in confining theories. One such mixed state entanglement measure is the entanglement of purification, which in holographic context is suggested to be given by the minimal area of the entanglement wedge cross-section \cite{Takayanagi:2017knl,Nguyen:2017yqw} \footnote{For some related discussion on the entanglement wedge cross-section and entanglement of purification, see \cite{Bhattacharyya:2018sbw,Bhattacharyya:2019tsi,Harper:2019lff,Hirai:2018jwy,Espindola:2018ozt,Umemoto:2018jpc,Bao:2018gck,Yang:2018gfq,Caputa:2018xuf,Liu:2019qje,
Amrahi:2020jqg,Liu:2020blk,Nomura:2018kji,BabaeiVelni:2019pkw}.}. However, as mentioned in the introduction, there are other holographic interpretations of the entanglement wedge as well (such as the holographic dual of the reflected entropy), and these different interpretations, unfortunately, do not exactly correlate with each other. In this work, we will mainly concentrate on the entanglement wedge cross-section, without worrying too much about its interpretational issues.

To compute the entanglement wedge cross-section holographically, we follow the prescription suggested in \cite{Takayanagi:2017knl,Nguyen:2017yqw}. We first consider
two subsystems $A$ and $B$ with no overlap on the $d$-dimensional boundary. The Ryu-Takayanagi
minimal surfaces for $A$, $B$ and their union $AB=A\cup B$ are denoted by $\Gamma^{min}_A$, $\Gamma^{min}_B$ and $\Gamma^{min}_{AB}$ respectively. The $d$-dimensional entanglement wedge $M_{AB}$ in the $(d+1)$-dimensional bulk  is then described as a region which is bounded by $A,B$ and $\Gamma^{min}_{AB}$, i.e. the entanglement wedge $M_{AB}$ is the bulk region whose boundary is
\begin{equation}
\partial M_{AB}=A\cup B\cup \Gamma^{min}_{AB}.
\label{EWboundary}
\end{equation}
One should note that the entanglement wedge is a bulk codimension zero region. However, here it turns out to be $d$-dimensional as we are considering static bulk configuration. Also note that if $A$ and $B$ are very small or have large separation between them then $M_{AB}$ will have a disconnected form.

We next divide $\Gamma_{AB}^{min}$ as follows
\begin{equation}
\Gamma_{AB}^{min} = \Gamma_{AB}^{(A)}\cup \Gamma_{AB}^{(B)}
\label{RTdiv}
\end{equation}

and we further define
\begin{eqnarray}
& & \tilde{\Gamma}_{A} = A\cup \Gamma_{AB}^{(A)}\,  \nonumber \\
& & \tilde{\Gamma}_{B} = A\cup \Gamma_{AB}^{(B)}\ .
\label{RTA}
\end{eqnarray}
Note that, as a result of the Eqs. \eqref{RTdiv} and \eqref{RTA}, the boundary of the entanglement wedge $M_{AB}$ is now divided into two parts:
\begin{equation}
\partial M_{AB}=\tilde{\Gamma}_{A}\cup \tilde{\Gamma}_{B}.
\end{equation}
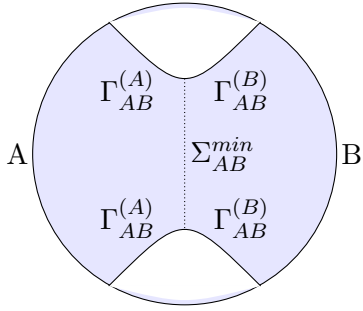
\begin{figure}[h]
\begin{center}
\begin{tikzpicture}
\filldraw[blue!10!, draw=black] (0,0) circle (2);
\draw[densely dotted] (0,1) -- (0,-1);
\filldraw[white!100!, draw=black] (-1,1.75) .. controls (0,0.75) .. (1,1.75);
\fill[white!100!] (-1,1.75) .. controls (0,2) .. (1,1.75);
\filldraw[white!100!, draw=black] (-1,-1.75) .. controls (0,-0.75) .. (1,-1.75);
\fill[white!100!] (-1,-1.75) .. controls (0,-2) .. (1,-1.75);
\node()at (-2.2,0){A};
\node()at (0.5,0){$\Sigma^{min}_{AB}$};
\node()at (-0.75,0.85){$\Gamma^{(A)}_{AB}$};
\node()at (-0.75,-0.85){$\Gamma^{(A)}_{AB}$};
\node()at (0.75,0.85){$\Gamma^{(B)}_{AB}$};
\node()at (0.75,-0.85){$\Gamma^{(B)}_{AB}$};
\node()at (2.2,0){B};
\end{tikzpicture}
\caption{The region in the blue colour is the entanglement wedge $M_{AB}$ corresponding to a pure state. The dotted surface is $\Sigma_{AB}$ which divides $M_{AB}$ into 2 parts.}=
\label{Entanglementwedgecrosspic}
\end{center}
\end{figure}
A pictorial representation of the holographic entanglement wedge is shown in Figure~\ref{Entanglementwedgecrosspic}. Next, look for the minimum surface $\Sigma_{AB}^{min}$ subjected to conditions
\begin{equation}
\begin{split}
&(i)\ \partial\Sigma_{AB}^{min}=\partial\tilde{\Gamma}_{A}=\partial\tilde{\Gamma}_{B}\ ,\\
&(ii)\ \Sigma_{AB}^{min}\ \mathrm{is\ homologous\ to}
\ \tilde{\Gamma}_{A}\ \mathrm{inside}\ M_{AB}.
\end{split}
\end{equation}
The holographic entanglement wedge cross-section $E_{W}(\rho_{AB})$ is then simply defined as the area of $\Sigma_{AB}^{min}$ divided by $4G_N^{d+1}$,
\begin{equation}\label{EW}
E_{W}(\rho_{AB}) = \min_{\Gamma_{AB}^{(A)}\subset\Gamma_{AB}^{min}}\left[\frac{\mathcal{A}(\Sigma_{AB}^{min})}{4G_N^{(d+1)}}\right].
\end{equation}
Note that for a given $A$ and $B$ subsystems, as we will also see later in this work, there can be many possible $\Sigma_{AB}^{min}$ surfaces. In those cases, the entanglement wedge cross-section is given by the surface that has the minimum area. To summarize, $E_{W}(\rho_{AB})$ is the minimal surface area of the entanglement wedge $M_{AB}$ connecting $A$ and $B$.

\subsection{Entanglement negativity}
In the previous section, we talked about the entanglement wedge cross-section (and its holographic definition) as a suitable
measure for the mixed state entanglement. Another quantum information quantity which is also known to capture mixed
state entanglement is the entanglement negativity. This is defined in quantum many body system as \cite{Vidal:2002zz,Horodecki1996},
\begin{equation}\label{EN}
\mathcal{N} = \frac{\|\rho^{T_2}\|-1}{2}\ ,
\end{equation}
where $\|\rho^{T_2}\|$ is the trace norm of the partially transposed reduced density matrix $\rho^{T_2}$ and this trace norm is generally given by the sum of the absolute eigenvalues of $\rho^{T_2}$. There also exists a close variant of the entanglement negativity, called logarithmic negativity, which is also frequently used in the quantum information community. This is defined as,
\begin{equation}\label{logen}
\mathcal{E} = \ln \|\rho^{T_2}\| = \ln \mathrm{Tr}|\rho^{T_2}|.
\end{equation}
The (logarithmic) entanglement negativity has been computed in a variety of quantum many body and field theory systems and has been widely used in condensed matter and quantum information community, see for example \cite{Calabrese:2012ew,Calabrese:2012nk,Alba,Calabrese:2013mi,Chung,Ruggiero:2016aqr,Ruggiero:2016yjt,Coser:2014gsa,Hoogeveen:2014bqa,Blondeau-Fournier:2015yoa,Castelnovo,Lee,Eisler,Wen:2015qwa,Wen:2016bla,Wen:2016snr,
Rangamani:2014ywa}. The primary reason for this is that it provides an upper bound on the distillable entanglement. Holographically, to the best of our knowledge, two different (yet equivalent) proposals for the entanglement negativity have been suggested. In the first proposal, the logarithmic negativity is suggested to be given by the area of an extremal cosmic brane that terminates on the boundary of the entanglement wedge \cite{Kudler-Flam:2018qjo,Kusuki:2019zsp}. In the second proposal, the logarithmic negativity is suggested to be given by certain combinations of the areas of co-dimension two minimal bulk surfaces \cite{Chaturvedi:2016rft,Chaturvedi:2016rcn,Jain:2017aqk,Jain:2017xsu,Jain:2017uhe,Jain:2018bai,Malvimat:2018txq,Malvimat:2018izs,Malvimat:2018cfe,Basak:2020bot}. Both these proposals have been tested in diverse physical situations and have shown to reproduce exact known results for the negativity in CFTs. Since the first proposal is very closely related to the entanglement wedge cross-section, which we will anyhow be going to investigate, it might be more informative, and at the time complimentary as well, if the second holographic proposal is adopted for the computation of entanglement negativity. It is also computationally slightly more straightforward to implement the second proposal, as opposed to the first proposal, which requires non-trivial cosmic brane backreaction calculation. Moreover, as we will see later on, the second proposal also advocates a universal result for the entanglement negativity in all holographic confining theories, which might be possible to check via lattice calculations.  For these reasons, we will take the second proposal for the entanglement negativity in this paper. It would certainly be interesting to explicitly compute the entanglement negativity in confining theories using the first proposal and find its similarities/differences with the second proposal. We leave this interesting exercise for future work.

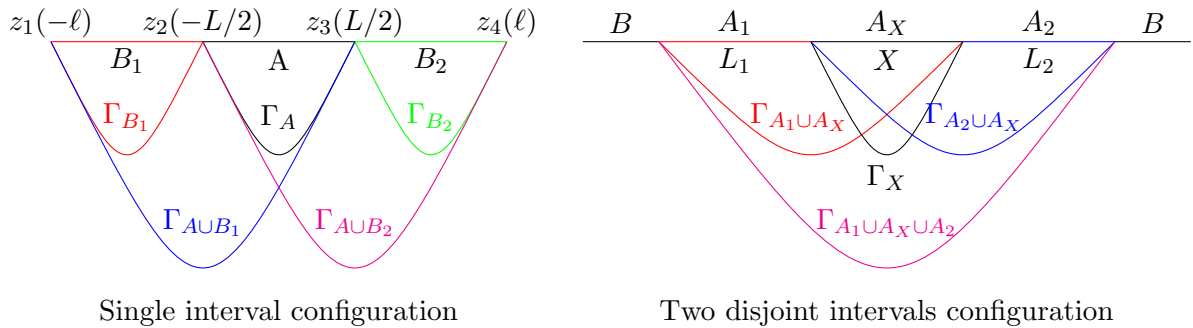
\begin{figure}[h]
\begin{tikzpicture}
\draw[red] (-20,0) -- (-18,0);
\draw (-18,0) -- (-16,0);
\draw[green] (-16,0) -- (-14,0);
\draw[red] (-20,0) .. controls (-19,-2) .. (-18,0);
\draw (-18,0) .. controls (-17,-2) .. (-16,0);
\draw[green] (-16,0) .. controls (-15,-2) .. (-14,0);
\draw[blue] (-20,0) .. controls (-18,-4) .. (-16,0);
\draw[magenta] (-18,0) .. controls (-16,-4) .. (-14,0);
\node()at (-17,-0.25){A};
\node()at (-19,-0.25){$B_1$};
\node()at (-15,-0.25){$B_2$};
\node()at (-20,0.25){$z_1(-\ell)$};
\node()at (-18,0.25){$z_2(-L/2)$};
\node()at (-16,0.25){$z_3(L/2)$};
\node()at (-14,0.25){$z_4(\ell)$};
\node()at (-17,-1){$\Gamma_A$};
\node[red]()at (-19,-1){$\Gamma_{B_1}$};
\node[green]()at (-15,-1){$\Gamma_{B_2}$};
\node[blue]()at (-18,-2.4){$\Gamma_{A\cup B_1}$};
\node[magenta]()at (-16,-2.4){$\Gamma_{A\cup B_2}$};
\node()at (-17,-3.6){Single interval configuration};
\draw (-13,0) -- (-12,0);
\draw[red] (-12,0) -- (-10,0);
\draw (-10,0) -- (-8,0);
\draw[blue] (-8,0) -- (-6,0);
\draw (-6,0) -- (-5,0);
\draw[red] (-12,0) .. controls (-10,-2) .. (-8,0);
\draw (-10,0) .. controls (-9,-2) .. (-8,0);
\draw[blue] (-10,0) .. controls (-8,-2) .. (-6,0);
\draw[magenta] (-12,0) .. controls (-9,-4) .. (-6,0);
\node()at (-9,0.25){$A_X$};
\node()at (-9,-0.25){$X$};
\node()at (-11,-0.25){$L_1$};
\node()at (-7,-0.25){$L_2$};
\node()at (-7,0.25){$A_2$};
\node()at (-12.5,0.25){$B$};
\node()at (-11,0.25){$A_1$};
\node()at (-5.5,0.25){$B$};
\node()at (-9,-1.8){$\Gamma_X$};
\node[red]()at (-10.15,-1){$\Gamma_{A_1\cup A_X}$};
\node[blue]()at (-7.85,-1){$\Gamma_{A_2\cup A_X}$};
\node[magenta]()at (-9,-2.4){$\Gamma_{A_1\cup A_X\cup A_2}$};
\node()at (-9,-3.6){Two disjoint intervals configuration};
\end{tikzpicture}
\caption{Pictorial representation of various minimal area surfaces that contribute in the entanglement negativity.}
\label{entanglenegativitypicture}
\end{figure}

To compute the holographic logarithmic negativity for a single interval using \cite{Chaturvedi:2016rft,Chaturvedi:2016rcn,Jain:2017aqk}, first, a bipartition of the system into $A$ and it's compliment $A^c$ is considered in $CFT_d$. Next, two finite length intervals $B_1$ and $B_2$ is taken adjacent on both sides of $A$ such that $B=B_1\cup B_2$. Refer Figure~\ref{entanglenegativitypicture}. If the corresponding codimension-two bulk static minimal surfaces in $AdS_{d+1}$ are denoted by $\Gamma_A$, $\Gamma_{B_1}$ and $\Gamma_{B_2}$, then the holographic entanglement negativity  for the bipartite system $(A \cup A_c)$ is suggested to be given by the following combination of the areas of the minimal bulk surfaces,
\begin{equation}\label{HEEsc}
\mathcal{E} = \lim_{B\rightarrow A^c}\frac{3}{16G_N^{(d+1)}}\left[2\mathcal{A}(\Gamma_A)+\mathcal{A}(\Gamma_{B_1})+ \mathcal{A}(\Gamma_{B_2})-\mathcal{A}(\Gamma_{A\cup B_1})-\mathcal{A}(\Gamma_{A\cup B_2})\right]\ ,
\end{equation}
where in the above equation it should be understood that both $B_1$ and $B_2$ are extending to infinity such that $B=B_1\cup B_2=A^c$. In the $AdS_3/CFT_2$ context, the above formula were deduced from how the four point twist correlation functions factorise in the large central charge limit. In particular, the entanglement negativity in $CFT_2$ is given by a specific four point twist correlator in the $n$-sheeted Riemann surface \cite{Calabrese:2012ew,Calabrese:2012nk}. This four point correlation function factorises into various combinations of two point functions in the large central charge limit, which can then be mapped, using the $AdS/CFT$ dictionary, to the length of the geodesic anchored on the boundary points and extending into the bulk. In this way, the authors of \cite{Chaturvedi:2016rft,Chaturvedi:2016rcn} arrived at the above formula for the entanglement negativity in $CFT_2$ (and its $d$-dimensional extension). This entanglement negativity formula, though is an unproven conjectured proposal, in contrast with the Ryu-Takayanagi entanglement entropy proposal, however, it does reproduce mixed and pure states results of the entanglement negativity in $CFT$. For more details on this entanglement negativity conjecture, see \cite{Chaturvedi:2016rft,Chaturvedi:2016rcn}.

One can also write the above formula in terms of the entanglement entropy using the Ryu-Takayanagi formula (Eq.~(\ref{hee})),
\begin{equation}\label{HEEsc1}
\mathcal{E} = \lim_{B\rightarrow A^c}\frac{3}{4}\left[2S_A+S_{B_1}+S_{B_2}-S_{A\cup B_1}-S_{A\cup B_2}\right]\,.
\end{equation}
In the subsequent sections, we will take the above formula to compute the entanglement negativity in a variety of top-down as well as bottom-up confining backgrounds. As we will see, the above entanglement negativity formula reduces to a simpler expression and suggests a universal relation in all confining theories.

For two disjoint intervals $A_1$ and $A_2$, the negativity is similarly suggested to be given by the following combination of the minimal area surfaces,
\begin{eqnarray}
\mathcal{E} = \frac{3}{4}\left[S_{A_1 \cup A_X} + S_{A_X \cup A_2} - S_{A_1 \cup A_2 \cup A_X}-S_{A_X} \right] \,.
\label{HEEsc1twostrip}
\end{eqnarray}
where $X$ is the separation between the two intervals. Refer Figure~\ref{entanglenegativitypicture} for more details on the surfaces appearing in Eq.~(\ref{HEEsc1twostrip}). Again, this is an unproven conjectured formula which is obtained by analysing the factorization of four point twist correlation function in terms of two point correlations in the large central charge limit \cite{Malvimat:2018txq,Basak:2020bot}. This conjectured formula, however, again reproduces the desirable entanglement negativity results of the boundary CFT.

\section{D4-branes on a circle}\label{D4system}
The first top-down holographic confining model we consider is obtained by putting $D4$-branes on a circle. As is well know, the low energy dynamics of $N_c$ coincidental $D4$ branes in type IIA string theory is given by  $(4+1)$ dimensional $U(N_c)$ supersymmetric Yang-Mills theory with 't Hooft coupling $\lambda=g_s N_c l_s$. This theory can be reduced to (3+1) dimensions with broken supersymmetry by compacting one of the directions along the brane, let's say $x^4$, on a circle with radius $R_4$ ($x^4\sim x^4+2\pi R_4$). The low energy dynamics of the reduced system is then given by the dimensionless parameter $\lambda_4=\lambda/R_4$ \cite{Witten:1998zw}. The condition $\lambda_4\gg1$ allows us to investigate this system using its dual gravitational picture. The near-horizon geometry of the $D4$-branes is \cite{Witten:1998zw},
\begin{equation}
\begin{split}
&ds^2 = \left(\frac{U}{R}\right)^{3/2}\left[\left(\frac{R}{U}\right)^{3}\frac{dU^2}{f(U)}+dx^{\mu}dx_{\mu}\right]+ R^{3/2}U^{1/2}d\Omega_4^2+\left(\frac{U}{R}\right)^{3/2}f(U)(dx^4)^2\ ,\\
&e^{-2\phi}=\left(\frac{R}{U}\right)^{3/2},\ f(U)=1-\left(\frac{U_0}{U}\right)^{3},\ U_0=\frac{4\pi\lambda}{9R_4^2},\ R^3=\pi\lambda.
\end{split}
\end{equation}
Here $R$ is the AdS length scale. Note that this geometry forms a cigar shape in $(U,x^4)$ coordinates, with radius of the $x^4$-circle goes to zero as $U\rightarrow U_0$. The radial value $U_0$ therefore introduces a mass gap in the theory.

\subsection{Entanglement entropy: one strip}
The entanglement entropy has already been computed for this system in \cite{Klebanov:2007ws}. Here, we will reproduce their results to set the stage for the later sections. To calculate the entanglement entropy, we consider a strip subsystem of length $L$ and define the subsystem domain as $-L/2\leq x_1=x\leq L/2$, $0\leq x_2\leq L_2$ and $0\leq x_3\leq L_3$. The parametrization $U=U(x)$ leads to the following expression for the entanglement entropy (\ref{heenc}),
\begin{eqnarray}\label{EED4}
S_A = \frac{L_2 L_3\omega_4 (2\pi R_4)}{4 G_N^{(10)}} \int dx \ R^{3/2}U^{5/2} \sqrt{f(U)+\left(\frac{R}{U}\right)^3 U'^2}
\end{eqnarray}
where $\omega_4$ is the area of unit four sphere. It turns out that there are actually two surfaces which minimise the above entanglement action: a ($U$-shaped) connected
and a disconnected surfaces. The expression of the entanglement entropy of the connected surface is simply given by
\begin{eqnarray}
S_A^{con} = \frac{L_2 L_3}{2 G_N^{(10)}}\frac{32 \pi^3 R^{9/2}}{9} \int_{U_*}^{U_\infty}dU \ \frac{U^{7/2}}{\sqrt{U_0}} \frac{\sqrt{f(U)}}{\sqrt{U^5 f(U)-U_*^{5}f(U_*)}} \,.
\label{d4Scon}
\end{eqnarray}
where $U_*$ is the turning point of the connected surface at which $U'(x)|_{U=U_*}=0$, and is related to the strip length $L$ in the following way
\begin{eqnarray}
L(U_*) = 2 R^{3/2} \int_{U_*}^{U_\infty}dU \ \frac{U_*^{5/2}}{U^{3/2}}\sqrt{\frac{f(U_*)}{f(U)}}\frac{1}{\sqrt{U^5 f(U)-U_*^{5}f(U_*)}} \,.
\label{d4stripl}
\end{eqnarray}
Similarly, the entanglement entropy for the disconnected surface is given by
\begin{eqnarray}
S_A^{discon} = \frac{L_2 L_3}{2 G_N^{(10)}}\frac{32 \pi^3 R^{9/2}}{9} \int_{U_0}^{U_\infty}dU \ \frac{U}{\sqrt{U_0}}\,.
\label{d4Sdiscon}
\end{eqnarray}
Note that the disconnected entanglement entropy does not depend on the strip length. This result will have profound implications in the entanglement structure of confined phase.

\begin{figure}[h!]
\begin{minipage}[b]{0.5\linewidth}
\centering
\includegraphics[width=2.8in,height=2.3in]{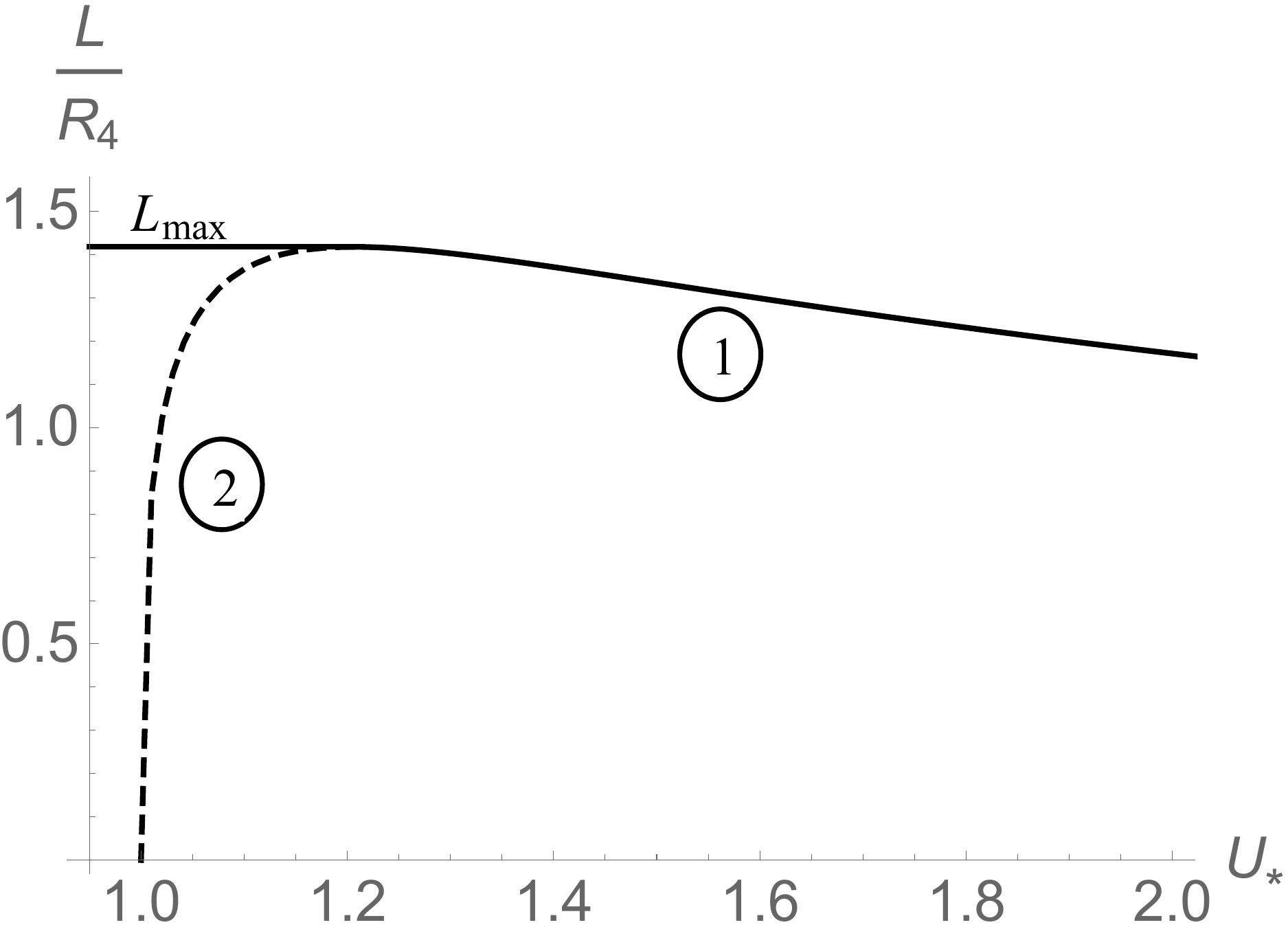}
\caption{ \small The behaviour of strip length $L$ as a function of $U_*$.}
\label{UsVsL1stripD4}
\end{minipage}
\hspace{0.4cm}
\begin{minipage}[b]{0.5\linewidth}
\centering
\includegraphics[width=2.8in,height=2.3in]{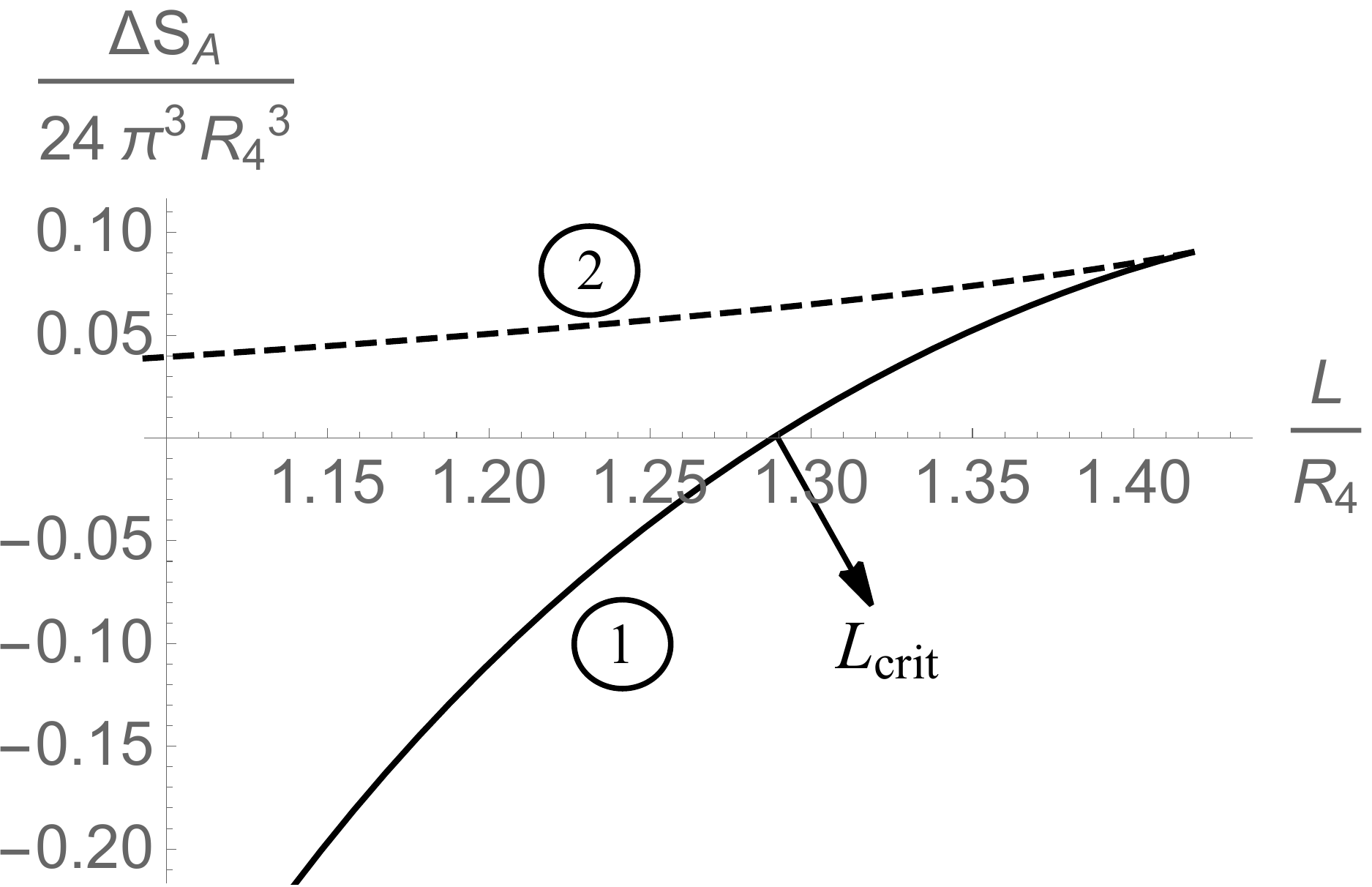}
\caption{\small $\triangle S_A= S_A^{con}-S_A^{discon}$ as a function of strip length $L$.}
\label{LVsdeltaS1stripD4}
\end{minipage}
\end{figure}

We now present the numerical results for the entanglement entropy. For the numerical purpose, we take $U_0=1$. The results are shown in
Figures~\ref{UsVsL1stripD4} and \ref{LVsdeltaS1stripD4}, where the variation of $L$ with respect to $U_*$ and connected and disconnected entanglement entropy difference ($\triangle S_A = S_A^{con} - S_A^{discon}$) with respect to $L$, respectively, are plotted \footnote{Here, and in the subsequent subsections, we have set $L_2 L_3/4 G_N^{(10)}=1$}. Note that at a given $L$, there exist three minimal area surfaces: one disconnected and two connected. The first connected surface $\circled{1}$ (shown by a solid line) is nearer to boundary compared to the second connected surface $\circled{2}$ (shown by a dashed line). These connected surfaces exist only below a maximum length $L_{max} \simeq 1.418 R_4$ and above $L_{max}$ they cease to exist. In particular, there is no solution for the connected surface above $L_{max}$ and only the disconnected surface remains. The second connected surface (dashed line) actually corresponds to a saddle point and its area (and hence the entanglement entropy) is always higher than the first connected surface (solid line).

Importantly, a change in sign in $\triangle S_A$ appears as the strip length $L$ varies. In particular, $\triangle S_A$ is negative for small $L$, indicating that the connected surface $\circled{2}$ has the lowest entanglement entropy for the small subsystem, whereas $\triangle S_A$ is positive for large $L$, indicating that the disconnected surface has the lowest entanglement entropy for the large subsystem. Therefore, a phase transition takes place between connected and disconnected entanglement entropies as $L$ increase. This phase transition occurs at $L=L_{crit}\simeq1.288 R_4$. Since the entanglement entropy of the disconnected surface is independent of $L$, it gives us the following important result,
\begin{eqnarray}
\frac{\partial S_{A}}{\partial L} &\propto &\frac{1}{G_N^{(10)}} = \mathcal{O}(N^2)\quad\text{for}\quad L < L_{crit}\,, \nonumber \\
&\propto& \frac{1}{[G_{N}^{(10)}]^0} = \mathcal{O}(N^0)\quad\text{for}\quad L > L_{crit}
\end{eqnarray}
The above type of connected/disconnected phase transition, where the order of the entanglement entropy changes from $\mathcal{O}(N^2)$ at small $L$ to $\mathcal{O}(N^0)$ at large $L$ was suggested to be reminiscent of the confinement/deconfinement transition in QCD \cite{Klebanov:2007ws}. Moreover, the fact that the order of colored gluon degrees of freedom ($\mathcal{O}(N^2)$) above the deconfinement temperature and color neutral confined degrees of freedom ($\mathcal{O}(N^0)$) below the deconfinement temperature match well with the connected and disconnected entanglement entropies, further naturally led to the interpretation of the critical subsystem size as the inverse deconfinement temperature ($T_D$) \textit{i.e.} $T_D\propto1/L_{crit}$. In \cite{Klebanov:2007ws}, the value of $L_{crit}$ was computed for different confining models and the relation $L_{crit}=\mathcal{O}(\Lambda_{IR}^{-1})$ was found. This further suggested that the entanglement entropy can act as a probe to diagnose confinement \footnote{It has been suggested recently that the entanglement entropy signals only the presence of a mass gap rather than the confinement \cite{Jokela:2020wgs}.}. This work was then generalised to many other top-down as well as bottom-up holographic QCD models and similar results were found in all the cases \cite{Dudal:2016joz,Dudal:2018ztm,Mahapatra:2019uql,Gursoy:2018ydr,Fujita0806,Kola1403,Georgiou:2015pia,
Ben-Ami:2014gsa,Kim,Lewkowycz,Ghodrati,Knaute:2017lll,Ali-Akbari:2017vtb,Anber:2018ohz,Arefeva:2020uec,Slepov:2019guc,Liu:2019npm,Fujita:2020qvp,Fu:2020oep}.

\subsection{Mutual information: two strips}
\begin{figure}[h!]
\begin{minipage}[b]{0.5\linewidth}
\centering
\includegraphics[width=2.8in,height=2.3in]{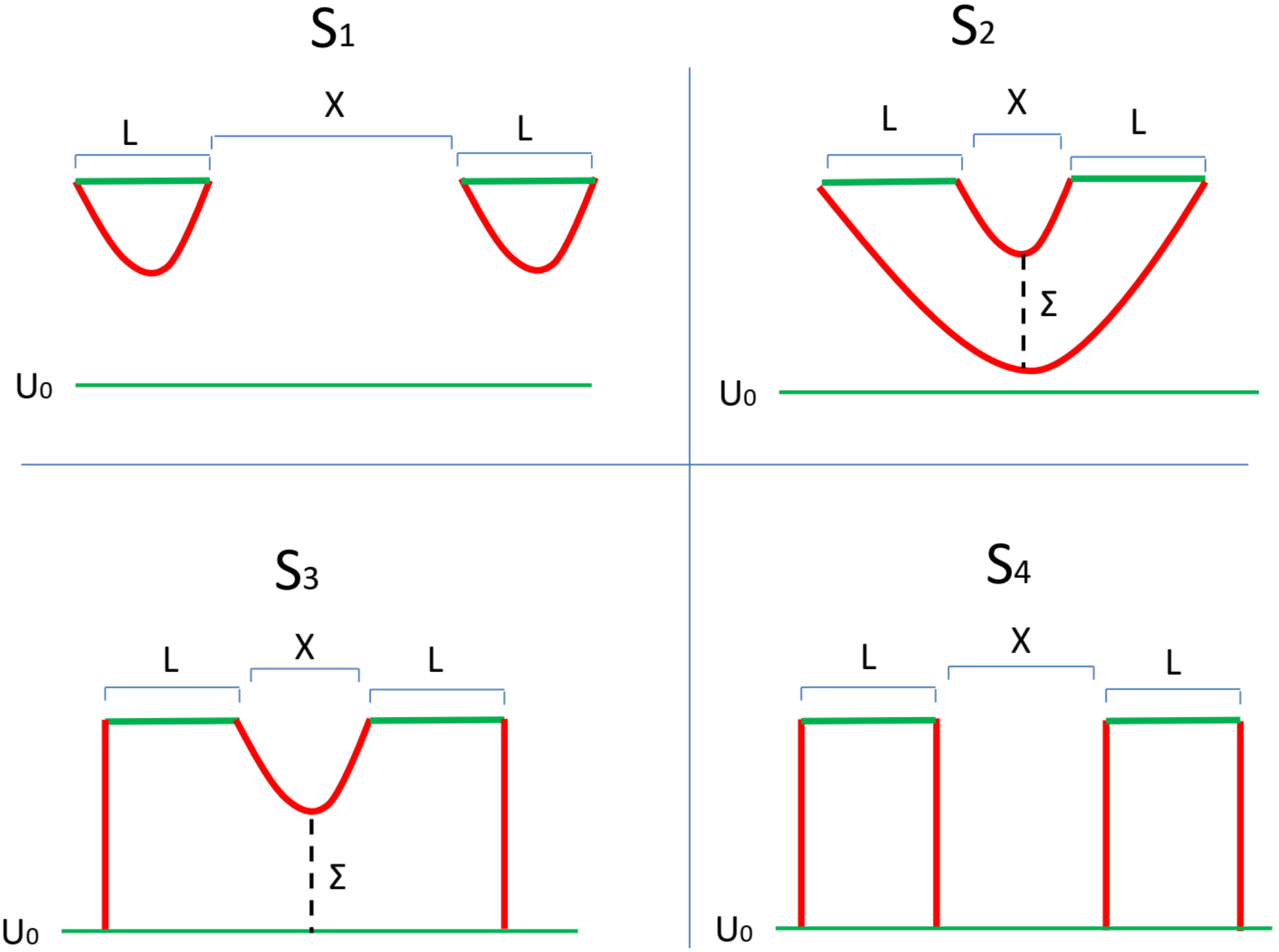}
\caption{ \small Sketch of four different minimal area surfaces that can occur for the cases of two parallel strips of equal length $L$ separated by a distance $X$ in the
confining background. The black dashed lines denote a non-zero connected entanglement wedge.}
\label{twostripphasediag}
\end{minipage}
\hspace{0.4cm}
\begin{minipage}[b]{0.5\linewidth}
\centering
\includegraphics[width=2.8in,height=2.3in]{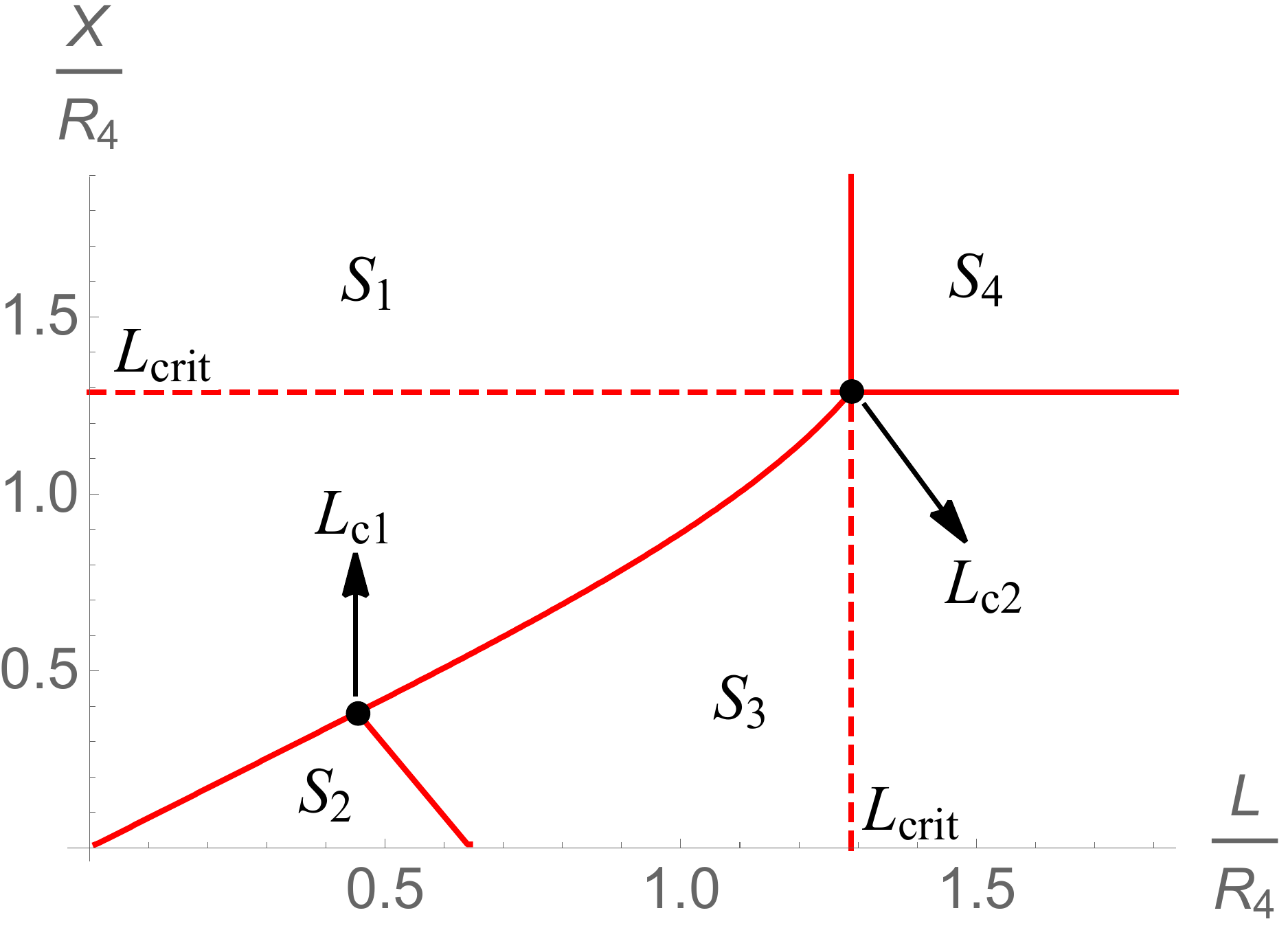}
\caption{\small The entanglement entropy phase diagram of various minimal area surfaces for the case of two strips
of equal length $L$ separated by a distance $X$ in the confining background. These four different phases correspond to the four bulk surfaces
of Figure \ref{twostripphasediag}.}
\label{twostripphasediagD4}
\end{minipage}
\end{figure}

Having discussed the entanglement structure with one strip, we now move on to discuss it with two strips. Here, for simplicity, we concentrate only on equal size strips ($L_1=L_2=L$). The entanglement structure with two strips is much more interesting than with one strip. In particular, depending upon the strips length $L$ and the distance between them $X$, there are now four different entangling surfaces which compete with each other. These surfaces are shown in Figure~\ref{twostripphasediag}. With two strips, both connected ($S_1$ and $S_2$) and disconnected ($S_4$) as well as a combination of connected and disconnected ($S_3$) surfaces can appear.

The entanglement entropy expressions of these surfaces are given as,
\begin{eqnarray}
& & S_1(L,X) = 2 S_A^{con}(L), \ \ \ \ \ \ \ \ \ \ \ \ \ \ \  S_2(L,X) = S_A^{con}(X)+ S_A^{con}(2L +X), \, \nonumber\\
& & S_3(L,X) = S_A^{con}(X) + S_A^{discon}, \ \ \   S_4(L,X) = 2S_A^{discon}\,,
\label{2stripEE}
\end{eqnarray}
where $S_A^{con}$ and $S_A^{discon}$ are as usual the entanglement entropies of the connected and disconnected surfaces with one strip.

Depending on the magnitude of  $L$ and $X$, there appear various phase transitions between $S_1$, $S_2$, $S_3$ and $S_4$. In Figure~\ref{twostripphasediagD4}, the complete phase diagram in the phase space of $(L,X)$ is shown. For small $X,L\ll L_{crit}$, it is usually the connected  $S_1$ surface which has the lowest area (and entanglement entropy). The area of the connected $S_2$ surface, however, becomes smaller as $L$ increases. With further increase in $L$, keeping $X(\ll L_{crit})$ small, the area of the $S_3$ becomes smaller and a phase transition from $S_2$ to $S_3$ takes place. For $X=0$, this phase transition occurs at $L=L_{crit}/2$. For a general value of $X$, the phase transition line between $S_2$ and $S_3$ is given by $2L+X=L_{crit}$, as is obvious from Eq.~(\ref{2stripEE}). Finally, the $S_4$ configuration becomes more favourable with a further increase in both $X,L>L_{crit}$.

Interestingly, with two strips, two tri-critical points appear where three minimal area entangling surfaces co-exist. These are denoted by black dots in Figure~\ref{twostripphasediagD4}. At the first tri-critical point $L_{c1}$, the phases ($S_1$, $S_2$ and $S_3$) coexist whereas at the second tri-critical point $L_{c2}$, the phases  ($S_1$, $S_3$ and $S_4$) coexist. The coordinates of these tri-critical points are ($L/R_4=0.454$, $X/R_4=0.381$) and ($L/R_4=1.288$, $X/R_4=1.288$), respectively. Importantly, the tri-critical points and two strip phase diagram again suggest non-analyticity in the entanglement structure.

Let us now probe the structure of mutual information in the above mentioned four entangling phases. In these phases, the mutual information simply reduces to
\begin{eqnarray}
 I_1  (L,X) &=& S_A^{con}(L) + S_A^{con}(L) - 2 S_A^{con}(L) = 0 \, \nonumber \\
 I_2 (L,X)  &=&  S_A^{con}(L) + S_A^{con}(L)-S_A^{con} (X) - S_A^{con} (2L+X) \geq 0  \, \nonumber \\
 I_3  (L,X) &=& S_A^{con}(L) + S_A^{con}(L) - S_A^{con} (X) - S_A^{discon} \geq 0 \, \nonumber \\
 I_4 (L,X)  &=& S_A^{discon} + S_A^{discon} - 2 S_A^{discon} =0 \,.
 \label{mutual2strips}
\end{eqnarray}
Here the range of $X$ and $L$ in $I_2$ and $I_3$ should be understood to be restricted in their respective phases. The above equations also implies
\begin{eqnarray}
\frac{\partial I_1}{\partial L} \propto \frac{1}{G_{N}^0} = \mathcal{O}(N^0), \ \ \ \ \ \ \frac{\partial I_2}{\partial L} \propto \frac{1}{G_N} = \mathcal{O}(N^2)  \nonumber \\
\frac{\partial I_3}{\partial L} \propto \frac{1}{G_N} = \mathcal{O}(N^2), \ \ \ \ \ \ \frac{\partial I_4}{\partial L} \propto \frac{1}{G_{N}^{0}} = \mathcal{O}(N^0) \,.
\end{eqnarray}

\begin{figure}[h!]
\begin{minipage}[b]{0.5\linewidth}
\centering
\includegraphics[width=2.8in,height=2.3in]{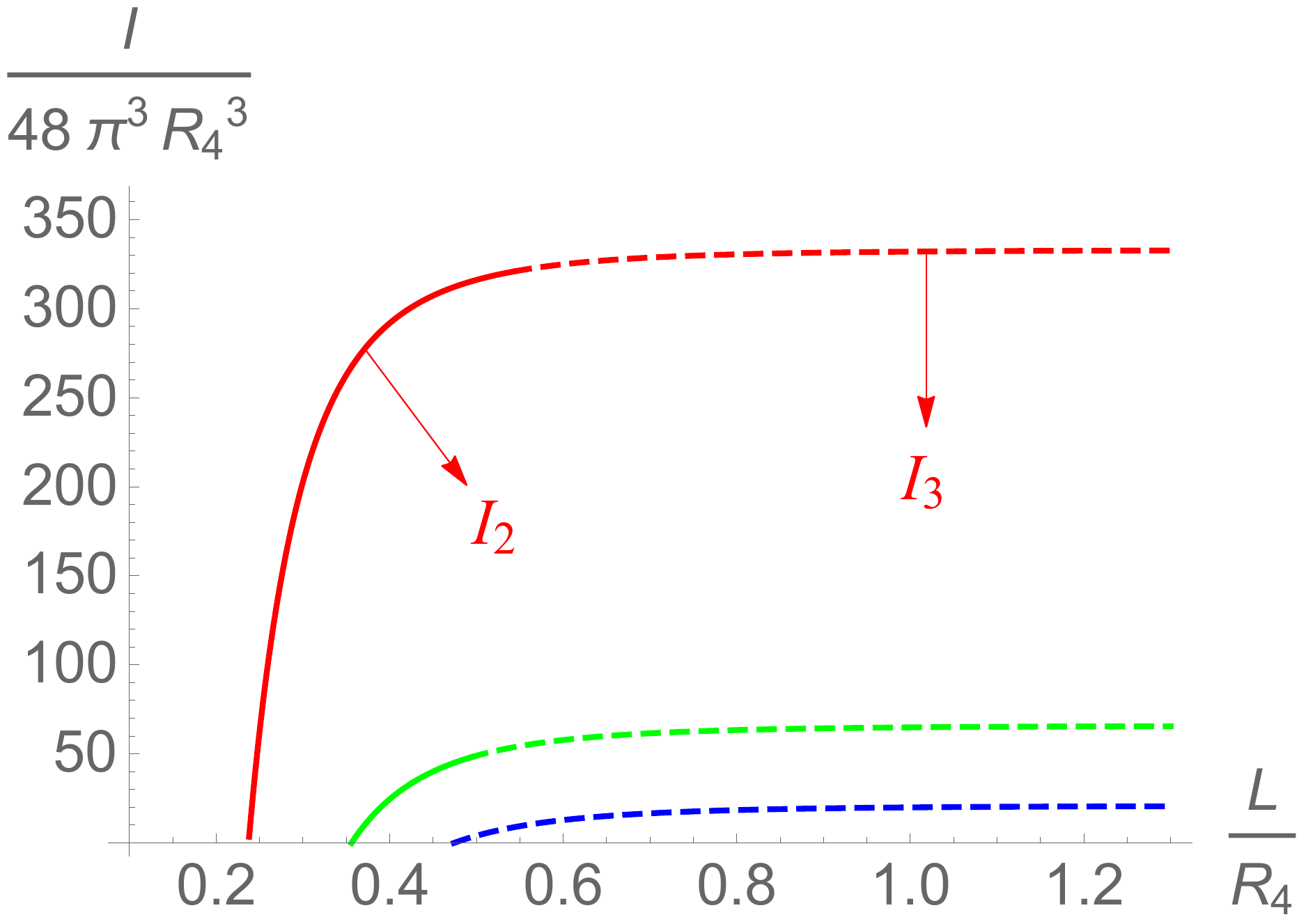}
\caption{ \small Mutual Information of $S_2$ and $S_3$ phases as a function of $L$. The solid lines correspond to
$I_{2}$ whereas the dashed lines corresponds to $I_{3}$.  The red, green and blue lines correspond to separation length $X/R_4=0.2$, $0.3$ and $0.4$ respectively.}
\label{LvsMIVsXD4}
\end{minipage}
\hspace{0.4cm}
\begin{minipage}[b]{0.5\linewidth}
\centering
\includegraphics[width=2.8in,height=2.3in]{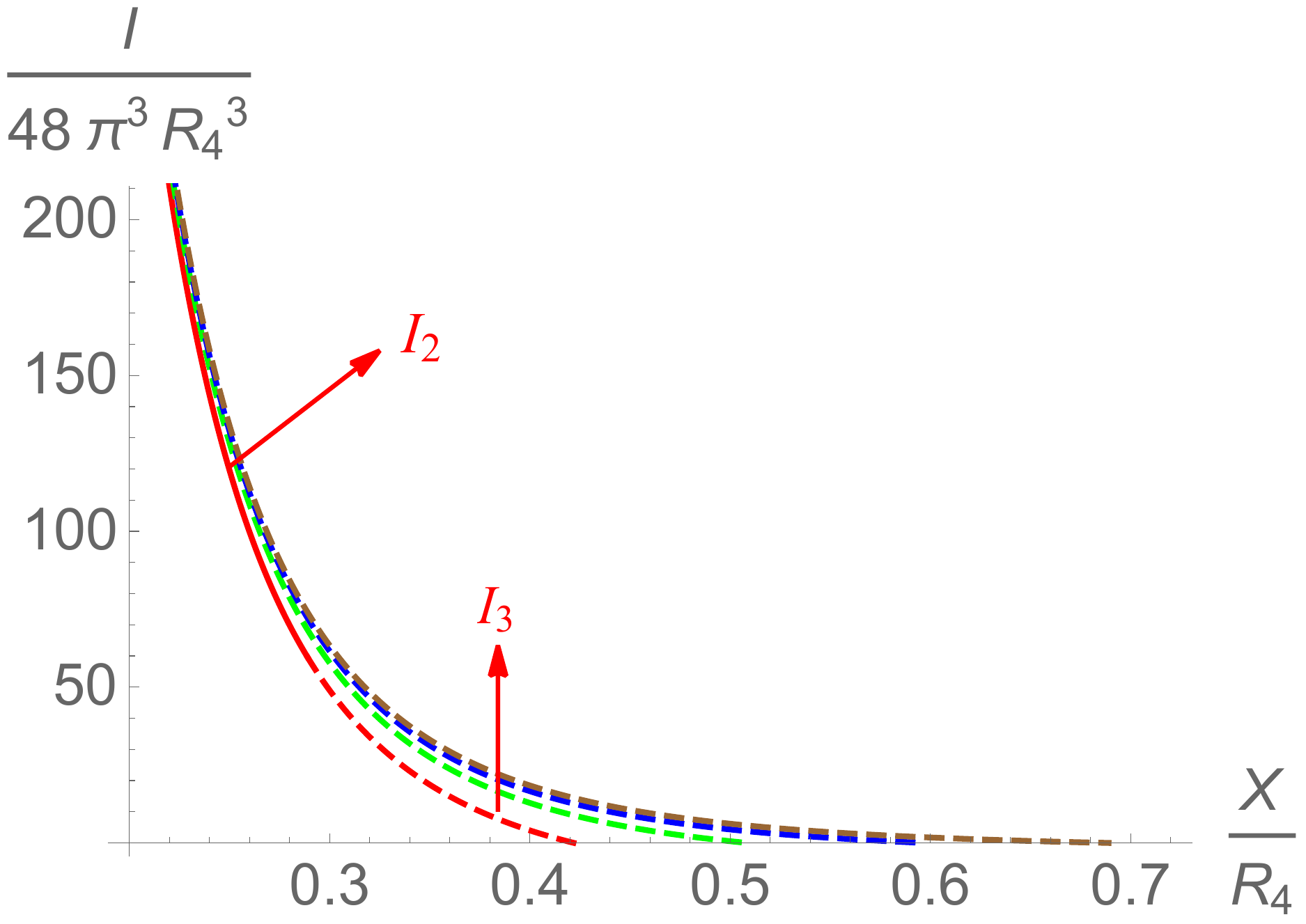}
\caption{\small Mutual Information of $S_2$ and $S_3$ phases as a function of $X$. The solid and dashed lines correspond to $I_{2}$ and $I_{3}$ respectively.
The red, green, blue and brown lines correspond to $L/R_4=0.5$, $0.6$, $0.7$ and $0.8$ respectively. }
\label{XVsMIVsLD4}
\end{minipage}
\end{figure}

As expected, the mutual information is zero for the phases containing two largely separated subsystems, \textit{i.e.} for $S_1$ and $S_4$ phases, whereas for the phases $S_2$ and $S_3$ it is non-zero and finite. The variation of mutual information with respect to strip length $L$ and separation length $X$ are shown in Figures~\ref{LvsMIVsXD4} and \ref{XVsMIVsLD4}. It turns out that the mutual information is not only a smooth function of $L$ and $X$ but also behaves smoothly as it passes from one phase to another. For instance, as shown in Figure~\ref{LvsMIVsXD4}, it connects smoothly between $S_2$ and $S_3$ phases and no discontinuity arises as the $S_2$/$S_3$ critical line is approached. Similarly, the mutual information also smoothly goes to zero as $S_1$ (or $S_4$) phase is approached from $S_2$ (or $S_3$) side. This is shown in Figure~\ref{XVsMIVsLD4}.

An important point to note is that, unlike the entanglement entropy, the order of the mutual information may or may not change as we go from one phase to another. For instance, a change in the order of mutual information ($\mathcal{O}(N^2)$ to $\mathcal{O}(N^0)$) occurs as we go from $S_2$ to $S_1$ phase (by increasing $X$), however, no such change in the order occurs from $S_2$ to $S_3$ phase (by increasing $L$). It all depends on the particular phases involved in the transition.

\subsection{Entanglement wedge cross-section}
We now move on to discuss the entanglement wedge cross-section $E_W$ in the current confining background. For this purpose, note that the surface which divides the entanglement wedge into two parts, associated with $A$ and $B$ (see Figure~\ref{Entanglementwedgecrosspic}), can be identified by symmetry consideration to be a vertical flat surface $\Sigma$. Therefore, the symmetry of the strip configuration ensures that the entanglement wedge cross-section in the confining background is given by the area of a constant-$x$ hypersurface located in the middle of the strips (see Figure~\ref{twostripphasediag}). The induced metric on $\Sigma$ (a surface defined with $t=\text{constant}$ and $x=\text{constant}$) is then given by
\begin{eqnarray}
(ds^{2})_{\Sigma}^{ind}=\biggl(\frac{U}{R}\biggr)^{3/2} \biggl[\biggl(\frac{R}{U}\biggr)^3 \frac{dU^2}{f(U)}+dx^i dx_i \biggr] + R^{3/2}U^{1/2}d\Omega_{4}^{2} + \biggl(\frac{U}{R}\biggr)^{3/2}f(U)(dx^4)^2 \,
\label{d4metricEWind}
\end{eqnarray}
where $i=1,2$. Correspondingly, the entanglement wedge cross-section for the dual confining theory is be given by
\begin{eqnarray}
E_W=\frac{1}{4 G_N^{(10)}}\int d^8\sigma e^{-2\phi} \sqrt{g_{\Sigma}^{ind}} \,,
\end{eqnarray}
here we have again used the fact that the induced metric in Eq.~(\ref{d4metricEWind}) is in string frame. As mentioned above, there are in fact four different minimal area surfaces for two parallel strips. Since there is no correlation between the two disjointed strips in $S_1$ and $S_4$ phases, the corresponding entanglement wedge cross-section is trivially zero. For the connected $S_2$ and $S_3$ phases, the entanglement wedge cross-section is non-zero. For $S_2$ phase, it is given by
\begin{eqnarray}
& & E_W^{2}=\frac{L_2 L_3 \omega_4 (2 \pi R_4)}{4 G_N^{(10)}}    \int_{U_*(2L+X)}^{U_*(X)} dU \ R^3 U \, \nonumber \\
& &  \ \ \ \ \ \ = \frac{L_2 L_3}{4 G_N^{(10)}} \frac{\left(12 \pi^3 R_4^{3} U_0 \right)}{2} \left[U_*^{2}(X) - U_*^{2}(2L+X) \right]
\label{EWS2D4}
\end{eqnarray}
Similarly, for the $S_3$ phase it is given by,
\begin{eqnarray}
& & E_W^{3}=\frac{L_2 L_3 \omega_4 (2 \pi R_4)}{4 G_N^{(10)}}    \int_{U_0}^{U_*(X)} dU \ R^3 U \, \nonumber \\
& &  \ \ \ \ \ \ = \frac{L_2 L_3}{4 G_N^{(10)}} \frac{\left(12 \pi^3 R_4^{3} U_0 \right)}{2} \left[U_*^{2}(X) - U_0^2 \right]
\label{EWS3D4}
\end{eqnarray}
From the above definition it is clear that $E_W^{2}$ and $E_W^{3}$ are both positive as $U_{*}(x) \geq U_{*}(2L+X) \geq U_0$.
\begin{figure}[h!]
\begin{minipage}[b]{0.5\linewidth}
\centering
\includegraphics[width=2.8in,height=2.3in]{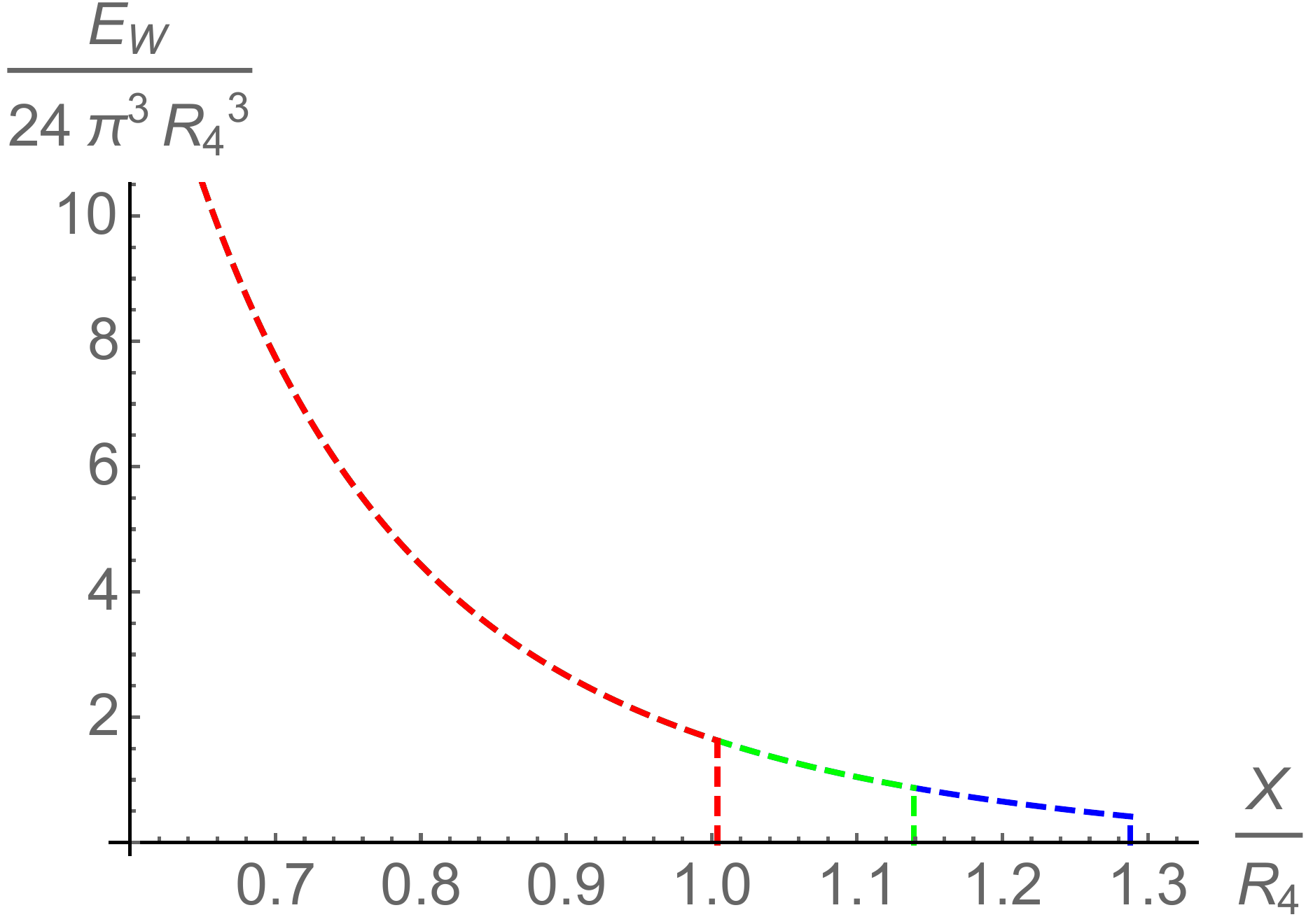}
\caption{ \small $E_W$ as a function of separation length $X$ for different values of strip length $L$. Here blue, green and red curves corresponds to $L/R_4 = 1.3 > L_{crit}$,  $1.2$ and $1.1$ respectively.}
\label{XVsEWVsLD4}
\end{minipage}
\hspace{0.4cm}
\begin{minipage}[b]{0.5\linewidth}
\centering
\includegraphics[width=2.8in,height=2.3in]{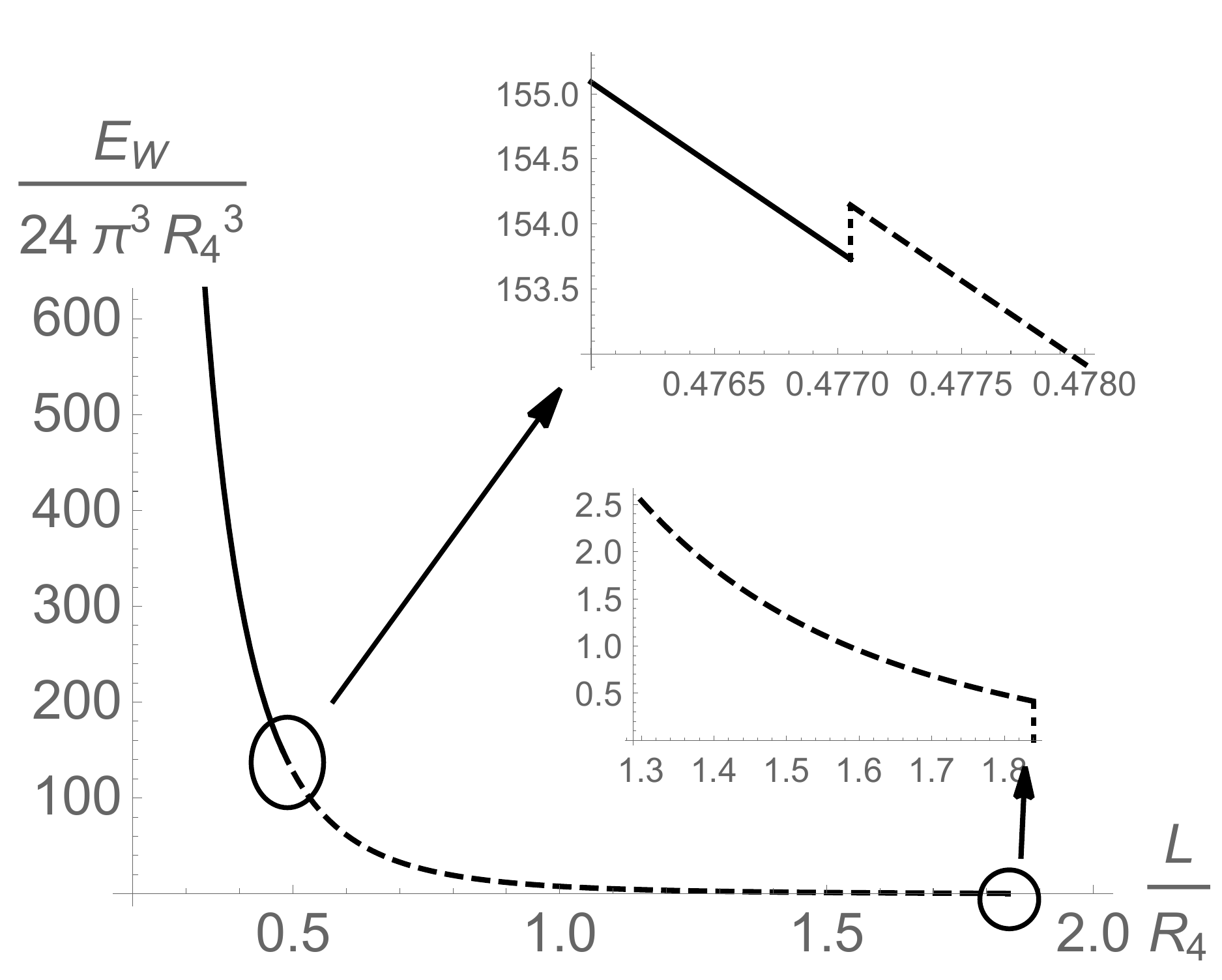}
\caption{\small $E_W$ as a function of separation length $L$ along a fixed line $X=0.7 L$. Here solid and dashed lines correspond to $E_W$ of $S_2$ and $S_3$ phases respectively.}
\label{LVsEWXPt7LD4}
\end{minipage}
\end{figure}

In Figure~\ref{XVsEWVsLD4}, we have shown the variation of entanglement wedge as a function of $X$ for a few values of $L$. Here, near critical values of $L$ are chosen so that $E_W$ behavior near the $S_1/S_3$ and $S_3/S_4$ phase transition points can be seen. We find that $E_W$ is a monotonic function of $X$ and is discontinuous at the transition point. In particular, $E_W$ does not go to zero as the $S_3/S_4$ critical line is approached from the $S_3$ side (blue curve). The discontinuous behavior can also be seen mathematically from Eq.~(\ref{EWS3D4}). Notice that for $E_W^{3}$ to be zero at the $S_3/S_4$ critical line, one requires $U_{*}(X=L_{crit})=U_0$. However, $U_{*}(L_{crit})$ is always larger than $U_0$ in the current confining background, which implies that $E_W^{3}$ is greater than zero at $X_{crit}$. Similarly, $E_W^{3}$ also does not go to zero at the $S_1/S_3$ transition line (red curve). This analysis suggests that, in the confining phase, the entanglement wedge cross-section vanishes discontinuously for large values of $X$ and $L$.

It is also interesting to investigate the behaviour of $E_W$ near the $S_2/S_3$ critical line, considering that $E_W$ is non-zero in both these phases. The results are shown in Figure~\ref{LVsEWXPt7LD4}. Here, we have considered a particular line, $X=0.7 L$, so that the behaviour of $E_W$ in $S_2$, $S_3$ and $S_4$ phases can be probed simultaneously. $E_W$ again turns out to be discontinuous at the $S_2/S_3$ transition line. This can also again be seen mathematically from Eq.~(\ref{EWS2D4}) and (\ref{EWS3D4}). In particular, the condition $U_*^{2}(2L+X)\neq U_0$ ensures that $E_W^{2}$ and $E_W^{3}$ do not correspond to a same value at the $S_2/S_3$ critical point. Moreover, as the $S_2/S_3$ critical point is approach from $S_2$ side, there is an upward jump in $E_W$ value \textit{i.e.} $E_W^{3}>E_W^{2}$, suggesting that the area of the entanglement wedge grows at the critical point. This result can again be traced back to the fact that $U_*(2L+X)> U_0$ and is also clear from Figure~\ref{twostripphasediag}. Therefore, it is clear that $E_W$ shows non-trivial features each time a phase transition between different phases occur.

\begin{figure}[h!]
\begin{minipage}[b]{0.5\linewidth}
\centering
\includegraphics[width=2.8in,height=2.3in]{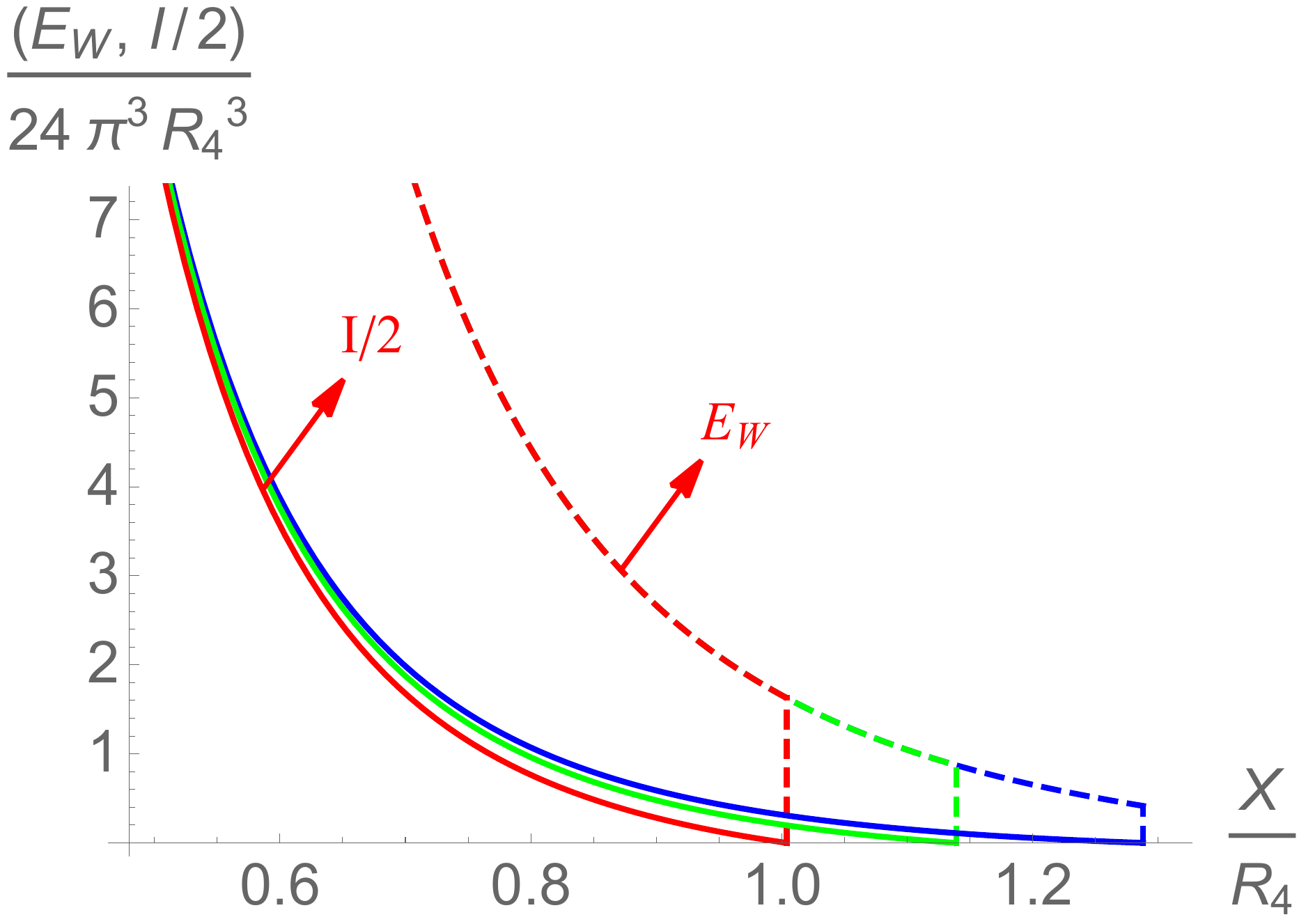}
\caption{ \small Mutual information $I$ and entanglement wedge $E_W$ as a function of $X$ for different values of $L$. The solid curves correspond to $I/2$ whereas the dashed curves correspond to $E_W$. Here Red, green and blue curves correspond to $L/R_4 = 1.1$,  $1.2$ and $1.3$ respectively.}
\label{XVsEWandMIVsLD4}
\end{minipage}
\hspace{0.4cm}
\begin{minipage}[b]{0.5\linewidth}
\centering
\includegraphics[width=2.8in,height=2.3in]{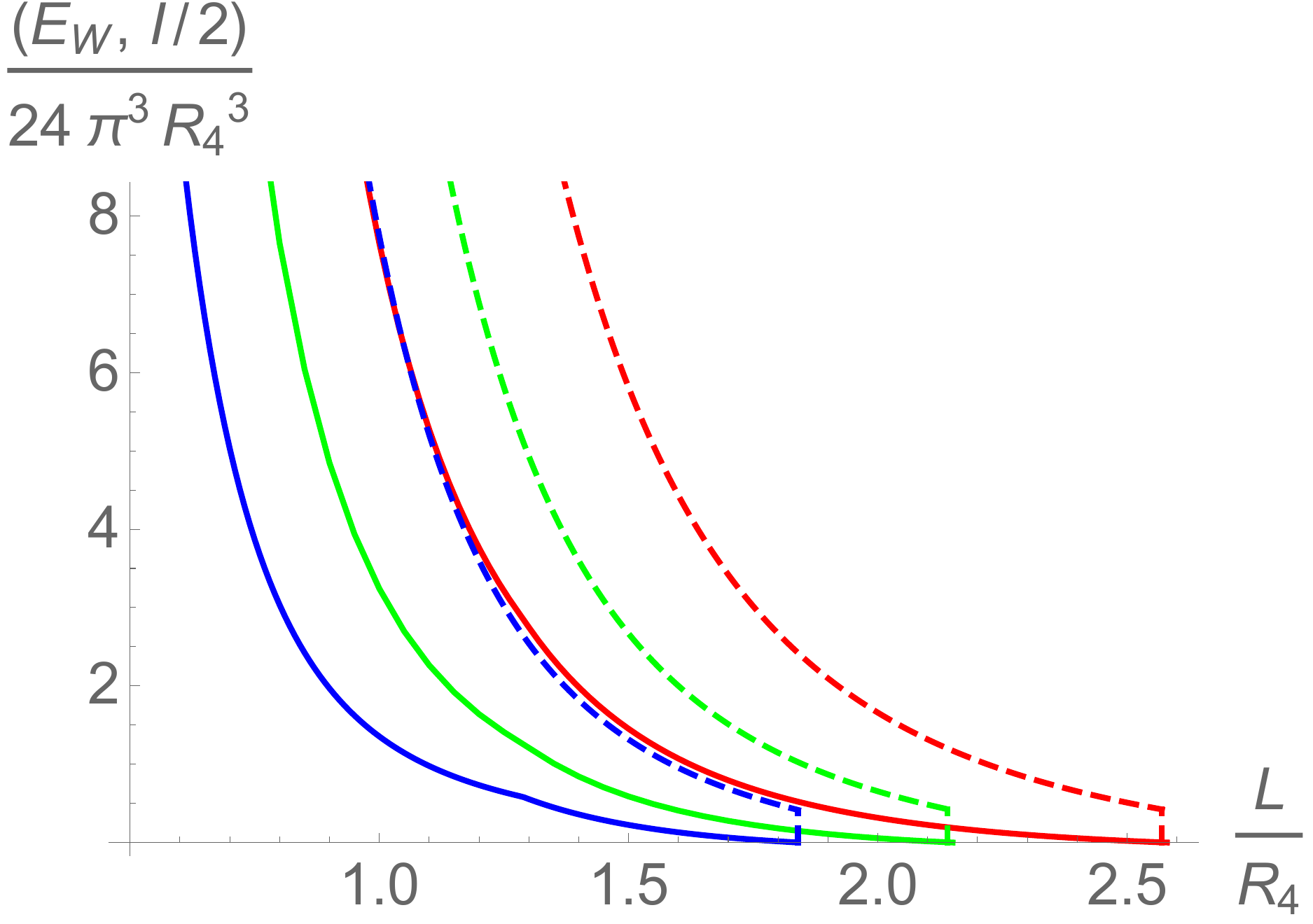}
\caption{\small Mutual information $I$ and entanglement wedge $E_W$ as a function of $L$ along a fixed line $X = \alpha L$. The solid curves correspond to $I/2$ whereas the dashed curves correspond to $E_W$. Here Red, green and blue curves correspond to $\alpha = 0.5$, $0.6$ and $0.7$ respectively.}
\label{LVsEWandMIXPt7LD4}
\end{minipage}
\end{figure}

The entanglement wedge is known to satisfy a few inequalities in holographic settings. In particular, it has been proved that $E_W$ always at least exceeds half the mutual information \textit{i.e.} $E_W \geq I/2$ \cite{Takayanagi:2017knl,Harper:2019lff}. In Figures~\ref{XVsEWandMIVsLD4} and \ref{LVsEWandMIXPt7LD4}, we made a comparison between $I/2$ and $E_W$ for diverse values of $X$ and $L$, and find that this inequality is always satisfied in the current holographic confining model. As we will show in later sections, this inequality will remain true for other holographic confining models as well, including a phenomenological bottom-up model. Therefore, the inequality $E_W \geq I/2$ appears to be a generic feature of all holographic models. It would certainly be interesting to see whether such an inequality exists in real QCD or not.

\subsection{Entanglement negativity}
We now move on to discuss the entanglement negativity in the current confining theory using the holographic prescription suggested in \cite{Chaturvedi:2016rft,Chaturvedi:2016rcn,Jain:2017aqk}. For a single interval, the entanglement negativity is given by Eq.~(\ref{HEEsc1}). Note that in the confined phase where the disconnected surface is more favorable for large strip length, the limiting condition $B\rightarrow A^c\rightarrow\infty$ ensures that
\begin{equation}
\mathcal{A}_{B_1}=\mathcal{A}_{B_2}=\mathcal{A}_{A\cup B_1}=\mathcal{A}_{A\cup B_2}=\mathcal{A}_{disconn}\ ,
\end{equation}
the above four terms cancel out in Eq.~(\ref{HEEsc1}) and we are left with
\begin{eqnarray}\label{HEEscD4}
& & \mathcal{E} = \lim_{B\rightarrow A^c}\frac{3}{4}\left[2S_A+S_{B_1}+S_{B_2}-S_{A\cup B_1}-S_{A\cup B_2}\right]\, \nonumber \\
& & \mathcal{E} = \frac{3}{2} S_A \,.
\end{eqnarray}
Interestingly, the entanglement negativity is just $3/2$ times of the entanglement entropy. Since $S_A$ depends on $L$ and exhibits a discontinuous behaviour at $L_{crit}$, this discontinuity of $S_A$ manifests itself in
$\mathcal{E}$ as well. Therefore, interestingly, just like the entanglement entropy, the order of the entanglement negativity also changes from
$\mathcal{O}(N^2)$ to $\mathcal{O}(N^0)$ at $L_{crit}$,
\begin{equation}
\begin{split}
\frac{\partial\mathcal{E}}{\partial L} = \mathcal{O}(N^2)\ \mathrm{for}\ L<L_{crit}\\
\frac{\partial\mathcal{E}}{\partial L} = \mathcal{O}(N^0)\ \mathrm{for}\ L>L_{crit}.
\end{split}
\end{equation}
The discontinuous behaviour of the entanglement negativity in the confined phase is a genuine new prediction from holography (strictly speaking, a prediction from the entanglement negativity proposal of \cite{Chaturvedi:2016rft,Chaturvedi:2016rcn,Jain:2017aqk}) and should be independently tested. Unfortunately, there are no lattice results for the negativity in real QCD to compare with yet. It would really be interesting if this discontinuous nature could be tested via lattice calculations in the near future.

\begin{figure}[h!]
\begin{minipage}[b]{0.5\linewidth}
\centering
\includegraphics[width=2.8in,height=2.3in]{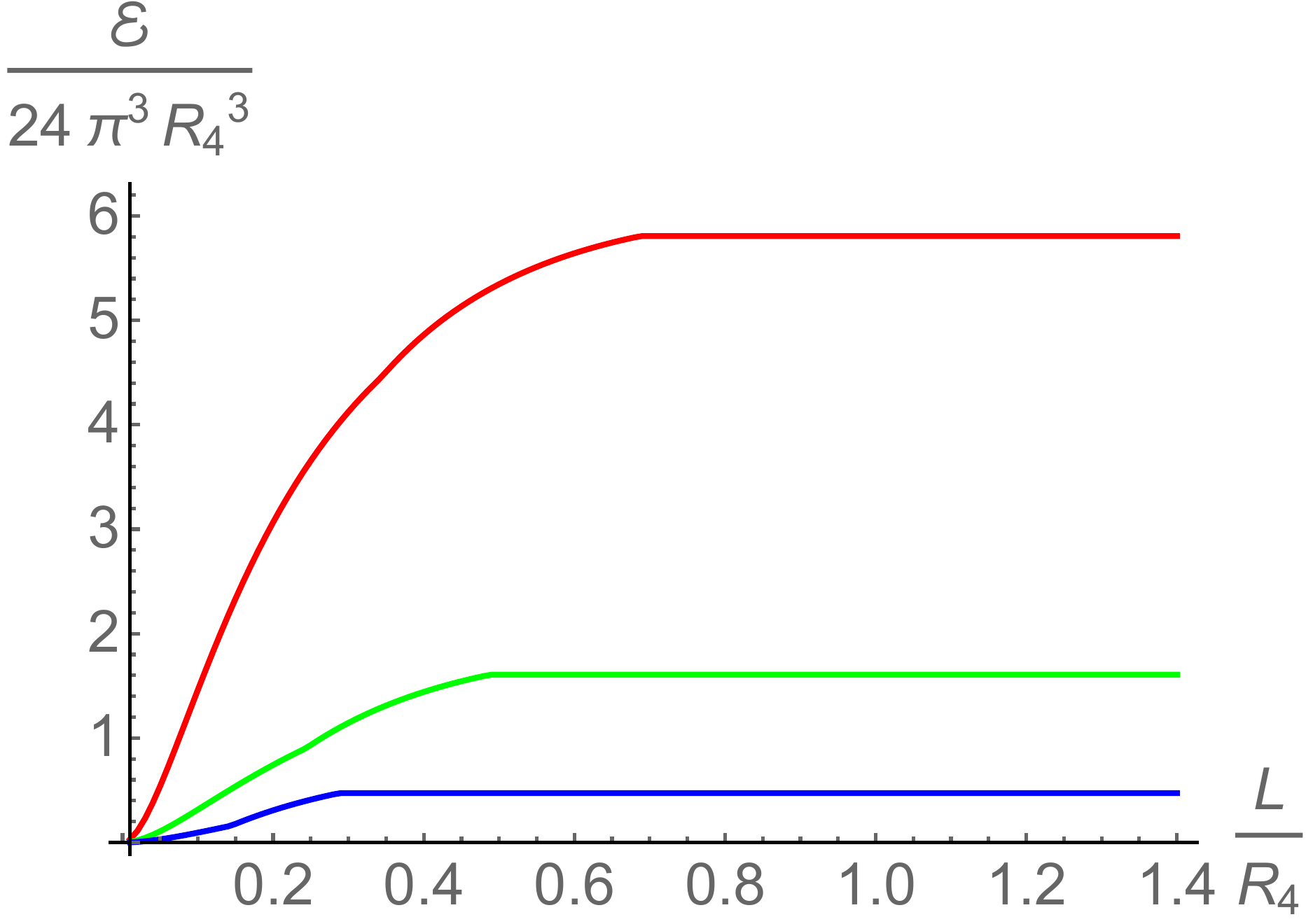}
\caption{ \small Entanglement negativity $\mathcal{E}$ of two parallel strips as a function of $L$ for different values of $X$. Here Red, green and blue curves correspond to $X/R_4 = 0.6$,  $0.8$ and $1.0$ respectively.}
\label{LVsENVsXtwostripD4}
\end{minipage}
\hspace{0.4cm}
\begin{minipage}[b]{0.5\linewidth}
\centering
\includegraphics[width=2.8in,height=2.3in]{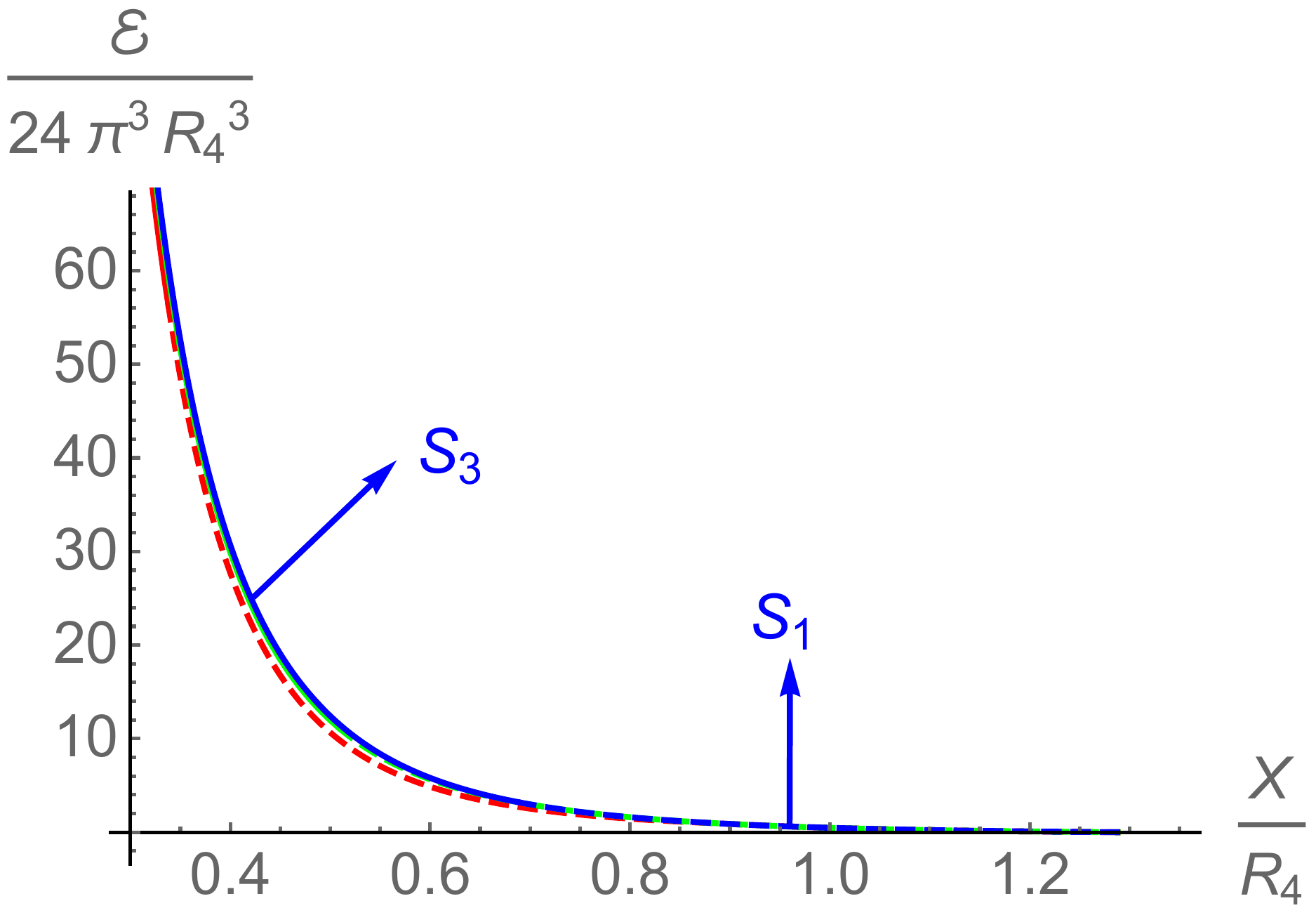}
\caption{\small Entanglement negativity $\mathcal{E}$ of two parallel strips as a function of $X$ for different values of $L$. Solid and dashed lines indicate $\mathcal{E}$ of $S_3$ and $S_1$ phases respectively.  Here Red, green and blue curves correspond to $L/R_4 = 0.4$,  $0.6$ and $0.8$ respectively.}
\label{XVsENVsLtwostripD4}
\end{minipage}
\end{figure}

Let us now discuss the entanglement negativity for two disjoint intervals \cite{Malvimat:2018txq,Basak:2020bot}. Compared to \cite{Malvimat:2018txq,Basak:2020bot}, we have
$l_s = X\ ,\ l_1=l_2=L$. So $\mathcal{E}$ is given as
\begin{equation}
\mathcal{E}=\frac{3}{4}\left[S_A(L+X)+S_A(L+X)-S_A(2L+X)-S_A(X)\right] \,,
\label{ENtwostripD4}
\end{equation}
where $S_A$ is again the entanglement entropy of a single strip of length $L$. Notice that for $X>L_{crit}$, $\mathcal{E}$ is trivially zero as every term in Eq.~(\ref{ENtwostripD4}) corresponds to $S_{A}^{discon}$. This implies that, just like the mutual information and entanglement wedge, $\mathcal{E}$ is zero in the $S_4$ phase as well. However, the same is not true for the $S_1$ phase. In particular, $\mathcal{E}$ is non-zero and monotonic function of $X$ and $L$ in $S_1$ phase. This is in sharp contrast with the mutual information and entanglement wedge behavior, which are zero in the $S_1$ phase as well. The results are shown in Figures~\ref{LVsENVsXtwostripD4} and \ref{XVsENVsLtwostripD4}. Note that $\mathcal{E}$ smoothly goes to zero at $X=L_{crit}$ whereas it remains finite for $L>L_{crit}$. Moreover, unlike the entanglement wedge, $\mathcal{E}$ is continuous across the various phase transition points but has a cusp (as is visible for Figure~\ref{LVsENVsXtwostripD4}). This is again a new prediction form holography and should be tested in lattice settings.

\section{D3-branes on a circle}\label{D3system}
Another top-down confining model can be obtained by wrapping $N_c$ $D3$-branes on a circle of radius $R_3$ with twisted boundary conditions for the fermions. Before the wrapping, the low energy dynamics of $N_c$ coincidental $D3$-branes is given by $\mathcal{N}=4$ Supersymmetric Yang-Mills theory with 't Hooft coupling $\lambda=g_s N_c$. This theory can be reduced to $(2+1)$ dimensional confining theory at long  distances by compactifying on a circle. The low energy dynamics of this system is then given by the dimensionless parameter $\lambda_3=\lambda/R_3$ \cite{Witten:1998zw}. In particular, for $\lambda \ll 1$, the theory is described $(2+1)$ dimensional Yang-Mills theory with coupling $\lambda_3$. For $\lambda\gg 1$, on the other hand, one can use the dual gravitational picture in terms of the near-horizon geometry
of the $N_c$ coincidental $D3$-branes,
\begin{eqnarray}
& & ds^2 = \left(\frac{U}{R}\right)^{2}\left[\left(\frac{R}{U}\right)^{4}\frac{dU^2}{f(U)}+dx^{\mu}dx_{\mu}\right]+ R^{2}d\Omega_5^2+\left(\frac{U}{R}\right)^{2}f(U)(dx^3)^2\,
\end{eqnarray}
where $R$ is the AdS length scale and,
\begin{eqnarray}
& & f(U)=1-\left(\frac{U_0}{U}\right)^{4},\ U_0^2=\frac{\pi\lambda}{R_3^2},\ R^4=4\pi\lambda.
\end{eqnarray}
Dilation is a constant and here we take it to be zero. Note that this geometry again forms a cigar shape in $(U,x^3)$ coordinates, with radius of the $x^3$-circle goes to zero as $U\rightarrow U_0$.

\subsection{Entanglement entropy: one strip}
The entanglement entropy for this top-down model has also been computed previously in \cite{Klebanov:2007ws}. Here, we first review their calculation. To compute the entanglement entropy, we consider the previous D4-branes setup by taking a strip subsystem of length $L$, with subsystem domain $-L/2\leq x_1=x\leq L/2$ and $0\leq x_2\leq L_2$. Using $U=U(x)$, we get the entanglement entropy (\ref{heenc}) as,
\begin{eqnarray}\label{EED4}
S_A = \frac{L_2\omega_5 (2\pi R_3)}{4 G_N^{(10)}} \int dx \ R^{2}U^{3} \sqrt{f(U)+\left(\frac{R}{U}\right)^4 U'^2}
\end{eqnarray}
where $\omega_5$ is the area of unit five sphere.
In this case also, there are two minimal area surfaces: a ($U$-shaped) connected surface and a disconnected surface.
The entanglement entropy of the connected surface turns out to be
\begin{eqnarray}\label{d3Scon}
S_A^{con} = \frac{L_2}{2 G_N^{(10)}}(\pi^4R^6)\int_{U_*}^{U_{\infty}}dU\ \frac{U^{4}}{U_0}\frac{\sqrt{f(U)}}{\sqrt{U^6f(U)-U_*^6f(U_*)}} \,.
\end{eqnarray}
where $U_*$ is the turning point for the above connected surface and $U'(x)|_{U=U_*}=0$. Further, the strip length $L$ as a function of $U_*$ is
\begin{eqnarray}\label{d3stripl}
L(U_*) = 2 R^{2}\int_{U_*}^{U_{\infty}}dU\ \frac{U_*^{3}}{U^{2}}\sqrt{\frac{f(U_*)}{f(U)}}\frac{1}{\sqrt{U^6f(U)-U_*^6f(U_*)}} \,.
\end{eqnarray}
and the expression of the entanglement entropy of the disconnected surface is
\begin{eqnarray}\label{d3Sdiscon}
S_A^{discon} = \frac{V_2}{2G_{10}}(\pi^4R^6)\int_{U_0}^{U_{\infty}}dU\ \frac{U}{U_0}\,.
\end{eqnarray}
It is worthwhile to note here that the entanglement entropy of the disconnected surface is again independent of the strip length. This information will be relevant and useful when studying the entanglement structure of the current $D3$-brane confined setup.

\begin{figure}[h!]
\begin{minipage}[b]{0.5\linewidth}
\centering
\includegraphics[width=2.8in,height=2.3in]{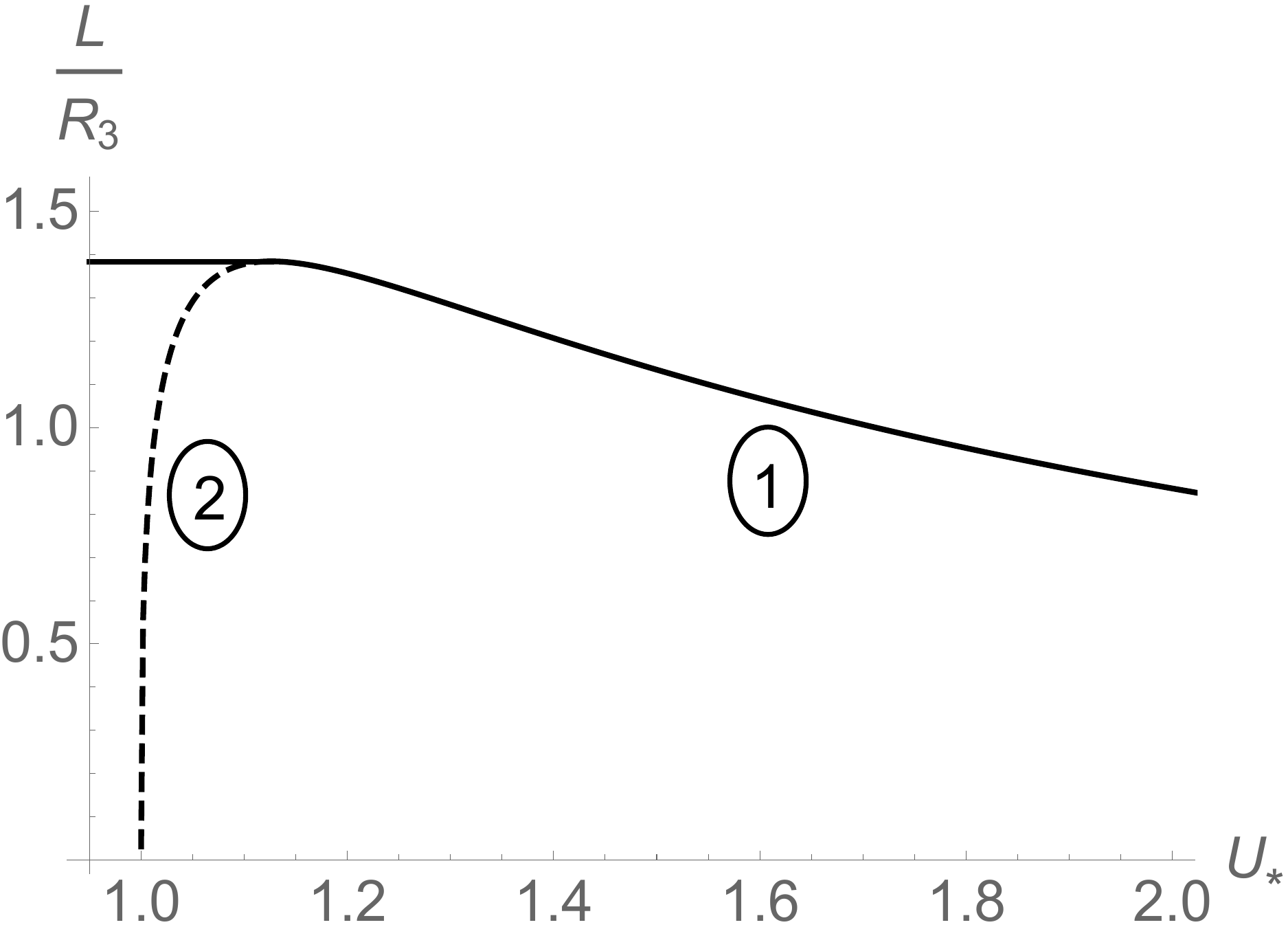}
\caption{ \small The behaviour of strip length $L$ as a function of $U_*$.}
\label{UsVsL1stripD3}
\end{minipage}
\hspace{0.4cm}
\begin{minipage}[b]{0.5\linewidth}
\centering
\includegraphics[width=2.8in,height=2.3in]{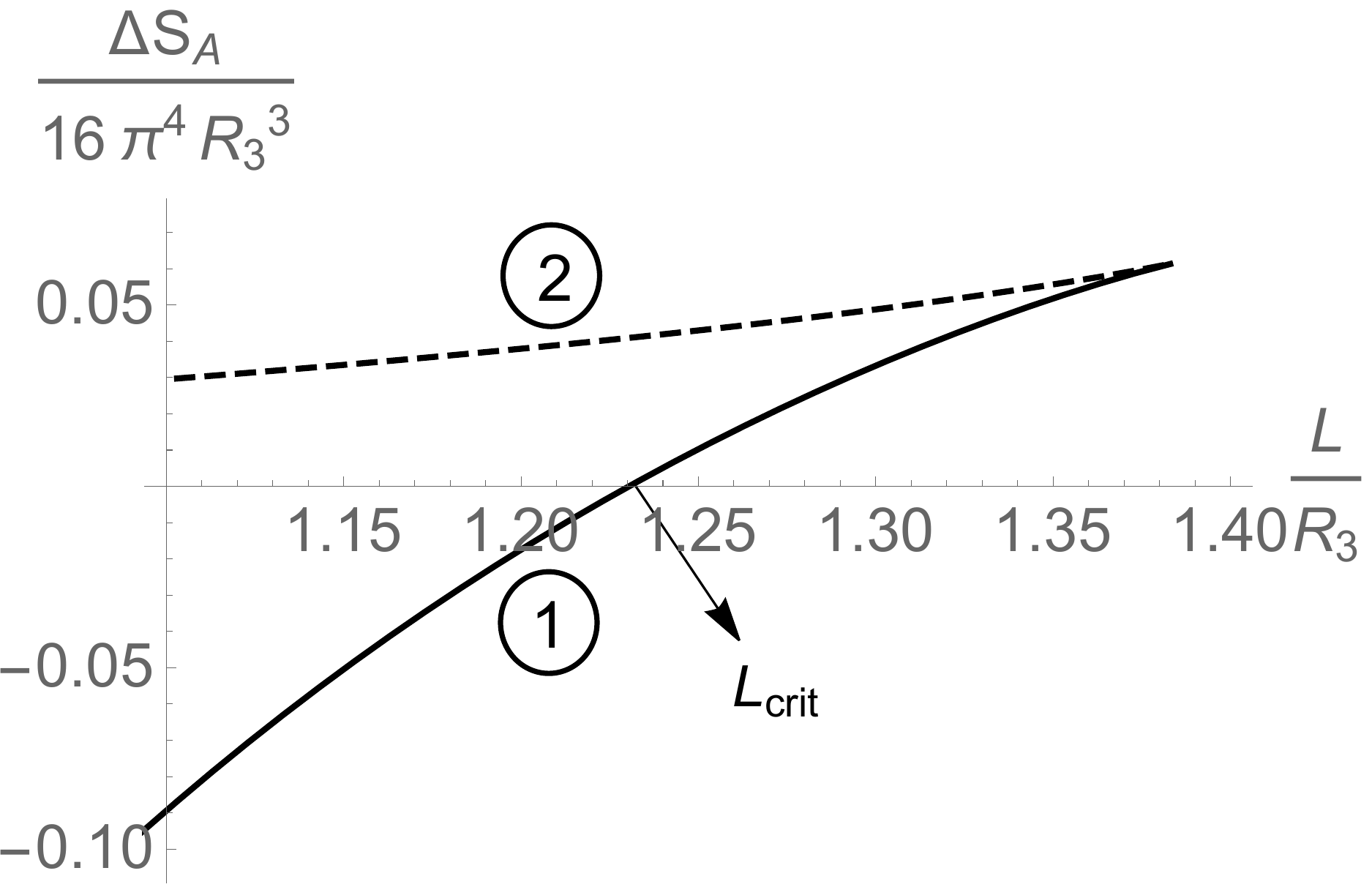}
\caption{\small $\triangle S_A= S_A^{con}-S_A^{discon}$ as a function of strip length $L$.}
\label{LVsdeltaS1stripD3}
\end{minipage}
\end{figure}

We now show the numerical results of the entanglement entropy in the current confining model. For numerical purpose, we again considered $U_0=1$. These numerical results are presented in the
Figures~\ref{UsVsL1stripD3} and \ref{LVsdeltaS1stripD3}, where in the first case we have plotted $L$ as a function of $U_*$ and in the second case we have plotted the difference between connected and disconnected entanglement entropies ($\triangle S_A = S_A^{con} - S_A^{discon}$) as a function of the strip length $L$ respectively \footnote{We have used $L_2/4G_{N}^{(10)}=1$ in the numerical calculations.}. Note that here also, like in the previous D4-brane model, we have three minimal area surfaces for a given $L$: one disconnected and two connected. In the plots below, the connected surface $\circled{1}$  (represented by a solid line) is closer to the boundary in comparison to the second connected surface $\circled{2}$ (represented by a dashed line). The second connected surface $\circled{2}$ again corresponds to a saddle point and its area is always higher than the first connected surface $\circled{1}$. Moreover, the existence of the connected surfaces is dictated by $L_{max}\simeq1.383R_3$, as these surfaces can occur only below $L_{max}$. For values of $L$ beyond $L_{max}$, we only have the disconnected surface.

We see from Figure~\ref{LVsdeltaS1stripD3} that $\triangle S_A$ changes sign with $L$. For small $L$, $\triangle S_A$ is negative. This reflects the fact that for small subsystem length the connected surface
$\circled{1}$ has the smallest area. On the other hand, for large
$L$, $\triangle S_A$ is positive, reflecting the fact that for large subsystem length the disconnected surface has the smallest area. Correspondingly, we have a phase transition at $L=L_{crit}\simeq1.23 R_3$ between connected  and disconnected entanglement entropies. As the minimal surface area for the disconnected surface is independent of $L$, its entanglement entropy is also independent of $L$. This connected/disconnected phase transition, therefore, leads to the following result (similar to the previous D4-brane setup),
\begin{eqnarray}
\frac{\partial S_{A}}{\partial L} &\propto &\frac{1}{G_N^{(10)}} = \mathcal{O}(N^2)\quad\text{for}\quad L < L_{crit}\,, \nonumber \\
&\propto& \frac{1}{[G_{N}^{(10)}]^0} = \mathcal{O}(N^0)\quad\text{for}\quad L > L_{crit}
\end{eqnarray}
Importantly, once again the order of the entanglement entropy changes as subsystem size is varied in the confined phase.

\subsection{Mutual information: two strips}
We now move on to discuss the entanglement structure of two disjoint strips in the current confining system. The two strip phase diagram of this system has been studied previously in \cite{Ben-Ami:2014gsa}. Here, we first reproduce their results to set the stage for the discussion of mutual information and entanglement wedge cross-section in this system later on.  For our purpose, we again consider equal strip lengths \textit{i.e.} ($L_1=L_2=L$) for simplicity. This leads to four entangling surfaces $\{S_1,S_2,S_3,S_4\}$, the expression of which are given in Eq.~(\ref{2stripEE}). See Figure~\ref{twostripphasediag} for the pictorial representation of these surfaces.

\begin{figure}[h!]
\centering
\includegraphics[width=2.8in,height=2.3in]{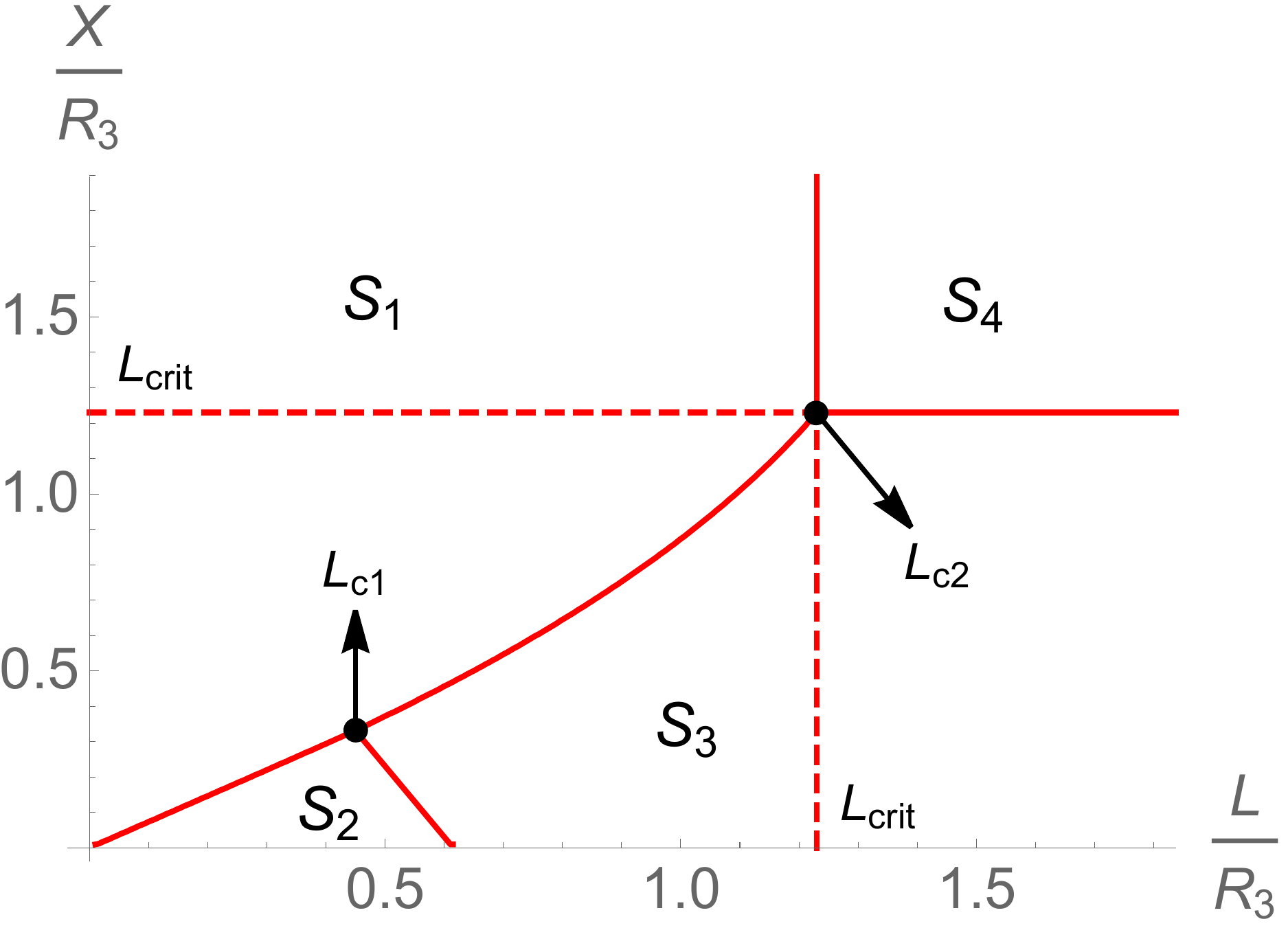}
\caption{\small The entanglement entropy phase diagram of various minimal surfaces for the case of two strips
of equal length $L$ separated by a distance $X$ in the confining background of D3-branes on a circle. These four different phases correspond to the four bulk surfaces
of Figure \ref{twostripphasediag}.}
\label{twostripphasediagD3}
\end{figure}

The two equal strip entanglement phase diagram is shown in Figure \ref{twostripphasediagD3}. The four phases $\{S_1,S_2,S_3,S_4\}$ again compete with each other, leading to a similar phase diagram as in the case of $D4$ branes on a circle confining system. For instance, for small values of $X,L\ll L_{crit}$, the $S_1$
configuration again have the lowest entanglement entropy, whereas for large values of $X,L > L_{crit}$, it is the $S_4$ configuration which has the lowest entanglement entropy. In between, we have $S_2$ and $S_3$ configurations. The $S_2$ to $S_3$ phase transition happens at $L=L_{crit}/2$ for $X=0$, whereas this phase transition is governed by the equation $2L+X=L_{crit}$ for a generic value of $X$.

The two tri-critical points, indicated by black dots in Figure \ref{twostripphasediagD3}, reflect the co-existence of three configurations together. The first
tri-critical point $L_{c1}$ denotes the co-existence of three configurations $\{S_1$, $S_2$, $S_3\}$ and its coordinates are ($L/R_3=0.45$, $X/R_3=0.331$) whereas the second tri-critical point $L_{c2}$ denotes the co-existence of $\{S_2$, $S_3$, $S_4\}$ configurations and its coordinates are ($L/R_3=1.23$, $X/R_3=1.23$). The two strip phase diagram, along with the presence of tri-critical points, again indicate the non-analytic nature of the entanglement structure in confining theories.

\begin{figure}[h!]
\begin{minipage}[b]{0.5\linewidth}
\centering
\includegraphics[width=2.8in,height=2.3in]{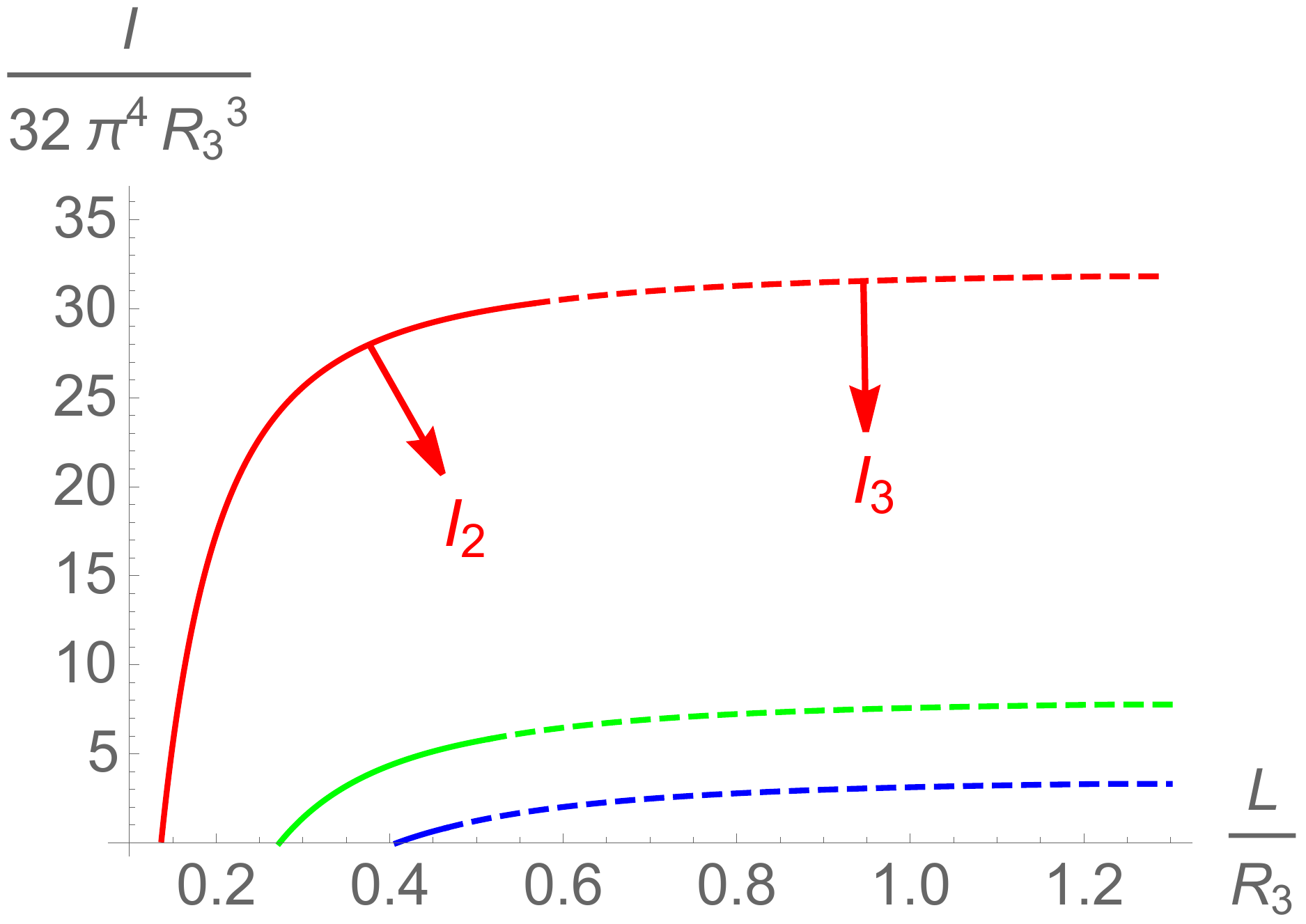}
\caption{ \small Mutual Information of $S_2$ and $S_3$ surfaces as a function of $L$. The solid lines correspond to
$I_{2}$ whereas the dashed lines correspond to $I_{3}$.  The red, green and blue lines correspond to separation length $X/R_3=0.1$, $0.2$ and $0.3$ respectively.}
\label{LvsMIVsXD3}
\end{minipage}
\hspace{0.4cm}
\begin{minipage}[b]{0.5\linewidth}
\centering
\includegraphics[width=2.8in,height=2.3in]{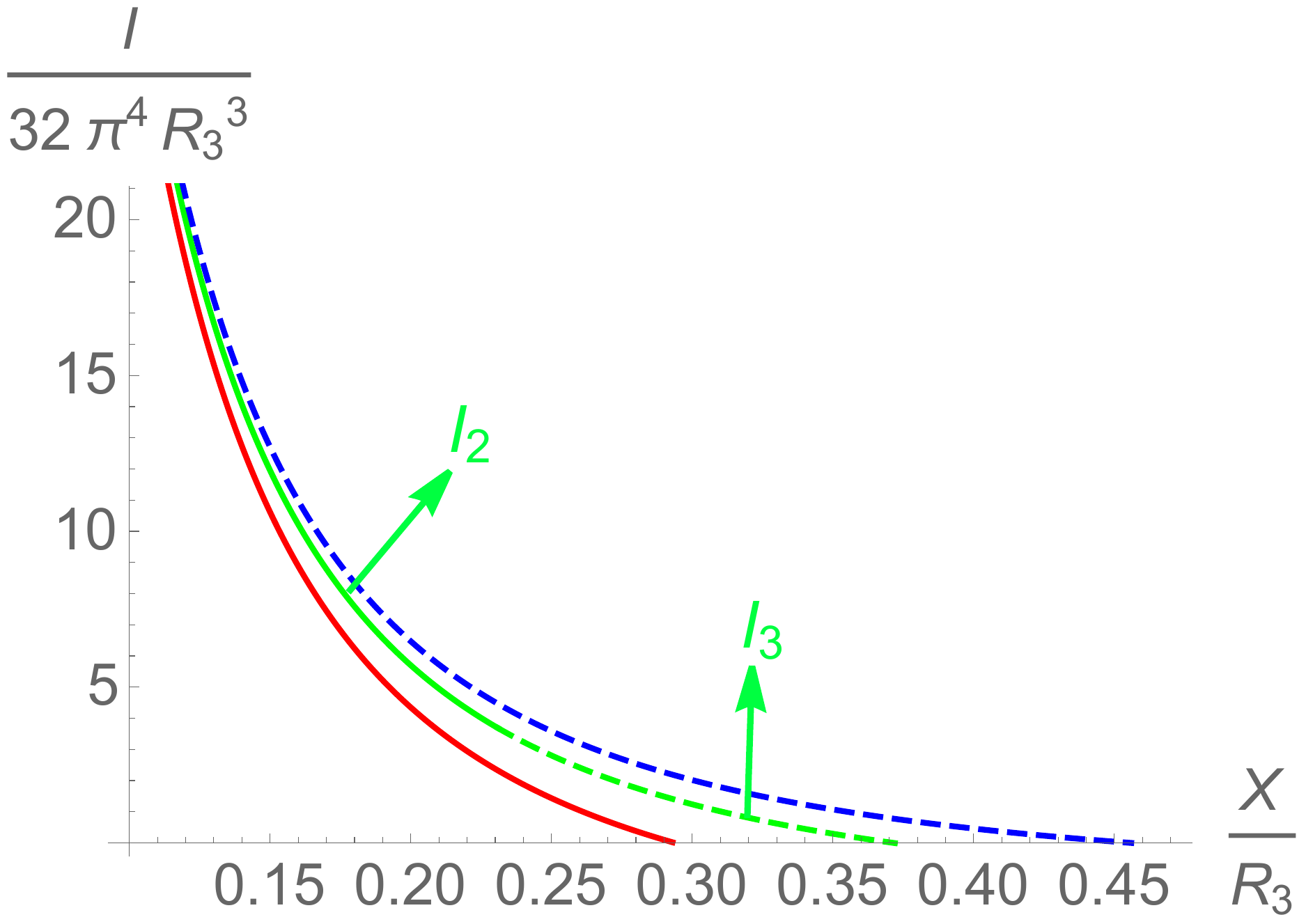}
\caption{\small Mutual Information of $S_2$ and $S_3$ surfaces as a function of $X$. The solid and dashed lines correspond to $I_{2}$ and $I_{3}$ respectively.
The red, green and blue lines correspond to $L/R_3=0.4$, $0.5$ and $0.6$ respectively. }
\label{XVsMIVsLD3}
\end{minipage}
\end{figure}
Similarly, one can compute the mutual information in these four phases. The relevant expressions are given in Eq.~(\ref{mutual2strips}). The mutual information is again zero when the two strips have a relatively large separation \textit{i.e.} for the $S_1$ and the $S_4$ configurations, whereas it is non-zero in the remaining $S_2$ and $S_3$ configurations. The behaviour of mutual information as a function of $L$ and $X$ is presented in Figures~\ref{LvsMIVsXD3} and \ref{XVsMIVsLD3}. Here the specific values of $L$ and $X$ are taken in order to move through the $S_2$ and $S_3$ configurations.
It turns out that the mutual information is a monotonic function of $L$ and $X$ and varies smoothly as these parameters are changed. Further, the mutual information also exhibits a smooth behaviour while passing from one configuration to another. An interesting aspect of this analysis is that there may or may not be a change in the order of mutual information. To be clearer, we see a change in the order of mutual information from $\mathcal{O}(N^2)$ to $\mathcal{O}(N^0)$ as we move from $S_2$ to $S_1$ by varying $X$. On the other hand, there is no such change in the order as we move from $S_2$ to $S_3$ by varying $L$. Therefore, we again have the following relations in the current confined phase
\begin{eqnarray}
\frac{\partial I_1}{\partial L} \propto \frac{1}{G_{N}^0} = \mathcal{O}(N^0), \ \ \ \ \ \ \frac{\partial I_2}{\partial L} \propto \frac{1}{G_N} = \mathcal{O}(N^2)  \nonumber \\
\frac{\partial I_3}{\partial L} \propto \frac{1}{G_N} = \mathcal{O}(N^2), \ \ \ \ \ \ \frac{\partial I_4}{\partial L} \propto \frac{1}{G_{N}^{0}} = \mathcal{O}(N^0) \,.
\end{eqnarray}

\subsection{Entanglement wedge cross-section}
Similar to the previous $D4$-brane setup, the symmetry of the strip configuration again dictates that the entanglement wedge cross-section in the current confining background is given by the area of a constant-$x$ hypersurface $\Sigma$, located in the middle of the strips (as depicted in Figure \ref{twostripphasediag}). From the following induced metric on $\Sigma$,
\begin{equation}\label{d3metricEWind}
(ds^2)^{ind}_{\Sigma} =\left(\frac{U}{R}\right)^{2}\left[\left(\frac{R}{U}\right)^{4}\frac{dU^2}{f(U)}+dx^{2}dx_{2}\right]+ R^{2}d\Omega_5^2+\left(\frac{U}{R}\right)^{2}f(U)(dx^3)^2\ .
\end{equation}
we can get the  expression of the Entanglement wedge cross-section as
\begin{equation}
E_{W}=\frac{1}{4G_{10}}\int(d^8\sigma)(e^{-2\phi})\sqrt{g_{\Sigma}^{ind}}\ .
\end{equation}
When considering the two disjoint strip setup, similar to the previous D4-brane case, the entanglement wedge cross-section is zero for the
$S_1$ and $S_4$ configurations while it is non-zero for the
$S_2$ and $S_3$ configurations. In particular, for the $S_2$ configuration it is given by:
\begin{equation}\label{EWS2D3}
\begin{split}
&E_W^{2}=\frac{L_2}{4G_{10}}(\pi^4R^6)\int_{U_*(2l+x)}^{U_*(x)}dU\ \frac{U}{U_0}\ ,\\
&\qquad=\frac{L_2}{4G_{10}}\left(\frac{\pi^4R^6}{2U_0}\right)\left[U_*^2(x)-U_*^2(2l+x)\right]\ ,\\
\end{split}
\end{equation}
whereas, for the $S_3$ configuration it is given by:
\begin{equation}\label{EWS3D3}
\begin{split}
&E_W^{3}=\frac{L_2}{4G_{10}}(\pi^4R^6)\int_{U_0}^{U_*(x)}dU\ \frac{U}{U_0}\ ,\\
&\qquad=\frac{L_2}{4G_{10}}\left(\frac{\pi^4R^6}{2U_0}\right)\left[U_*^2(x)-U_0^2\right].
\end{split}
\end{equation}
From the above equations one can again conclude that both $E_W^{2}$ and $E_W^{3}$ are positive, as $U_{*}(x) \geq U_{*}(2L+X) \geq U_0$.

\begin{figure}[h!]
\begin{minipage}[b]{0.5\linewidth}
\centering
\includegraphics[width=2.8in,height=2.3in]{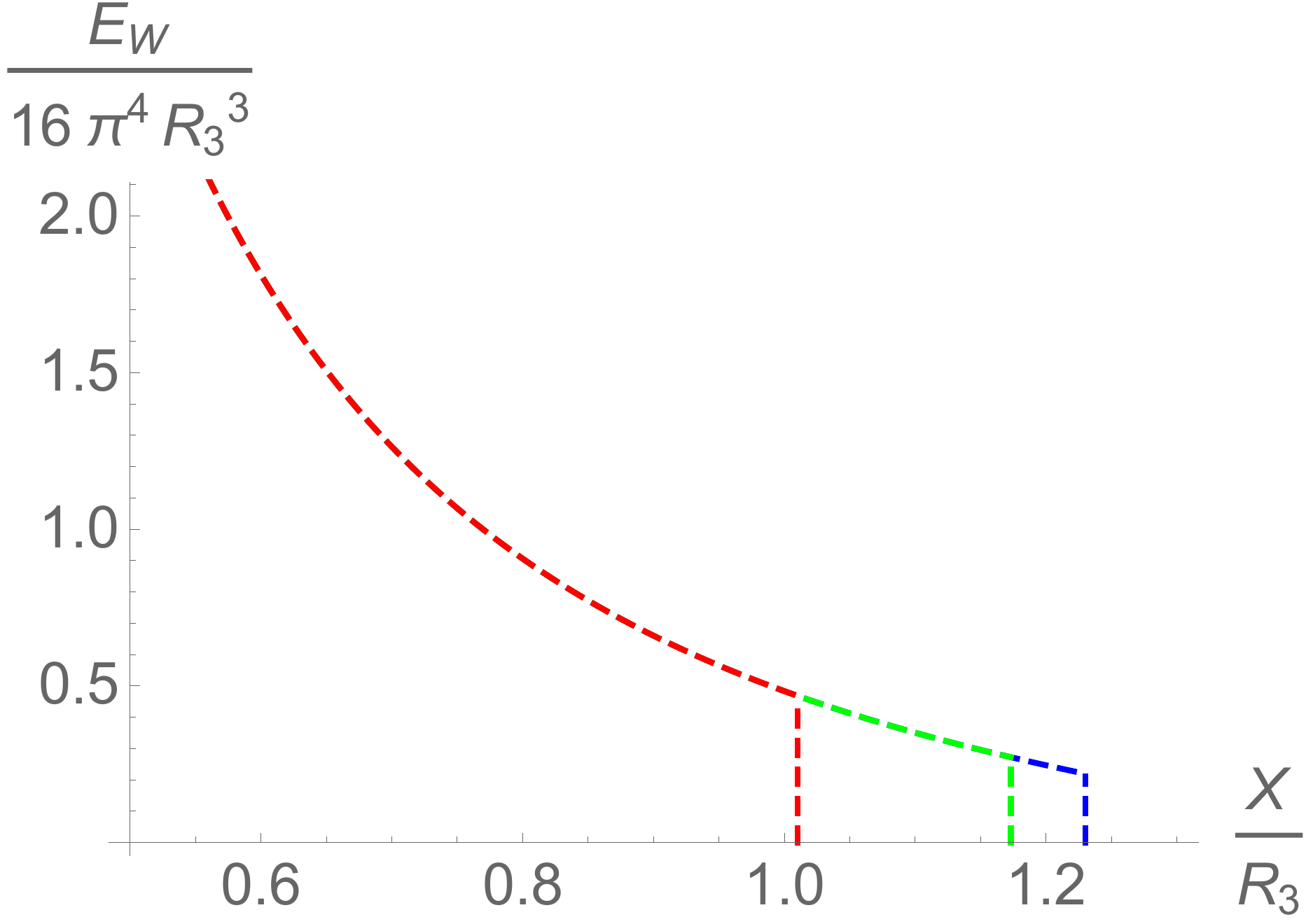}
\caption{ \small $E_W$ as a function of separation length $X$ for different values of strip length $L$. Here blue, green and red curves corresponds to $L/R_3 = 1.3 > L_{crit}$,  $1.2$ and $1.1$ respectively.}
\label{XVsEWVsLD3}
\end{minipage}
\hspace{0.4cm}
\begin{minipage}[b]{0.5\linewidth}
\centering
\includegraphics[width=2.8in,height=2.3in]{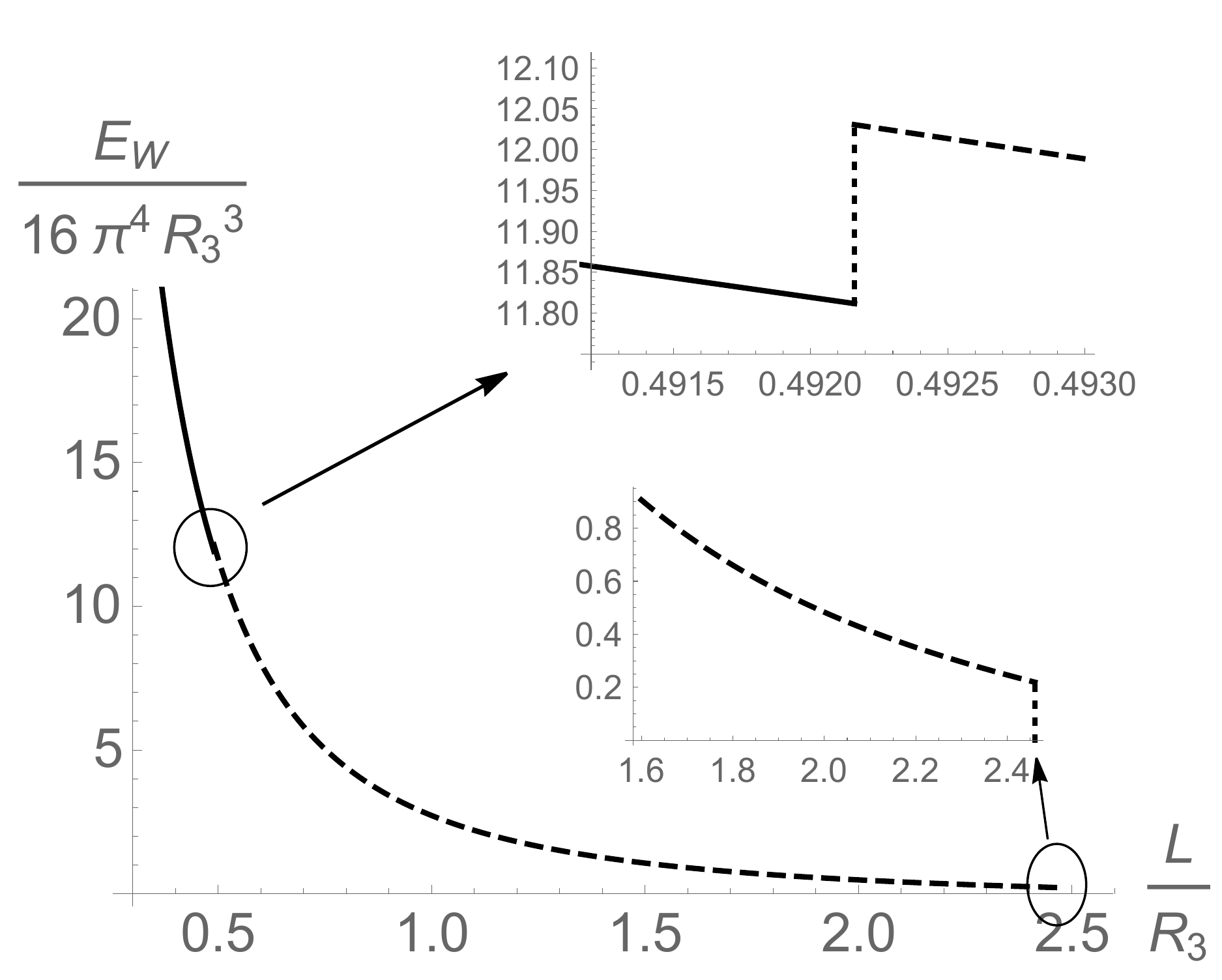}
\caption{\small $E_W$ as a function of separation length $L$ along a fixed line $X=0.5 L$. Here solid and dashed lines correspond to $E_W$ of $S_2$ and $S_3$ phases respectively.}
\label{LVsEWXPt5LD3}
\end{minipage}
\end{figure}

The behaviour of entanglement wedge as a function of $X$ for specific values of $L$ is shown in Figure~\ref{XVsEWVsLD3}. These specific values of $L$ are again chosen so that we can see the nature of $E_W$ close to the $S_1/S_3$ and $S_3/S_4$ phase transition points. From this analysis we see that $E_W$ changes monotonically with respect to $X$ and shows discontinuity at the transition point. The discontinuity in $E_W$ at the $S_3/S_4$ transition line can also be inferred from Eq.~(\ref{EWS3D3}). In particular, the condition $U_{*}(X=L_{crit})>U_0$ ensures that $E_W^{3}>0$ at $X_{crit}$ (as opposed to $E_W^{4}$, which is zero at $X_{crit}$). In the same fashion $E_W^{3}$ picks up a non-zero value at the
$S_1/S_3$ transition line (as indicated by a red curve). Our entire exercise suggest that the entanglement wedge cross-section disappears in a discontinuous fashion when the values of $X,L$ are large in the case of confining backgrounds.

We have also studied the nature of $E_W$ close to the $S_2/S_3$ critical line. This analysis is shown in Figure~\ref{LVsEWXPt5LD3}. We have chosen a specific value of $X=0.5 L$ in order to study $E_W$ for $S_2,S_3$ and $S_4$ configurations simultaneously. We find a discontinuity in $E_W$ at the $S_2/S_3$ transition line as well. This discontinuous behaviour can also be inferred from Eq.~(\ref{EWS2D3}) and (\ref{EWS3D3}). In particular, the condition $U_*(2L+X)\neq U_0$ ascertain that $E_W^{2}$ and $E_W^{3}$ cannot have the same value at the $S_2/S_3$ critical point. There is also a positive jump in the value of $E_W$ as the $S_2/S_3$ critical point is approached from the $S_2$ side, indicating an increase in the area of entanglement wedge at the critical point. This increment in area can also be extracted from the condition $U_*(2L+X)> U_0$. To conclude, we see that $E_W$ shows a non-trivial behaviour whenever a phase transition happens.
\begin{figure}[h!]
\begin{minipage}[b]{0.5\linewidth}
\centering
\includegraphics[width=2.8in,height=2.3in]{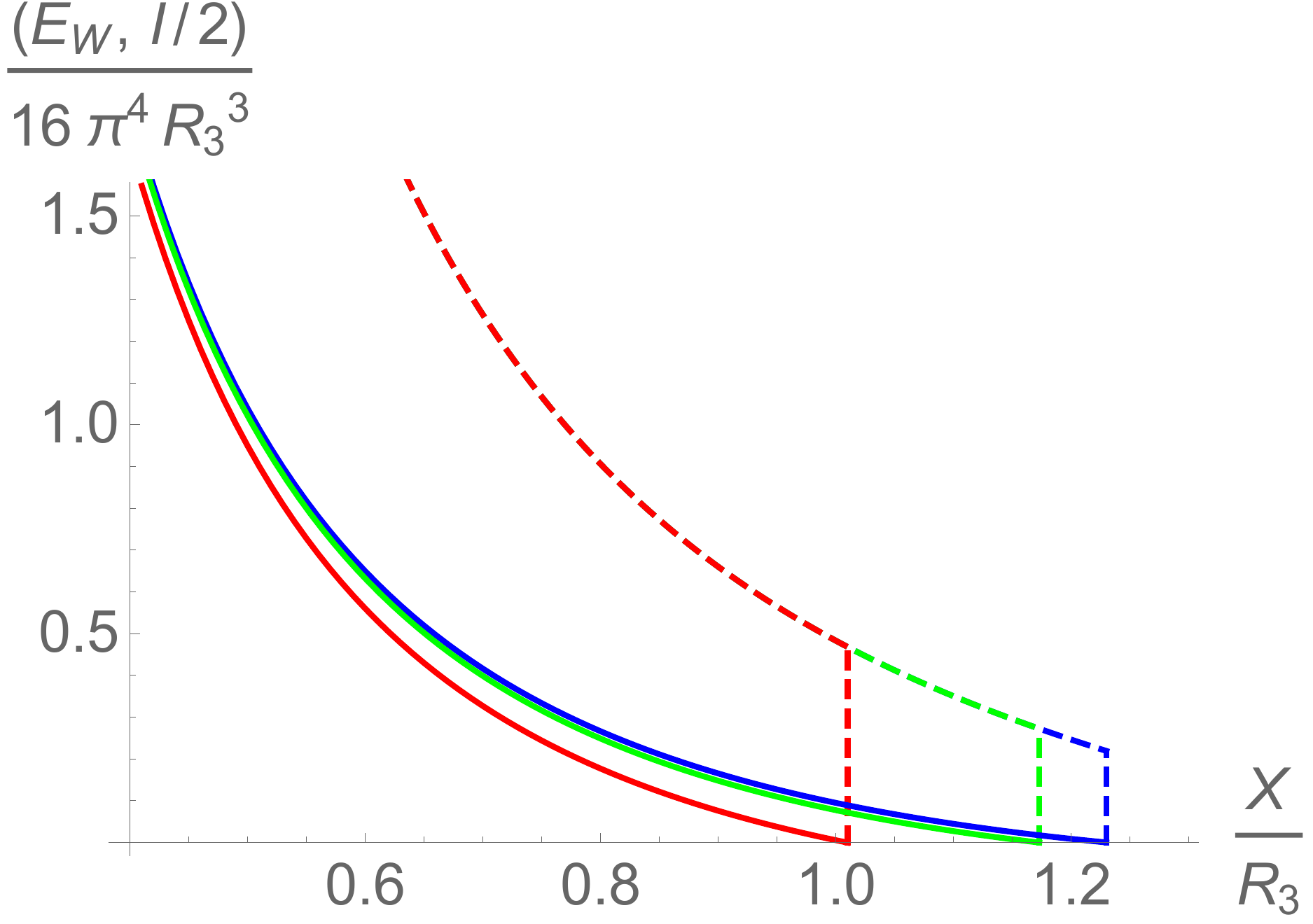}
\caption{ \small Mutual information $I$ and entanglement wedge $E_W$ as a function of $X$ for different values of $L$. The solid curves correspond to $I/2$ whereas the dashed curves correspond to $E_W$. Here red, green and blue curves correspond to $L/R_3 = 1.3$,  $1.2$ and $1.1$ respectively.}
\label{XVsEWandMIVsLD3}
\end{minipage}
\hspace{0.4cm}
\begin{minipage}[b]{0.5\linewidth}
\centering
\includegraphics[width=2.8in,height=2.3in]{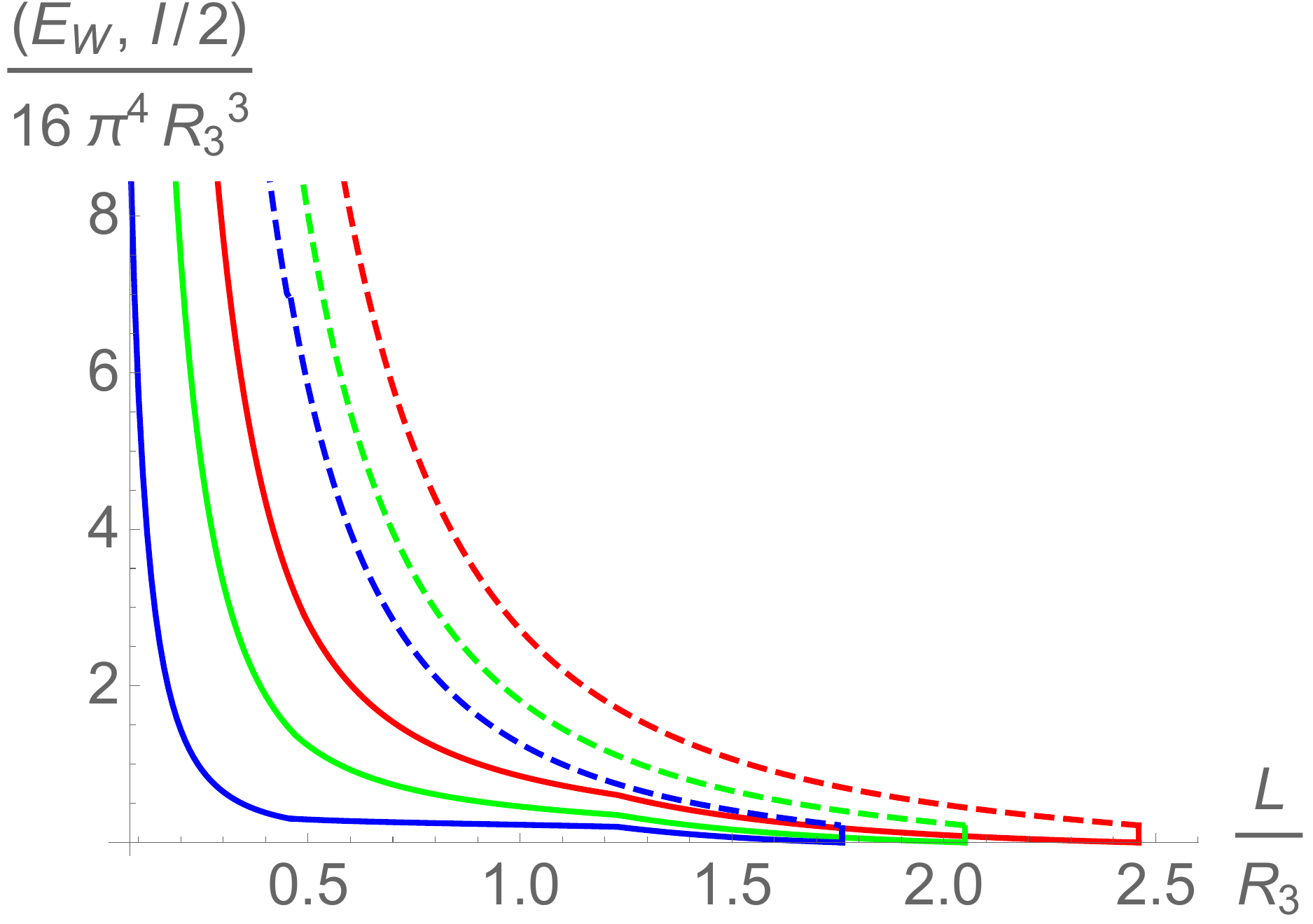}
\caption{\small Mutual information $I$ and entanglement wedge $E_W$ as a function of $L$ along a fixed line $X = \alpha L$. The solid curves correspond to $I/2$ whereas the dashed curves correspond to $E_W$. Here red, green and blue curves correspond to $\alpha = 0.5$, $0.6$ and $0.7$ respectively.}
\label{LVsEWandMIVsXLD3}
\end{minipage}
\end{figure}

We further tested the inequality $E_W \geq I/2$ in the current confining setup. We have studied this inequality numerically for various values of $X$ and $L$ and found it to be always true. The results for this exercise are presented in the Figures~\ref{XVsEWandMIVsLD3} and \ref{LVsEWandMIVsXLD3}.

\subsection{Entanglement negativity}
The holographic entanglement negativity for a single interval is given by the Eq.~(\ref{HEEsc1}). In the limit $B\rightarrow A^c\rightarrow\infty$, we again have
\begin{equation}
\mathcal{A}_{B_1}=\mathcal{A}_{B_2}=\mathcal{A}_{A\cup B_1}=\mathcal{A}_{A\cup B_2}=\mathcal{A}_{disconn}\ ,
\end{equation}
which leads to the following result for the negativity
\begin{eqnarray}\label{HEEscD3}
& & \mathcal{E} = \lim_{B\rightarrow A^c}\frac{3}{4}\left[2S_A+S_{B_1}+S_{B_2}-S_{A\cup B_1}-S_{A\cup B_2}\right]\, \nonumber \\
& & \mathcal{E} = \frac{3}{2} S_A \,.
\end{eqnarray}
This again tells us that the entanglement negativity is discontinuous at $L_{crit}$. Consequently, in this confining model as well, a change in the order of entanglement negativity (from
$\mathcal{O}(N^2)$ to $\mathcal{O}(N^0)$ or vice versa) appears at $L_{crit}$.

\begin{figure}[h!]
\begin{minipage}[b]{0.5\linewidth}
\centering
\includegraphics[width=2.8in,height=2.3in]{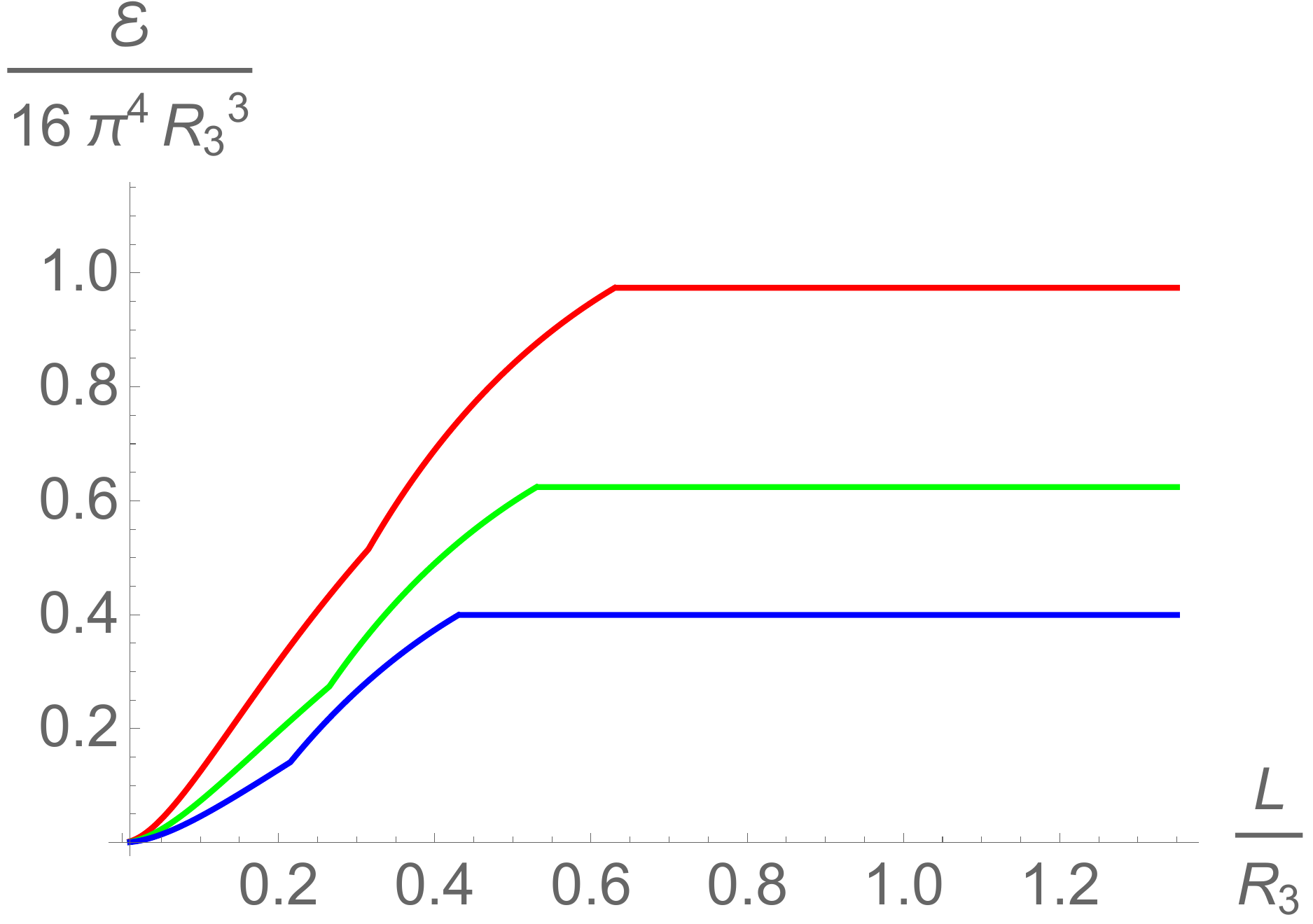}
\caption{ \small Entanglement negativity $\mathcal{E}$ of two parallel strips as a function of $X$ for different values of $L$. Here Red, green and blue curves correspond to $X/R_3 = 0.6$,  $0.7$ and $0.8$ respectively.}
\label{LVsENVsXtwostripD3}
\end{minipage}
\hspace{0.4cm}
\begin{minipage}[b]{0.5\linewidth}
\centering
\includegraphics[width=2.8in,height=2.3in]{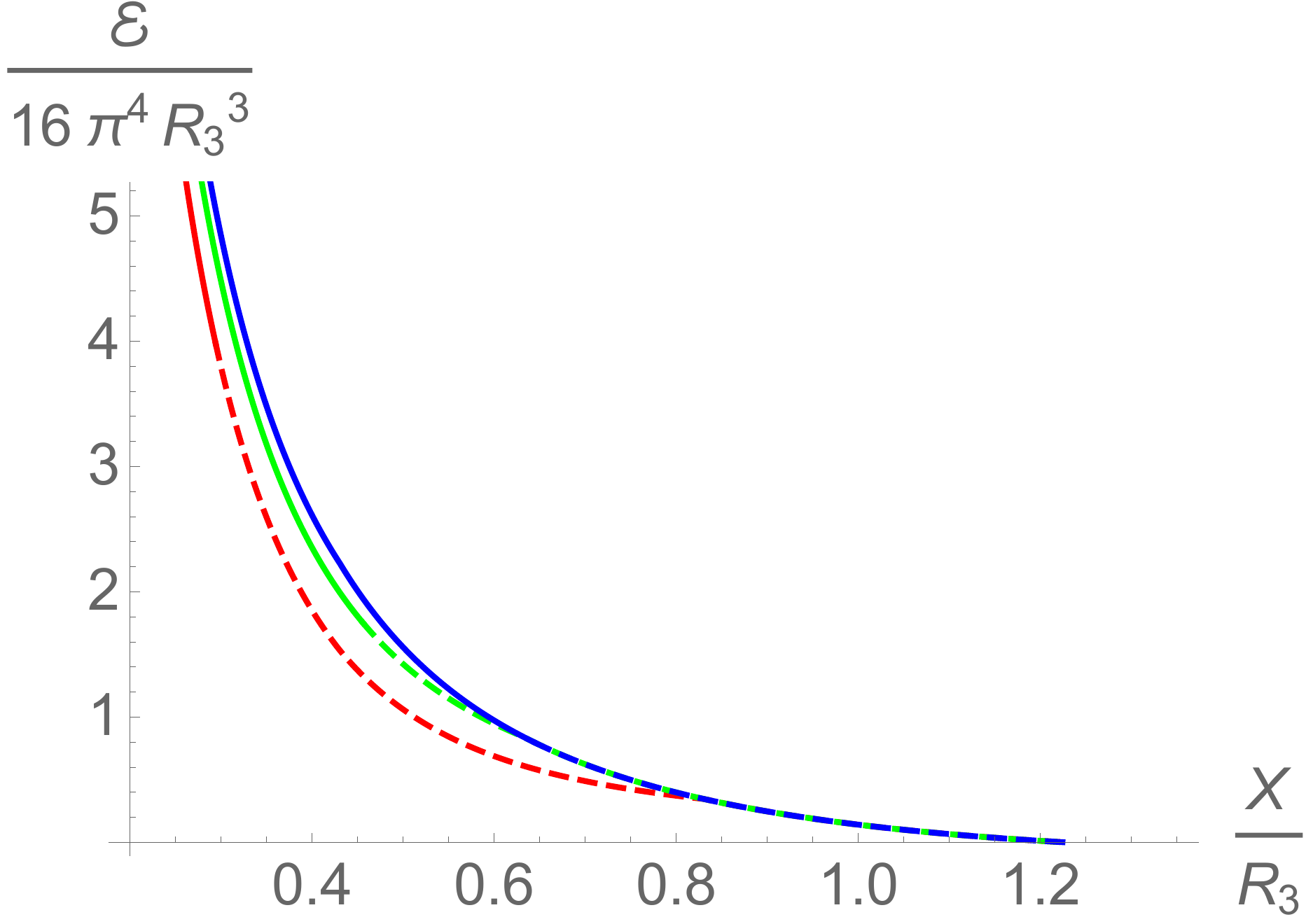}
\caption{\small Entanglement negativity $\mathcal{E}$ of two parallel strips as a function of $L$ for different values of $X$. Here Red, green and blue curves correspond to $L/R_4 = 0.4$,  $0.6$ and $0.8$ respectively.}
\label{XVsENVsLtwostripD3}
\end{minipage}
\end{figure}

The entanglement negativity for the two disjoint strips is again given by \cite{Malvimat:2018txq,Basak:2020bot},
\begin{equation}\label{ENtwostripD3}
\mathcal{E}= \frac{3}{4}\left[S_A(L+X)+S_A(L+X)-S_A(2L+X)-S_A(X)\right] \,,
\end{equation}
where $S_A$ as the entanglement entropy of a single interval. For the $S_4$ configuration, where $X>L_{crit}$, $\mathcal{E}$ is zero as all the terms in Eq.~(\ref{ENtwostripD3}) become $S_A^{discon}$. Whereas, $\mathcal{E}$ is non-zero in all other configurations, in particular, in $S_1$ configuration. This behaviour is different from the mutual information and entanglement wedge cross-section behaviour, which were in $S_1$ configuration as well.  From the Figures~\ref{LVsENVsXtwostripD3} and \ref{XVsENVsLtwostripD3}, we can infer that $\mathcal{E}$ in general (except for the $S_4$ case) has a non-zero value and behaves monotonically with respect to $X$ and $L$. Further, $\mathcal{E}$ is finite for $L>L_{crit}$
and it goes to zero in a smooth fashion at $X=L_{crit}$. In contrast to the entanglement wedge, $\mathcal{E}$ shows a continuous behaviour during the phase transition but exhibits a cusp (see Figure~\ref{LVsENVsXtwostripD3}). \\

Before we end our discussion on top-down models, we further like to add that we have performed similar analysis in other top-down confining models as well. In particular, we have also computed entanglement entropy, mutual information, entanglement wedge cross-section and entanglement negativity in the Klebanov-Strassler and Klebanov-Tseytlin confining backgrounds \cite{Klebanov:2000hb}, and results similar to those presented here are found for these entanglement measures.

\section{Bottom-up holographic confining model}\label{bottomupsystem}
Having discussed the entanglement measures in top-down holographic confining models, we now move on to discuss them in a bottom-up confining model. As is well known, the top-down holographic QCD models generally exhibit undesirable features whose analogue in real QCD do not exist. In particular, the dual boundary theory of these top-down holographic models usually contains non-running coupling constants, additional Hilbert states (coming from the Kaluza-Klein modes of extra dimensions), problematic conformal symmetries  etc. Whereas the phenomenological bottom-up holographic models, although lack strong AdS/CFT justification and generally constitute in an ad-hoc manner to reproduced desirable QCD like features of the dual boundary theory, can overcome most of the difficulties of the top-down models. Therefore, it is interesting to see how different entanglement measures behave in bottom-up confining models as well.

Here, we consider a particular bottom-up confining model suggested in \cite{Dudal:2017max,Dudal:2018ztm}. This model is based on the Einstein-Maxwell-dilaton (EMD) gravity action. Importantly, this gravity model can be solved exactly and the closed loop expressions of the spacetime metric can be found \footnote{Analytical solutions can be found for the Einstein-Maxwell-dilaton (EMD) gravity system using the potential reconstruction method. For more details on this method see \cite{Mahapatra:2018gig,Mahapatra:2020wym}.}. Further, this model predicts a thermal-AdS/black hole phase transition, which on the dual boundary theory corresponds to the confined/deconfined phase transition. The model, moreover, exhibits a linear Regge trajectory for heavy meson spectrum. Here, we briefly describe the analytic expression of the relevant spacetime metric, which will be needed for our discussion in later sections and more details can be found in \cite{Dudal:2017max,Dudal:2018ztm}.

In this model, the dual spacetime geometry for the confined phase is,
\begin{eqnarray}
& & ds^2=\frac{R^2 e^{2 A(z)}}{z^2}\biggl(- dt^2 + dz^2 + dy_{1}^2+dy_{2}^2+dy_{3}^2 \biggr) \,.
\label{metric}
\end{eqnarray}
Where $A(z)=-a z^2$, with $a>0$, is the scale function and $R$ is the AdS length scale. The radial coordinate $z$ runs from $z=0$ (asymptotic boundary) to $z=\infty$ (deep bulk). This solution asymptotes to AdS near the boundary ($z\rightarrow0$) and has a negative curvature throughout the spacetime. The parameter $a=0.145 \ GeV^2$ is fixed by demanding the thermal-AdS/black hole (or the dual confined/deconfined) phase transition to be around $270 \ MeV$, as is observed in large $N$ lattice QCD in the pure glue sector.

\subsection{Entanglement entropy: one strip}
The behaviour of entanglement entropy in this confining model has already been studied in \cite{Dudal:2018ztm,Mahapatra:2019uql}. Here, we briefly mention their results. For a strip subsystem $A$ of domain  \{$-L/2\leq y_1 \leq L/2$, $0\leq y_2 \leq L_2$, $0\leq y_3 \leq L_3$ \}, the entanglement entropy of the connected surface is given by
\begin{eqnarray}
S_{A}^{con} (L)=\frac{L_2 L_2 R^3}{4 G_N^{(5)}} \int_{0}^{z_*} dz \ \frac{2 z_{*}^3}{z^3} \frac{e^{3 A(z)-3 A(z_*)}}{\sqrt{[z_{*}^6 e^{-6A(z_*)}-z^6 e^{-6A(z)}]}} \,,
\label{SEEcon}
\end{eqnarray}
where $z_*$ is the turning point of the connected minimal area surface and is related to the strip length $L$ in the following way
\begin{eqnarray}
L=2\int_{0}^{z_*} dz \ \frac{z^3 e^{-3 A(z)}}{\sqrt{[z_{*}^6 e^{-6A(z_*)}-z^6 e^{-A(z)}]}}\,.
\label{lengthSEEcon}
\end{eqnarray}
Whereas, the entanglement entropy for the disconnected surface is given by
\begin{eqnarray}
S_{A}^{discon}=\frac{L_2 L_3 R^3}{4 G_N^{(5)}} \int_{0}^{\infty} dz \ \frac{2 e^{3 A(z)}}{z^3} \,.
\label{SEEdiscon}
\end{eqnarray}
Note that the entanglement entropy of the disconnected surface is again independent of the strip length $L$.
\begin{figure}[h!]
\begin{minipage}[b]{0.5\linewidth}
\centering
\includegraphics[width=2.8in,height=2.3in]{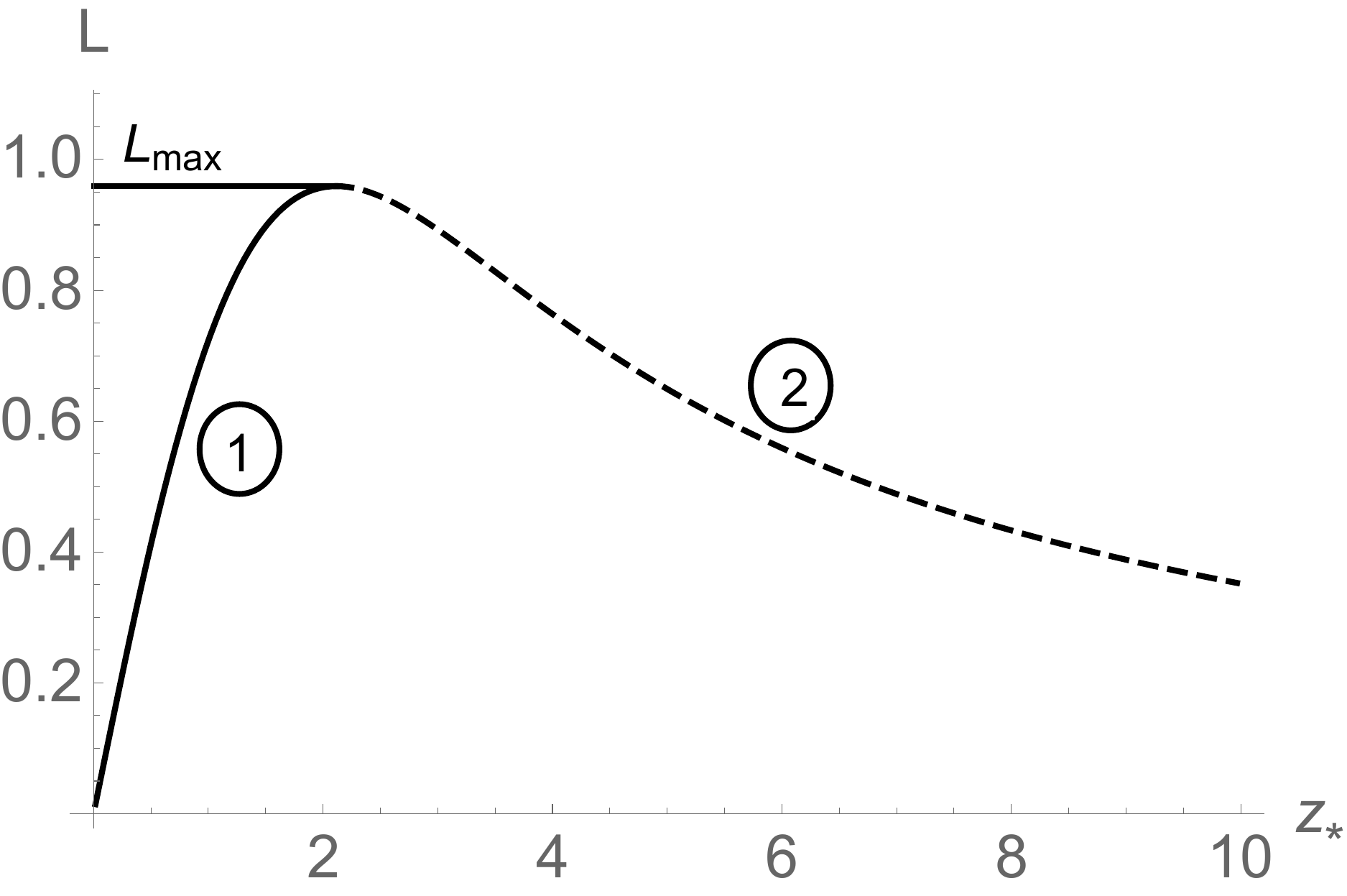}
\caption{ \small The behaviour of strip length $L$ as a function of $z_*$.}
\label{zsVsL1stripEMD}
\end{minipage}
\hspace{0.4cm}
\begin{minipage}[b]{0.5\linewidth}
\centering
\includegraphics[width=2.8in,height=2.3in]{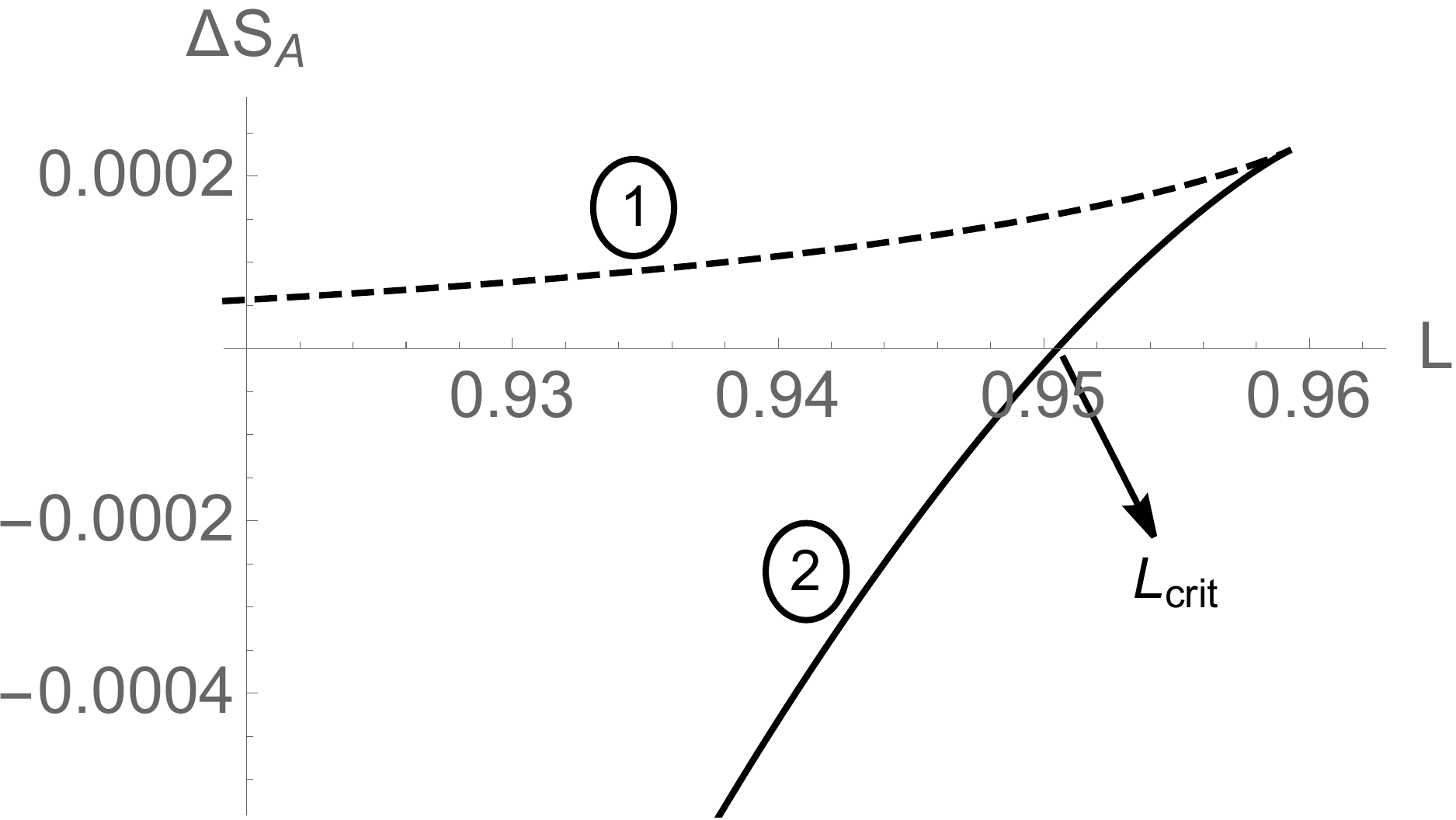}
\caption{\small $\triangle S_A= S_A^{con}-S_A^{discon}$ as a function of strip length $L$.}
\label{LVsdeltaS1stripEMD}
\end{minipage}
\end{figure}

The numerical results for the entanglement entropy are shown in Figures~\ref{zsVsL1stripEMD} and \ref{LVsdeltaS1stripEMD} \footnote{Here, and in the subsequent subsections, we have used $\frac{L_2 L_3 R^3}{4 G_N^{(5)}}=1$ for the numerical purposes.}. Similar to the top-down confining models, here again, a maximum length appears ($L_{max} \simeq 0.959$) above which no solution for the connected surface exists. Below $L_{max}$, there are two solutions for the connected surface, from which the surface which is nearer to the boundary always has the smallest area (shown by a solid line). The entanglement entropy again undergoes a phase transition from connected to disconnected surface as $L$ increases. This transition happens at $L_{crit} \simeq 0.951<L_{max}$, at which the area of the connected surface becomes larger than the disconnected surface. Correspondingly, in this bottom-up confining system as well, the entanglement entropy becomes independent of $L$ for large $L$. The discontinuous nature of the entanglement entropy, therefore, appears to be a generic feature of all holographic confining theories.

\subsection{Mutual information: two strips}
\begin{figure}[h!]
\center
\includegraphics[width=2.8in,height=2.3in]{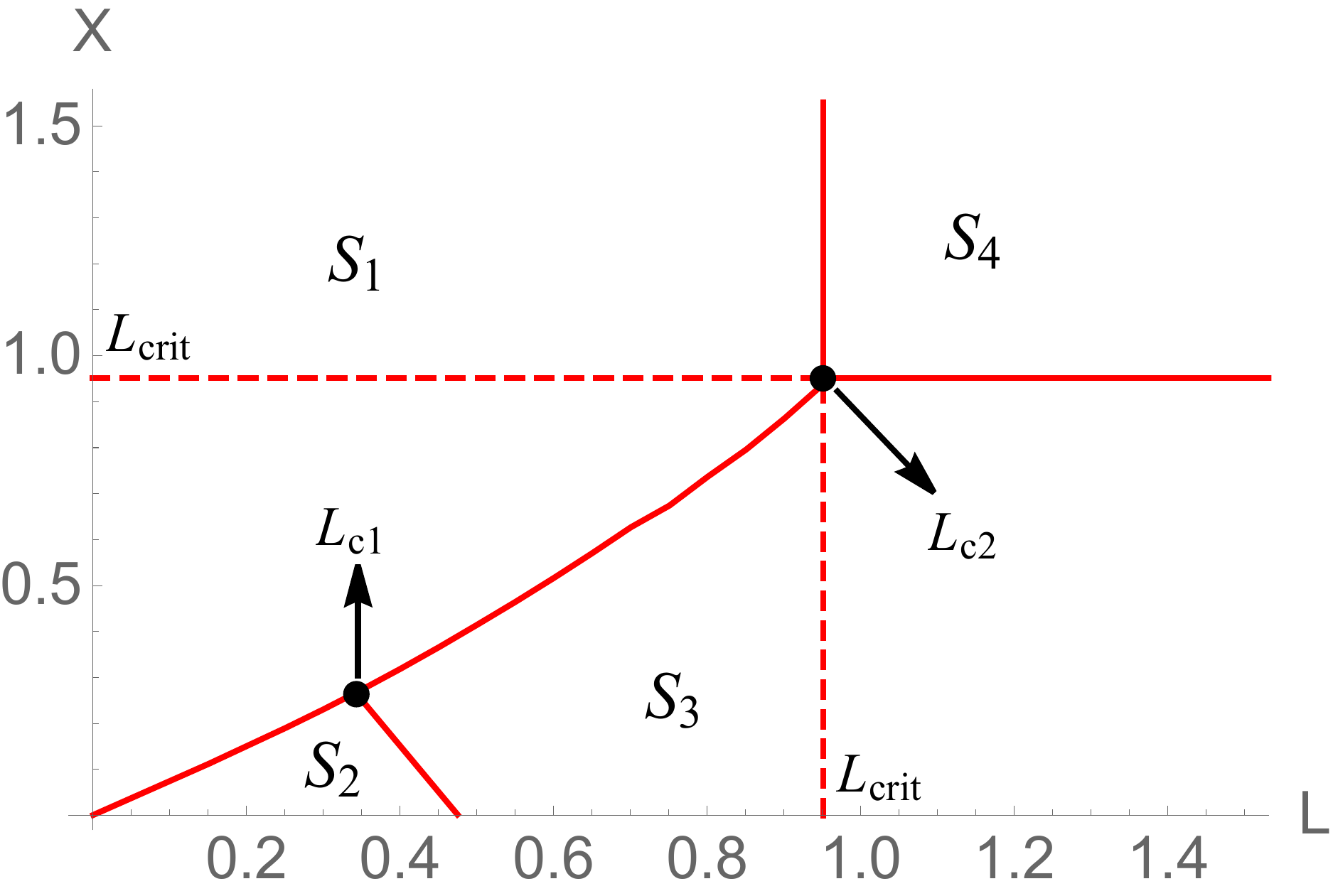}
\caption{\small The two equal strip entanglement phase diagram for the EMD bottom-up confining background. The four different phases correspond to the four bulk surfaces
of Figure \ref{twostripphasediag}. Two black dots indicate the two tri-critical points}
\label{twostripphasediagEMD}
\end{figure}

The two equal strip phase diagram of the EMD bottom-up confining background is shown in Figure~\ref{twostripphasediagEMD} \footnote{The two equal strip entanglement phase diagram and mutual information  of this bottom-up confined system have also been investigated in \cite{Mahapatra:2019uql}.}. It's two strip entanglement phase structure is quite similar to the top-down confining models. In particular, depending upon the magnitude of $X$ and $L$, four different entangling surfaces $\{S_1,S_2,S_3,S_4\}$ again dominate the entanglement structure. The expressions of $\{S_1,S_2,S_3,S_4\}$ are the same as in Eq.~(\ref{2stripEE}).

There are again two tri-critical points, where three phases co-exist. The coordinates of these tri-critical points are $(L=0.343,X=0.266)$ and $(L=0.951,X=0.951)$. These tri-critical points, as well as other critical lines, again suggest various non-analyticities in the entanglement structure of the confined phase.

\begin{figure}[h!]
\begin{minipage}[b]{0.5\linewidth}
\centering
\includegraphics[width=2.8in,height=2.3in]{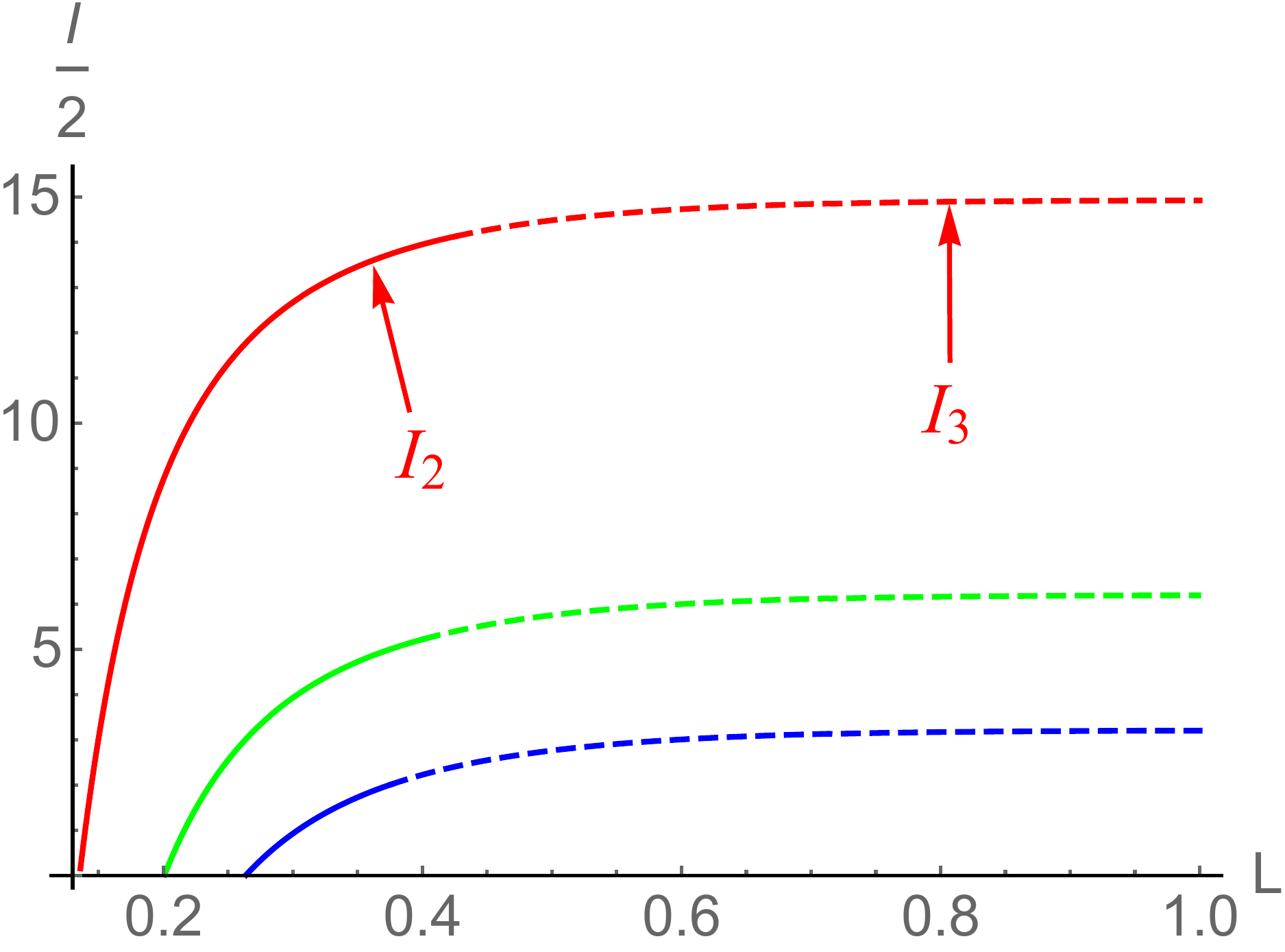}
\caption{ \small Mutual Information of $S_2$ and $S_3$ phases as a function of $L$. The solid and dashed lines correspond to $I_{2}$ and $I_{3}$ respectively.  The red, green and blue lines correspond to $X=0.10$, $0.15$ and $0.20$ respectively.}
\label{LvsMIVsXEMD}
\end{minipage}
\hspace{0.4cm}
\begin{minipage}[b]{0.5\linewidth}
\centering
\includegraphics[width=2.8in,height=2.3in]{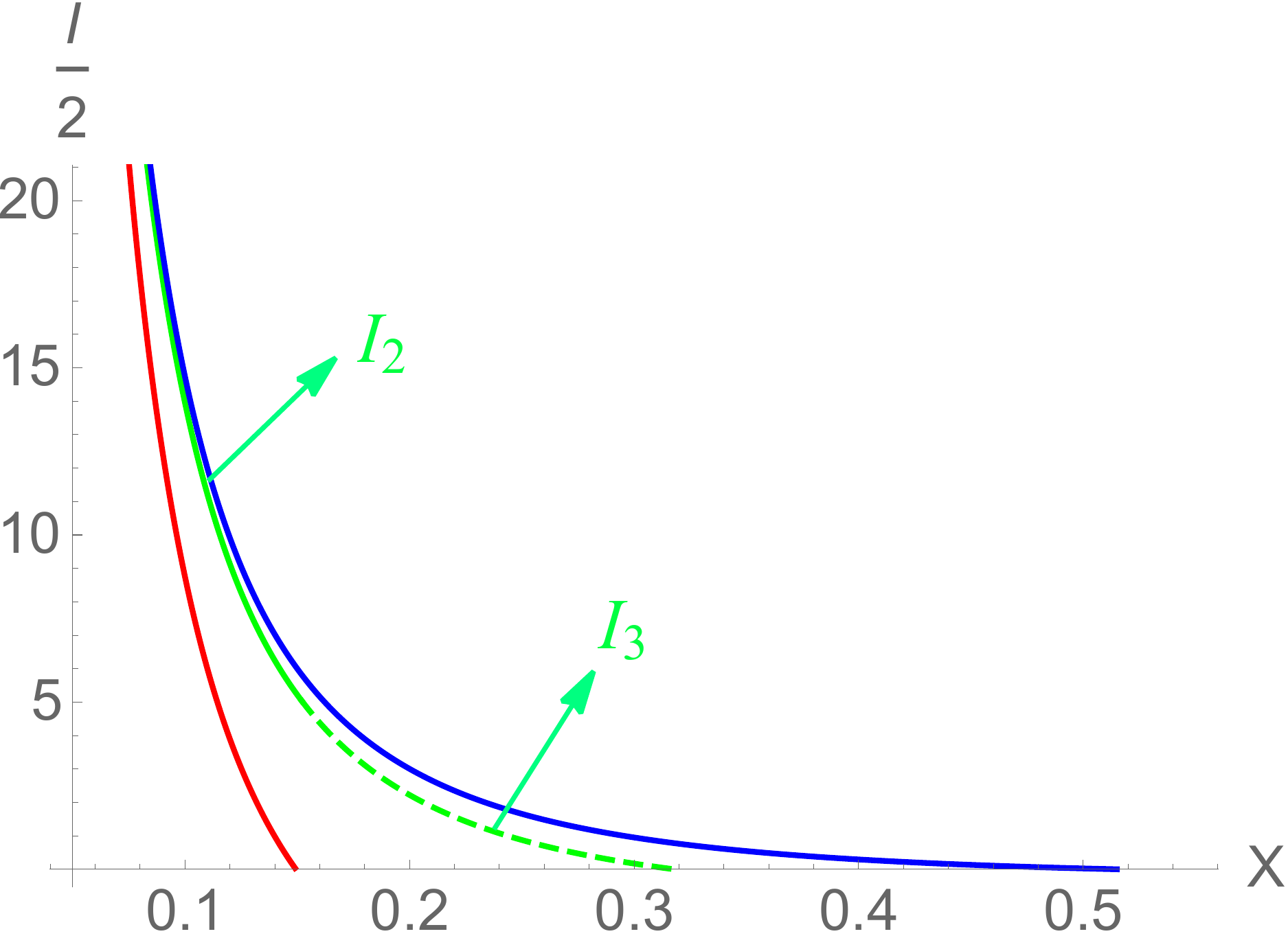}
\caption{\small Mutual Information of $S_2$ and $S_3$ phases as a function of $X$. The solid and dashed lines correspond to $I_{2}$ and $I_{3}$ respectively.
The red, green and blue lines correspond to $L=0.2$, $0.4$ and $0.6$ respectively. }
\label{XVsMIVsLEMD}
\end{minipage}
\end{figure}

Similarly, the mutual information in these four entangling phases is given by Eq.~(\ref{mutual2strips}). It again goes to zero in $S_1$ and $S_4$ phases, whereas it remain finite and positive for $S_2$ and $S_3$ phases. Therefore, the mutual information is again of order $\mathcal{O}(N^2)$ in $S_2$ and $S_3$ phases, whereas it is of order $\mathcal{O}(N^0)$ in $S_1$ and $S_4$ phases. In Figures~\ref{LvsMIVsXEMD} and \ref{XVsMIVsLEMD}, we have shown the variation of mutual information in $S_2$ and $S_3$ phases as a function of $X$ and $L$. It again connects smoothly between $S_2$ and $S_3$ phases. Moreover, as we approach $S_1$ and $S_4$ phases from $S_2$ and $S_3$ phases by changing $X$ and $L$, it again goes to zero in a continuous manner. It is clear that apart from a qualitative change in the magnitude of the critical points, the overall mutual information structure of this bottom-up EMD confining model remains the same as in the above discussed top-down confining models.

\subsection{Entanglement wedge cross-section}
We now calculate the entanglement wedge cross-section of the dual confining theory of Eq.~(\ref{metric}). It is given by the minimal area of $t=\text{constant}$ and $y_1=\text{constant}$ surface. The induced metric on this surface is
\begin{eqnarray}
& & ds^2=\frac{R^2 e^{2 A(z)}}{z^2}\biggl( dz^2 + dy_{2}^2+dy_{3}^2 \biggr) \,,
\label{metric}
\end{eqnarray}
which implies,
\begin{eqnarray}
E_W=\frac{L_2 L_3 R^3}{4 G_N^{(5)}}\int dz \frac{e^{-3A(z)}}{z^3} \,.
\end{eqnarray}
As earlier, out of the four surfaces, the disjoint surfaces $S_1$ and $S_4$ again have zero entanglement wedge cross-section, whereas the connected surfaces $S_2$ and $S_3$ have a non-zero entanglement wedge cross-section. For $S_2$ phase, it is given by
\begin{eqnarray}
& & E_W^2=\frac{L_2 L_3 R^3}{4 G_N^{(5)}}\int_{z_*(X)}^{z_*(2L+X)} dz \frac{e^{-3A(z)}}{z^3} \, \nonumber\\
& & \ \ \ \ \ \ = - \frac{L_2 L_3 R^3}{4 G_N^{(5)}} \bigg\rvert \frac{e^{-3az^2}}{2z^2}+\frac{3a}{2} Ei(-3 a z^2)\bigg\rvert_{z_*(X)}^{z_*(2L+X)} \,.
\label{EWS2EMD}
\end{eqnarray}
Where $Ei(x)$ is an exponential integral function.  Similarly, for the $S_3$ phase
\begin{eqnarray}
& & E_W^3=\frac{L_2 L_3 R^3}{4 G_N^{(5)}}\int_{z_*(X)}^{\infty} dz \frac{e^{-3A(z)}}{z^3} \, \nonumber\\
& & \ \ \ \ \ \ =  \frac{L_2 L_3 R^3}{4 G_N^{(5)}} \left[ \frac{e^{-3 a z_*^2(X)}}{2 z_*^2(X)} -3 a \Gamma\left[0,3az_*^2(X)\right]  \right] \,.
\label{EWS2EMD}
\end{eqnarray}
Where $\Gamma\left[a,b\right]$ is the incomplete gamma function. Note that $ E_W^3$ is actually independent of strip length $L$ and depends only on $X$. This would imply that $E_W$ profiles for different values of $L$ will overlap in $S_3$ phase. Also, from the definition itself, it is explicitly clear that both $E_W^2$ and $E_W^3$ are positive.

\begin{figure}[h!]
\begin{minipage}[b]{0.5\linewidth}
\centering
\includegraphics[width=2.8in,height=2.3in]{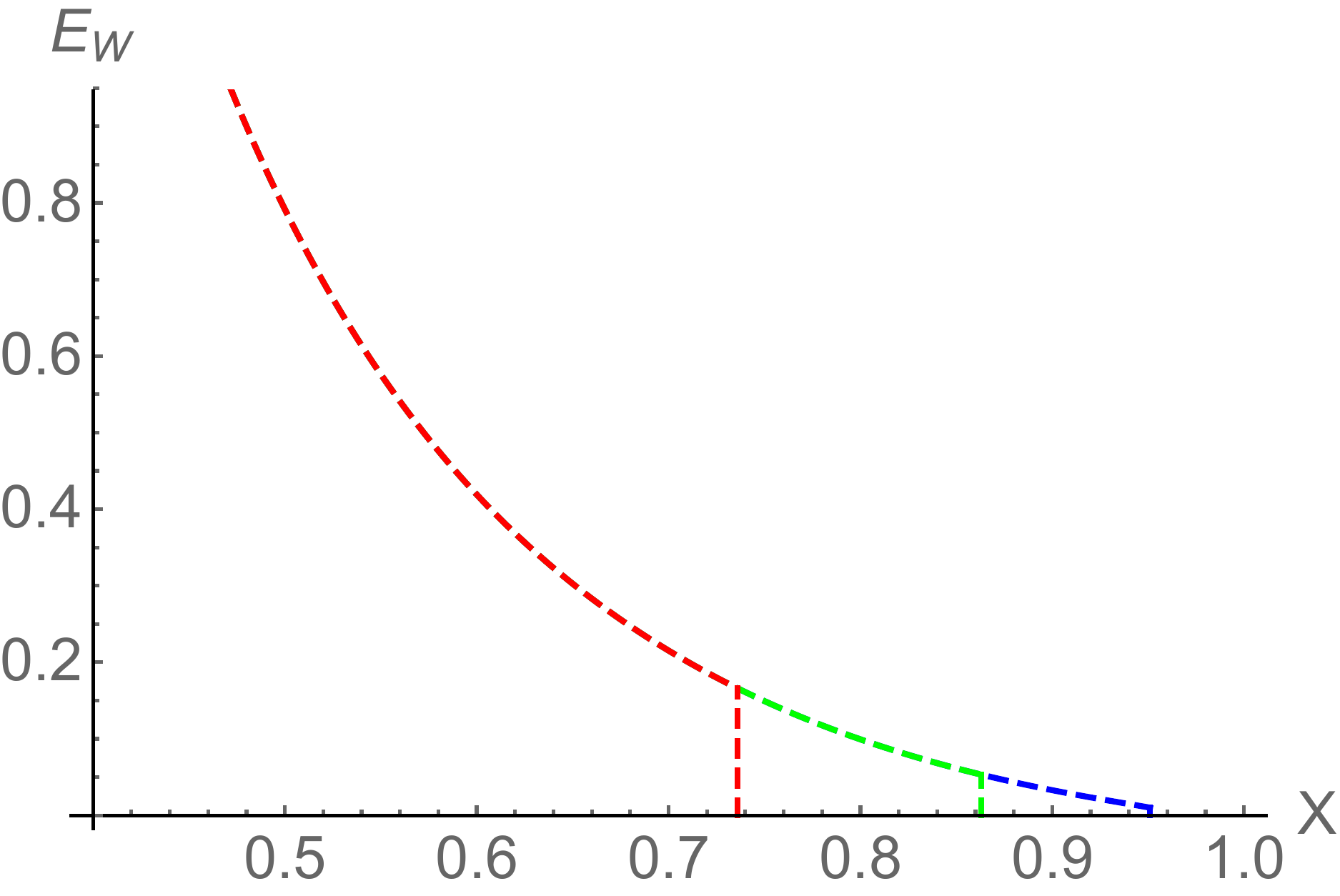}
\caption{ \small $E_W$ as a function of $X$ for different values of $L$. Here red, green and blue curves corresponds to $L = 0.8$,  $0.9$ and $1.0>L_{crit}$ respectively.}
\label{XVsEWVsLEMD}
\end{minipage}
\hspace{0.4cm}
\begin{minipage}[b]{0.5\linewidth}
\centering
\includegraphics[width=2.8in,height=2.3in]{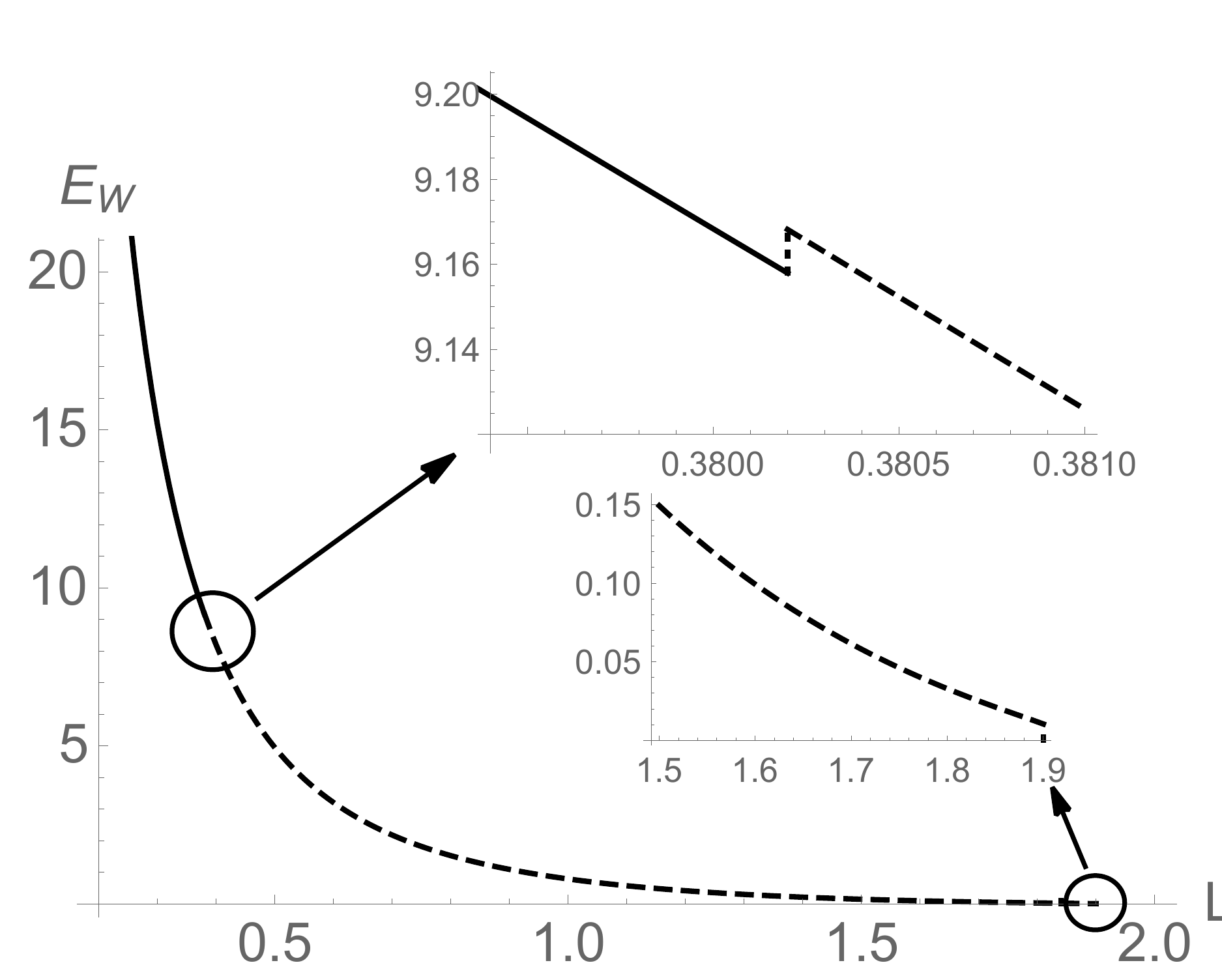}
\caption{\small $E_W$ as a function of $L$ along a fixed line $X=0.5 L$. Here solid and dashed lines correspond to $E_W$ of $S_2$ and $S_3$ phases respectively.}
\label{LVsEWXPt5LEMD}
\end{minipage}
\end{figure}
In Figure~\ref{XVsEWVsLEMD}, the variation of $E_W$ as a function of $X$ is shown. We find that $E_W$, again, is not only a monotonically decreasing function of $X$, but is also discontinuous at the critical points. In particular, $E_W$ does not go to zero as the $S_1/S_3$ and $S_3/S_4$ critical lines are approached from the $S_3$ side. Therefore, just like in the top-down confining models, the entanglement wedge also vanishes discontinuously for large values of $X$ and $L$ in the EMD confining model.

\begin{figure}[h!]
\begin{minipage}[b]{0.5\linewidth}
\centering
\includegraphics[width=2.8in,height=2.3in]{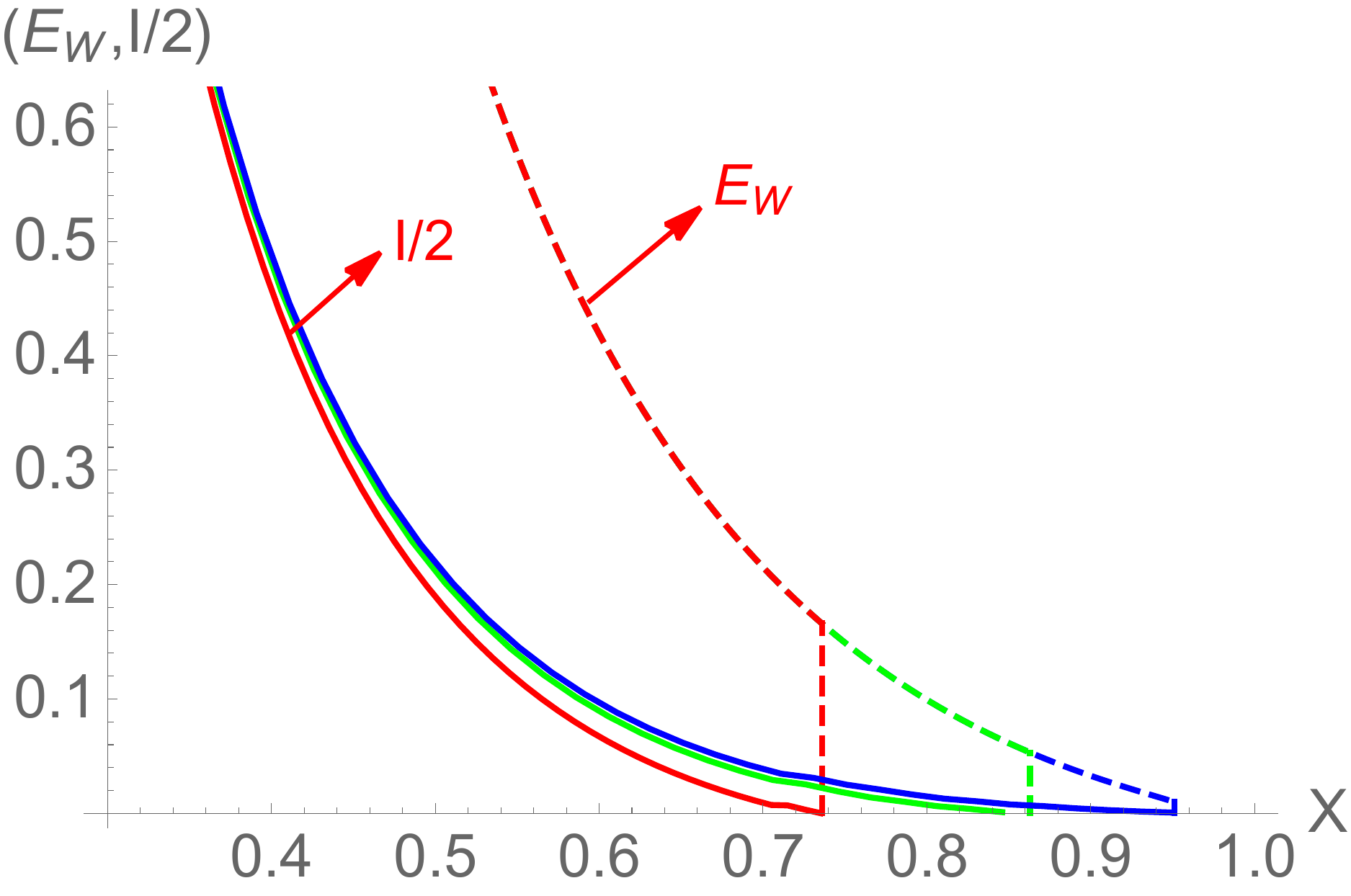}
\caption{ \small Mutual information $I$ and entanglement wedge $E_W$ as a function of $X$ for different values of $L$. The solid curves correspond to $I/2$ whereas the dashed curves correspond to $E_W$. Here Red, green and blue curves correspond to $L = 0.8$,  $0.9$ and $1.0$ respectively.}
\label{XVsEWandMIVsLEMD}
\end{minipage}
\hspace{0.4cm}
\begin{minipage}[b]{0.5\linewidth}
\centering
\includegraphics[width=2.8in,height=2.3in]{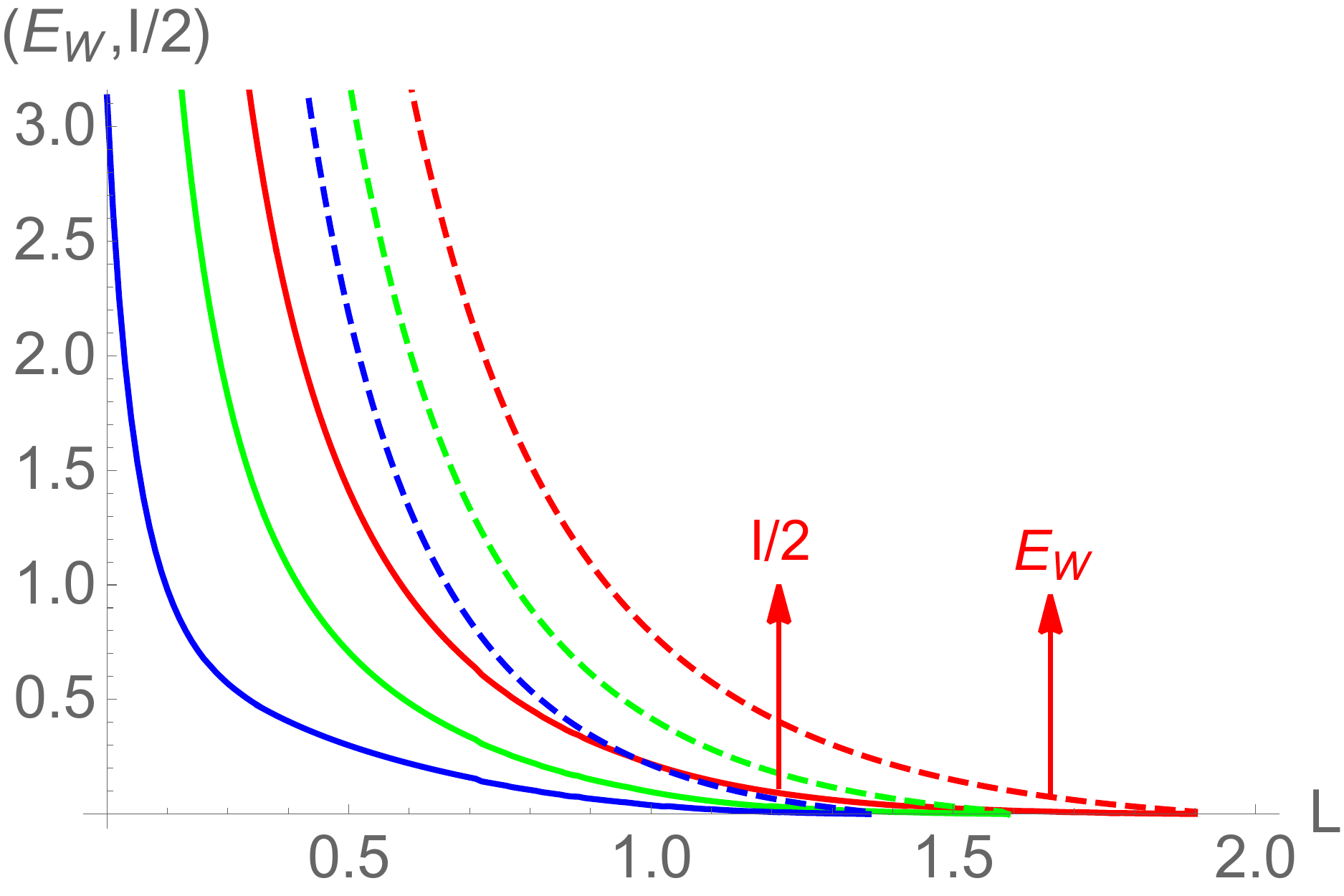}
\caption{\small Mutual information $I$ and entanglement wedge $E_W$ as a function of $L$ along a fixed line $X = \alpha L$. The solid curves correspond to $I/2$ whereas the dashed curves correspond to $E_W$. Here Red, green and blue curves correspond to $\alpha = 0.5$, $0.6$ and $0.7$ respectively.}
\label{LVsEWandMIXPt5LEMD}
\end{minipage}
\end{figure}

To further explore the discontinues nature of $E_W$, we investigate its behaviour near the $S_2/S_3$ critical points. This is interesting considering that $E_W$ is non-zero in both these phase. The results are shown in Figure~\ref{LVsEWXPt5LEMD}. Here $E_W$ is evaluated along a particular line $X=0.5L$, which allow us probe its behaviour in three different phases $\{S_2,S_3, S_4\}$ simultaneously. We again find a discontinuous jump in the area of the entanglement wedge at the  $S_2/S_3$ critical point. In particular, $E_W^2$ and $E_W^3$ values do not match at the $S_2/S_3$ critical point. We checked this behaviour for many other $X=\alpha L$ lines and found a similar discontinuous pattern at the $S_2/S_3$ critical points. Moreover, as also mentioned above, $E_W$ again exhibited a discontinuous pattern at the $S_3/S_4$ critical points. Since similar kind of results appeared in the top-down confining models as well, these findings advocate for the case that the non-analyticity in the structure of $E_W$ is a universal feature of all holographic confining theories.

We further tested the entanglement wedge and mutual information inequality ($E_W\geq I/2$) in current EMD confining model. The results are shown in Figures~\ref{XVsEWandMIVsLEMD} and \ref{LVsEWandMIXPt5LEMD}. We find that this inequality is again satisfied. We have numerically checked this inequality for many different values of $X$ and $L$, and find that $E_W$ is always greater than $I/2$. The inequality saturates only at the critical points, at which $I/2$ continuously goes to zero whereas $E_W$ exhibits a sharp drop to zero.

\subsection{Entanglement negativity}
We now calculate the entanglement negativity in the current EMD confining model. For a single interval subsystem, it is given by Eq.~(\ref{HEEsc1}). Since the disconnected entanglement entropy, which is independent of the strip length, is more favorable at large strip lengths in this confining model as well, the limiting condition $B\rightarrow A^c\rightarrow \infty$ again ensures that
\begin{eqnarray}
S_{B_{1}}=S_{B_{2}}=S_{A\cup B_1}=S_{A\cup B_2}=S_A^{discon} \,.
\end{eqnarray}
This implies,
\begin{eqnarray}\label{HEEscEMD}
& & \mathcal{E} = \lim_{B\rightarrow A^c}\frac{3}{4}\left[2S_A+S_{B_1}+S_{B_2}-S_{A\cup B_1}-S_{A\cup B_2}\right]\, \nonumber \\
& & \mathcal{E} = \frac{3}{2} S_A \,.
\end{eqnarray}
Again, the entanglement negativity is just $3/2$ times of the entanglement entropy. Therefore, the discontinuous nature of the entanglement entropy at $L_{crit}$ again implies a non-analytic behaviour of the entanglement negativity at the same critical length. Correspondingly, the entanglement negativity exhibits an order change (form $\mathcal{O}(N^2)$ to $\mathcal{O}(N^0)$ or vice versa) at $L_{crit}$ in this confining background as well.

For the two disjoint equal strip intervals, the entanglement negativity is given by Eq.~(\ref{ENtwostripD4}),
\begin{equation}
\mathcal{E}=\frac{3}{4}\left[S_A(L+X)+S_A(L+X)-S_A(2L+X)-S_A(X)\right].
\label{ENtwostripEMD}
\end{equation}

\begin{figure}[h!]
\begin{minipage}[b]{0.5\linewidth}
\centering
\includegraphics[width=2.8in,height=2.3in]{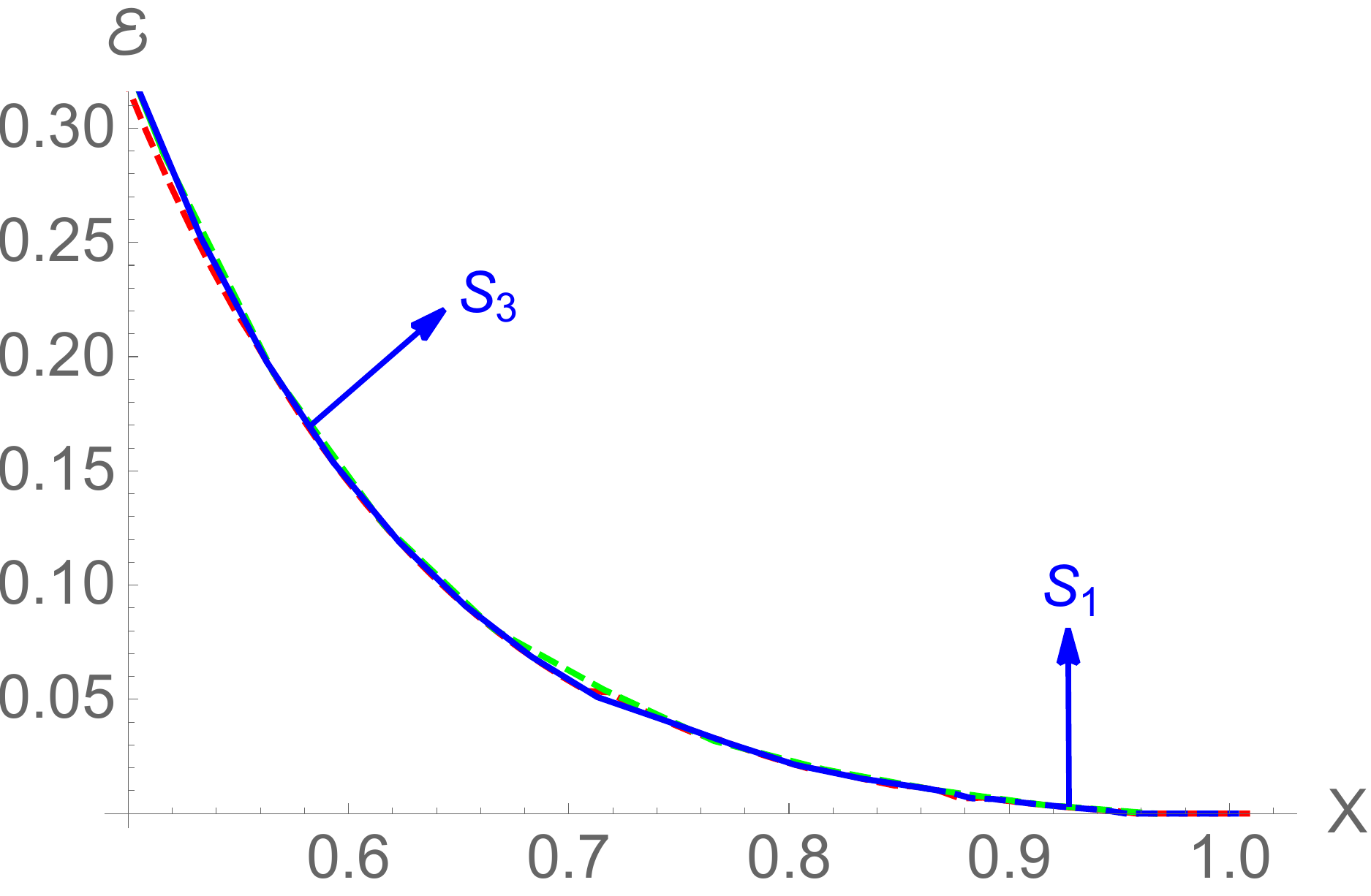}
\caption{\small Entanglement negativity $\mathcal{E}$ of two parallel strips as a function of $X$ for different values of $L$. Solid and dashed lines indicate $\mathcal{E}$ of $S_3$ and $S_1$ phases respectively. Here Red, green and blue curves correspond to $L = 0.4$,  $0.6$ and $0.8$ respectively.}
\label{XVsENVsLtwostripEMD}
\end{minipage}
\hspace{0.4cm}
\begin{minipage}[b]{0.5\linewidth}
\centering
\includegraphics[width=2.8in,height=2.3in]{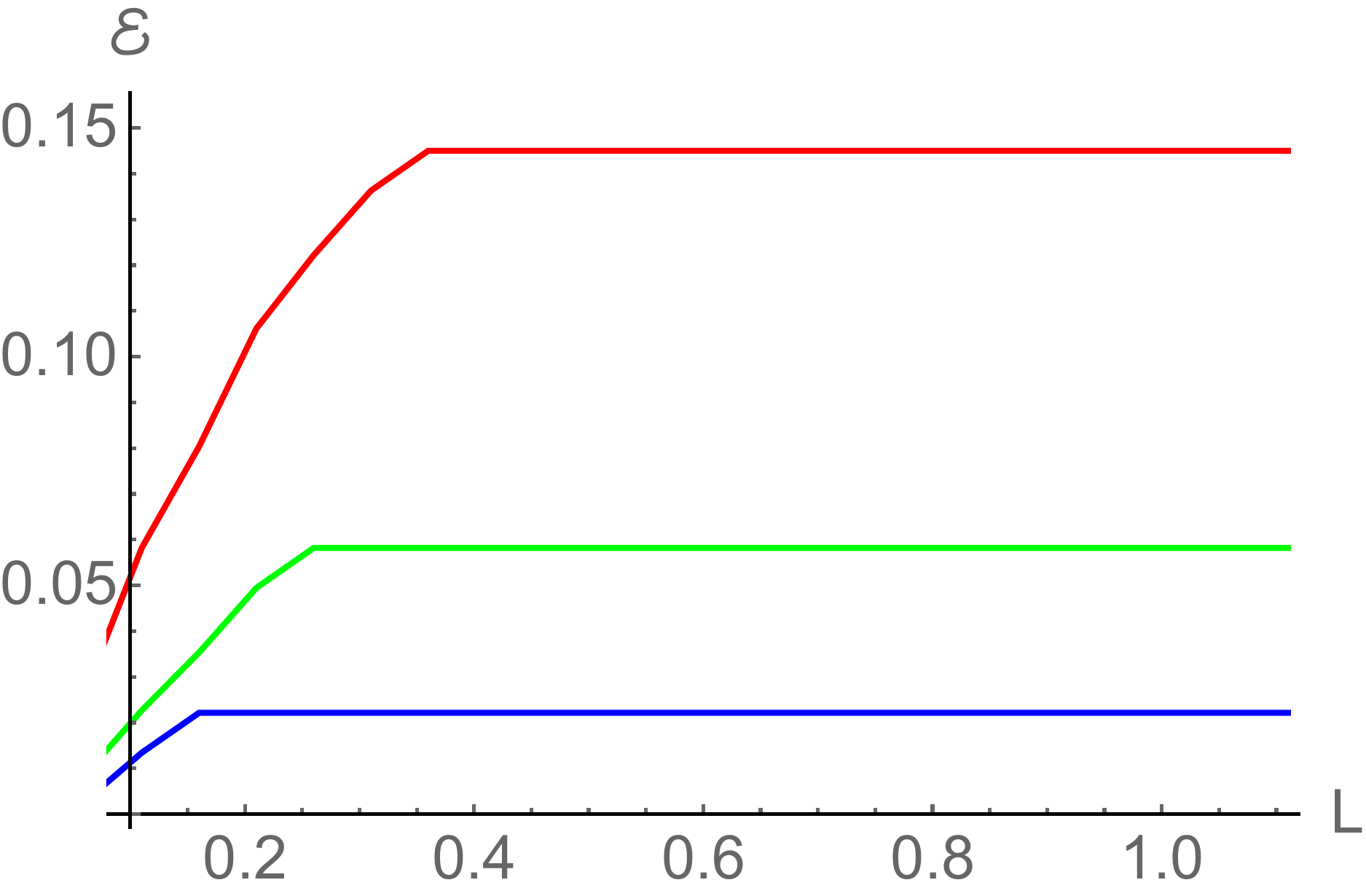}
\caption{\small Entanglement negativity $\mathcal{E}$ of two parallel strips as a function of $L$ for different values of $X$. Here Red, green and blue curves correspond to $X = 0.6$,  $0.7$ and $0.8$ respectively.}
\label{LVsENVsXtwostripEMD}
\end{minipage}
\end{figure}

The numerical results of the entanglement negativity for two strips are shown in Figures~\ref{XVsENVsLtwostripEMD} and \ref{LVsENVsXtwostripEMD}. It again turns out to be a monotonic function of $X$ and $L$. Notably, unlike the entanglement wedge and mutual information, it remains non-zero in the $S_1$ phase. In particular, the entanglement negativity is zero only when the separation between the two strips is larger than $L_{crit}$. This also implies entanglement negativity is trivially zero in $S_4$ phase (\textit{i.e.} $X>L_{crit}$, $L>L_{crit}$). In $S_2$ and $S_3$ phases, entanglement negativity is finite and positive. Importantly, the entanglement negativity remains continuous across the various phase transitions (though again exhibits a cusp) and it smoothly goes to zero at $X=L_{crit}$.

\section{Discussion and Conclusion}\label{DiscussionandConclusion}
In this work, we did a comprehensive analysis of pure and mixed state entanglement measures, such as the entanglement entropy, mutual information, entanglement wedge cross-section and entanglement negativity, in top-down as well as bottom-up holographic confining models. For the top-down case, we considered two models that are obtained by compactifying $D3$ and $D4$ branes on a circle whereas, for the bottom-up case, we considered the Einstein-Maxwell-dilaton model.

We first reproduced the known results of the entanglement entropy in these models. In particular, we reproduced the fact that with a single strip, there are two minimal area surfaces (connected and disconnected) which exchanged dominance as the size of the subsystem is varied. This provided a phase transition in the entanglement entropy at a critical length $L_{crit}$, at which the order of the entanglement entropy changed from $\mathcal{O}(N^2)$ to $\mathcal{O}(N^0)$. We then studied two equal strip entanglement phase diagram in the parameter space of $L$ and $X$, and found four distinct phases $\{S_1,S_2,S_3,S_4\}$. These four phases exchanged dominance as $L$ and $X$ are varied. The mutual information turns out to be a monotonic function of $X$ and $L$ in $S_2$ and $S_3$ phases and it smoothly goes to zero in $S_1$ and $S_4$ phase. Importantly, the order of mutual information may or may not change as the critical points are crossed. The entanglement wedge cross-section $E_W$, on the other hand, vanished discontinuously for large values of $X$ and $L$, and displayed non-analytic behaviour every time a critical point is crossed. This suggests that there can be other non-trivial length scales (apart from the natural length scale $L_{crit}$, that comes from the discontinuity in entanglement entropy) where non-analytic entanglement structure can appear in the confining theories. Unfortunately, unlike the entanglement entropy, lattice results for $E_W$ are not available yet. These results therefore might be considered as genuine predictions from holography. We moreover tested the inequality involving mutual information and entanglement wedge cross-section, and found that the latter always exceeds half of the former. We further discussed the entanglement negativity with one and two intervals using the prescription suggested in \cite{Chaturvedi:2016rft}. A straightforward implementation of this prescription in the confined phase suggested that the entanglement negativity is proportional to the entanglement entropy and, therefore, undergoes an order change at $L_{crit}$.  This suggests that, like the entanglement entropy, the entanglement negativity can also be used to probe confinement. This is again a new result and it needs to be independently verified by lattice calculations.

We end this discussion by pointing out a few directions in which the present work can be extended. It would be certainly interesting to compute the entanglement negativity in confined phases using the first holographic proposal \cite{Kudler-Flam:2018qjo,Kusuki:2019zsp} and independently check the validity of the above mentioned results. This would be a bit non-trivial, as one first needs to compute the backreaction of cosmic brane on the spacetime geometry. On the application side, it would also be interesting to compute the entanglement wedge cross-section and negativity after a global quantum quench in the confined phase and study the corresponding thermalization process, as this might provide useful information about the QGP formation in QCD. Another interesting direction to extend our work is to discuss the anisotropic effects on the entanglement wedge and negativity by including a background magnetic field in the lines of \cite{Bohra:2019ebj,Bohra:2020qom,Arefeva:2018hyo}, and use these entanglement measures to investigate (inverse) magnetic catalysis.

\section*{Acknowledgments}
The work of S.~M.~is supported by the Department of Science
and Technology, Government of India under the Grant Agreement number IFA17-PH207 (INSPIRE Faculty Award).


\end{document}